\documentclass[sigconf, screen, authorversion,nonacm]{acmart}

\input{macros}

\AtBeginDocument{%
  }

\setcopyright{acmcopyright}
\copyrightyear{2018}
\acmYear{2018}
\acmDOI{XXXXXXX.XXXXXXX}

\acmConference[Conference acronym 'XX]{Make sure to enter the correct
  conference title from your rights confirmation emai}{June 03--05,
  2018}{Woodstock, NY}
\acmPrice{15.00}
\acmISBN{978-1-4503-XXXX-X/18/06}

\begin{document}

\title{Diagrammatic Algebra of First Order Logic}

\author{Filippo Bonchi}
\email{filippo.bonchi@unipi.it}
\orcid{0000-0002-3433-723X}             %
\author{Alessandro Di Giorgio}
\email{alessandro.digiorgio@phd.unipi.it}
\orcid{0000-0002-6428-6461}
\affiliation{
  \institution{University of Pisa}            %
  \city{Pisa}
  \country{Italy}                    %
}

\author{Nathan Haydon}
\email{nathan.haydon@taltech.ee}
\orcid{0000-0002-1604-9832}
\author{Pawe\l{} Soboci\'{n}ski}
\email{pawel.sobocinski@taltech.ee}
\orcid{0000-0002-7992-9685}
\affiliation{
  \institution{Tallinn University of Technology}
  \city{Tallinn}
  \country{Estonia}
}

\begin{abstract}
  We introduce the calculus of neo-Peircean relations, a string diagrammatic extension of the calculus of binary relations that has the same expressivity as first order logic and comes with a complete axiomatisation. The axioms are obtained by combining two well known categorical structures: cartesian and linear bicategories.
\end{abstract}



\received{20 February 2007}
\received[revised]{12 March 2009}
\received[accepted]{5 June 2009}

\maketitle

\section{Introduction}

The modern understanding of first order logic ($\FOL$) %
is the result of an evolution with contributions from many philosophers and mathematicians. %
Amongst these, particularly relevant for our exposition is the calculus of relations ($\CR$) by Charles S. Peirce~\cite{peirce1897_the-logic-of-relatives}.
Peirce, inspired by De Morgan~\cite{de1860syllogism}, proposed a relational analogue of Boole's algebra~\cite{boole1847mathematical}: a rigorous mathematical language for combining relations with  operations governed by algebraic laws.

With the rise of first order logic, Peirce's calculus was forgotten until Tarski, who in~\cite{tarski1941calculus} recognised its algebraic flavour.
In the introduction to~\cite{tarski1988formalization}, written shortly before his death, Tarski put much emphasis on two key features of $\CR$: (a) its lack of quantifiers and (b) its sole deduction rule of substituting equals by equals. The calculus, however, comes with two great shortcomings: (c) it is strictly less expressive than $\FOL$ and (d) it is \emph{not} axiomatisable.

Despite these limitations, $\CR$ had ---and continues to have--- a great impact in computer science, e.g., in the theory of databases~\cite{codd1983relational} and in the semantics of programming languages~\cite{pratt1976semantical,hoare1986weakest,lassen1998relational,bird1996algebra,DBLP:journals/pacmpl/LagoG22a}.
Indeed, the lack of quantifiers avoids the usual burden of bindings, scopes and capture-avoid substitutions (see~\cite{DBLP:journals/fac/GabbayP02,pitts2013nominal,pfenning1988higher,hofmann1999semantical,fiore1999abstract,goncharov2023towards} for some theories developed to address specifically the issue of bindings). This feature, together with purely equational proofs, makes $\CR$ particularly suitable for proof assistants~\cite{pous2013kleene, pous2016automata, krauss2012regular}.

Less influential in computer science, there are two others quantifiers-free alternatives to $\FOL$ that are worth mentioning: first, \emph{predicate functor logic} ($\PF$)~\cite{QUINE1971309} that was thought by Quine as the first order logic analogue of combinatory logic~\cite{curry1958combinatory} for the $\lambda$-calculus; %
second, Peirce's \emph{existential graphs} ($\EG$s)~\cite{roberts1973_the-existential-graphs-of-charles-s.-peirce} and, in particular, its fragment named \emph{system $\beta$}. In this system $\FOL$ formulas are \emph{diagrams} and the deduction system consists of rules for their manipulation.  Peirce's work on $\EG$s remained unpublished during his lifetime.

Diagrams have been used as formal entities since the dawn of computer science, e.g.\
in the %
B\"ohm-Jacopini theorem~\cite{bohm1966flow}. More recently,
the spatial nature of mobile computations led Milner to move from the traditional term-based syntax of process calculi to bigraphs~\cite{milner2009space}. Similarly, the impossibility of copying quantum information and, more generally, the paradigm-shift of treating data as a physical resource (see e.g.~\cite{orchard2019quantitative,gaboardi2016combining}), has led to the use~\cite{BaezErbele-CategoriesInControl,DBLP:journals/pacmpl/BonchiHPSZ19,Bonchi2015,CoeckeDuncanZX2011,Fong2015,DBLP:journals/corr/abs-2009-06836,Ghica2016,DBLP:conf/lics/MuroyaCG18,Piedeleu2021,DBLP:journals/jacm/BonchiGKSZ22} of \emph{string diagrams}~\cite{joyal1991geometry,Selinger2009} as syntax.
 String diagrams, formally arrows of a freely generated  symmetric (strict) monoidal category, combine the rigour of traditional terms with a visual and intuitive graphical representation.
 Like traditional terms, they can be equipped with a compositional semantics.

\smallskip

In this paper, we introduce the calculus of \emph{neo-Peircean relations}, a string diagrammatic account of $\FOL$ that has several key features:

\begin{enumerate}
\item Its diagrammatic syntax is closely related to Peirce's $\EG$s, but it can also be given through a context free grammar  equipped with an elementary type system;
\item It is quantifier-free and, differently than $\FOL$, its compositional semantics can be given by few simple rules: see \eqref{fig:semantics};
\item Terms and predicates are not treated as separate syntactic and semantic entities;
\item Its sole deduction rule is substituting equals by equals, like $\CR$, but differently, it features a complete axiomatisation;
\item The axioms are those of well-known algebraic structures, also occurring in different fields such as linear algebra \cite{interactinghopf} or quantum foundations \cite{CoeckeDuncanZX2011};
\item It allows for compositional encodings of $\FOL$, $\CR$ and $\PF$;
\item String diagrams disambiguate interesting corner cases where traditional $\FOL$ encounters difficulties. One perk is that we allow empty models ---forbidden in classical treatments--- leading to (slightly) more general G\"odel completeness;
\item The corner case of empty models coincides with \emph{propositional} models and in that case our axiomatisation simplifies to the deep inference Calculus of Structures~\cite{DBLP:phd/de/Brunnler2003,guglielmi2007system}. %
\end{enumerate}
By returning to the algebraic roots of logic we preserve $\CR$'s benefits (a) and (b) while overcoming its limitations (c) and (d). %

\begin{figure*}
		\begin{minipage}{0.68\textwidth}
		$
		\begin{array}{ll}
		  \Circ{c}  \, \Coloneqq\!\!\!\!  & \copierCirc[+]       \mid
										 \discardCirc[+]      \mid
														 \boxCirc[+]{R}       \mid
														 \codiscardCirc[+]    \mid
														 \cocopierCirc[+]     \mid
														 \emptyCirc[+]        \mid
														 \idCirc[+]           \mid
														 \symmCirc[+]         \mid
														 \seqCirc[+]{c}{c}    \mid
														 \tensorCirc[+]{c}{c} \mid  \notag \\[14pt]
						& \copierCirc[-]        \mid
										  \discardCirc[-]       \mid
											\boxOpCirc[-]{R}        \mid
											\codiscardCirc[-]     \mid
											\cocopierCirc[-]      \mid
											\emptyCirc[-]         \mid
											\idCirc[-]            \mid
											\symmCirc[-]          \mid
											\seqCirc[-]{c}{c}     \mid
											\tensorCirc[-]{c}{c}
		\end{array}
		$
		\end{minipage}
		\hfill
		\begin{minipage}{0.3\textwidth}
		\scalebox{0.8}{$
		\xymatrix@C=10pt@R=10pt{
			\copierCirc[+]\;\; \discardCirc[+]  \ar@{<->}[dd]^{\text{lin. adj.}} \ar[rr]_{\text{left adj.}} & &
			\cocopierCirc[+] \; \;  \codiscardCirc[+]    \ar@{<->}[dd]_{\text{lin. adj.}}    \ar@{--}@(ul,ur)[ll]_{\text{spec. Frob.}}
			\\ \\
				\cocopierCirc[-] \; \;  \codiscardCirc[-] \ar@{--}@/^1pc/[uu]^{\text{lin. Frob.}} &&
				\copierCirc[-]\;\; \discardCirc[-] \ar[ll]_{\text{right adj.}}  \ar@{--}@(dl,dr)[ll]^{\text{spec. Frob.}} \ar@{--}@/_1pc/[uu]_{\text{lin. Frob.}}
		}
		$}
		\end{minipage}
	\caption{Diagrammatic syntax of $\NPR$ (left) and a summary of its axioms (right)}\label{fig:diagsyntax}
\end{figure*}
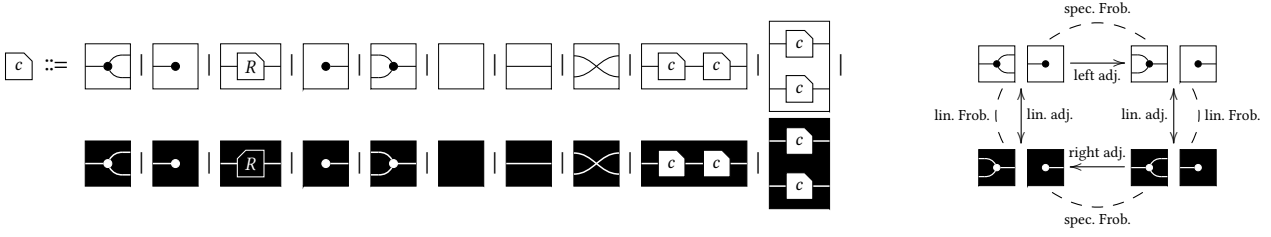

\paragraph{Cartesian syntax} To ease the reader into this work, we show how traditional terms appear as string diagrams. Consider a signature $\sign$ consisting of a unary symbol $f$ and two binary symbols $g$ and $h$. The term $h(\,g(f(x_3),f(x_3)), x_1\,)$ corresponds to the string diagram on the left below.
\[ 
    \InputIfFileExists{cartesianTerm.tikz}{}{\input{tikz/cartesianTerm.tikz}}
 \qquad \qquad 
    \InputIfFileExists{cartesianTerm2.tikz}{}{\input{tikz/cartesianTerm2.tikz}}
 \]
A difference wrt traditional syntax tree is the explicit treatment of copying and discarding. The discharger $\scalebox{0.8}{\discardCirc[+]}$ informs us that the variable $x_2$ does not appear in the term; the copier $\scalebox{0.8}{\copierCirc[+]}$ makes clear that the variable $x_3$ is shared by two sub-terms. The string diagram on the represents the same term if one admits the equations
\begin{equation}\label{eq:natintro}\tag{Nat}

    \InputIfFileExists{axioms/cb/plus/copierLaxNat1.tikz}{}{\input{tikz/axioms/cb/plus/copierLaxNat1.tikz}}
 = 
    \InputIfFileExists{axioms/cb/plus/copierLaxNat2.tikz}{}{\input{tikz/axioms/cb/plus/copierLaxNat2.tikz}}
\; \text{ and }\; 
    \InputIfFileExists{axioms/cb/plus/discardLaxNat.tikz}{}{\input{tikz/axioms/cb/plus/discardLaxNat.tikz}}
 = \discardCirc[+]\text{ .}
\end{equation}
Fox~\cite{fox1976coalgebras} showed that \eqref{eq:natintro} together with axioms asserting that copier and discard form a \emph{comonoid} (\eqref{ax:comPlusAssoc}, \eqref{ax:comPlusUnit}, \eqref{ax:comPlusComm} in Fig.~\ref{fig:cb axioms}) force the monoidal category of string diagrams to be  \emph{cartesian} ($\otimes$ is the categorical product): actually, it is the \emph{free} cartesian category on $\sign$. %

\paragraph{Functorial semantics} The work of Lawvere~\cite{LawvereOriginalPaper} illustrates the deep connection of syntax with semantics, explaining why cartesian syntax is so well-suited to functional structures, but also hinting at its limitations %
when denoting other structures, e.g. relations. Given an algebraic theory $\T{T}$ in the universal algebraic sense, i.e., a signature $\sign$ with a set of equations $E$, one can freely generate a cartesian category $\Cat{L}_{\T{T}}$. %
\emph{Models} --in the standard algebraic
sense-- are in one-to-one correspondence with cartesian functors $\mathcal{M}$ from  $\Cat{L}_{\T{T}}$ to $\Cat{Set}$, the category of sets and functions. More generally, models of the theory in any cartesian category $\Cat{C}$ are cartesian functors $\mathcal{M}\colon \Cat{L}_{\T{T}} \to \Cat{C}$.
By taking $\Cat{C}$ to be $\Relp$, the category of sets and relations, one could wish to use the same approach for relational theories but any such attempt fails immediately since the cartesian product of sets is not the categorical product in $\Relp$.

\paragraph{Cartesian bicategories} An evolution of Lawvere's approach for relational structures is developed in~\cite{DBLP:journals/corr/abs-1711-08699,GCQ,seeber2020logical}. Departing from cartesian syntax, it uses string diagrams generated by the \emph{first} row of the grammar in Fig.~\ref{fig:diagsyntax},
where $R$ is taken from a monoidal signature $\sign$ -- a set of symbols equipped with both an arity and also a \emph{coarity} -- and can be thought of as akin to relation symbols of $\FOL$. The diagrams are subject to the laws of cartesian bicategories~\cite{carboni1987cartesian} in Fig.~\ref{fig:cb axioms}: $\scalebox{0.8}{\copierCirc[+]}$ and $\scalebox{0.8}{\discardCirc[+]}$ form a comonoid, but the category of diagrams is not cartesian since the equations in \eqref{eq:natintro} hold laxly (\eqref{ax:comPlusLaxNat}, \eqref{ax:discPlusLaxNat}). The diagrams $\scalebox{0.8}{\cocopierCirc[+]}$ and $\scalebox{0.8}{\codiscardCirc[+]}$ form a \emph{monoid} (\eqref{ax:monPlusAssoc}, \eqref{ax:monPlusUnit}, \eqref{ax:monPlusComm}) and are right \emph{adjoint} to copier and discard. Monoids and comonoids together satisfy \emph{special Frobenius} equations (\eqref{ax:plusSpecFrob},\eqref{ax:plusFrob}). The category of diagrams $\CB$ is the free cartesian bicategory generated by $\sign$ and, like in Lawvere's functorial semantics, models are morphisms of cartesian bicategories $\mathcal{M}\colon \CB \to \Relp$. Importantly, the laws of cartesian bicategories provide a complete axiomatisation for $\Relp$, meaning that $c,d$ in $\CB$ are provably equal with the laws of cartesian bicategories iff $\mathcal{M}(c)=\mathcal{M}(d)$ for all models $\mathcal{M}$.

\begin{wrapfigure}{R}{0.10\textwidth}
    \vspace{-12pt}
    $
    \InputIfFileExists{regularFormula.tikz}{}{\input{tikz/regularFormula.tikz}}
$
    \vspace{-12pt}
\end{wrapfigure}
The (co)monoid structures allow one to express existential quantification: for instance, the $\FOL$ formula $\exists x_2. P(x_1,x_2) \wedge Q(x_2)$ is depicted as the diagram on the right. The expressive power of $\CB$ is, however, limited to the existential-conjunctive fragment of $\FOL$. %

\paragraph{Cocartesian bicategories} To express the universal-disjunctive fragment, we consider the category $\CCB$  of string diagrams generated by the \emph{second} row of the grammar in Fig.~\ref{fig:diagsyntax} %
and subject to the laws of cocartesian bicategories in Fig.~\ref{fig:cocb axioms}:  those of cartesian bicategories but with the reversed order $\geq$. The diagrams of $\CCB$ are photographic negative of those in $\CB$.
 To explain this change of colour, note that sets and relations form \emph{another} category: $\Relm$.  Composition $\seq[-]$ in $\Relm$ is the De Morgan dual of the usual relational composition: $R\seq[+]S \defeq \{(x,z) \mid \exists y . (x,y)\in R \wedge (y,z)\in S\}$ while $R\seq[-]S \defeq \{(x,z) \mid \forall y . (x,y)\in R \vee (y,z)\in S\}$. While $\Relp$ is a cartesian bicategory, $\Relm$ is \emph{cocartesian}. Interestingly, the ``black'' composition $\seq[-]$ was used in Peirce's approach~\cite{peirce1883_studies-in-logic.-by-members-of-the-johns-hopkins-university} to relational algebra.%

\medskip

Just as $\CB$ is complete with respect to $\Relp$, dually, $\CCB$ is complete wrt $\Relm$. The former accounts for the existential-conjunctive fragment of $\FOL$; the latter for its universal-disjunctive fragment. This raises a natural question:

\smallskip
\begin{center}
How do the white and black structures combine to form a complete account of first order logic?
\end{center}

\paragraph{Linear bicategories} %
Although $\Relp$ and $\Relm$ have the same objects and arrows, there
are two different compositions ($\seq[+]$ and $\seq[-]$). The appropriate categorical structures to deal with these situations are \emph{linear bicategories} introduced in \cite{cockett2000introduction} as a horizontal categorification of linearly distributive categories \cite{de1991dialectica,cockett1997weakly}. The laws of linear bicategories are in Fig.~\ref{fig:closed lin axioms}: the key law is \emph{linearly distributivity} of $\seq[+]$ over $\seq[-]$ (\eqref{ax:leftLinDistr}, \eqref{ax:rightLinDistr}), that was already known to hold for relations since the work of Peirce \cite{peirce1883_studies-in-logic.-by-members-of-the-johns-hopkins-university}.
Another crucial property observed by Peirce is that for any relation $R\subseteq X \times Y$, the relation $\rla{R}\subseteq Y \times X\defeq \{(y,x) \mid (x,y) \notin R\}$ is its \emph{linear adjoint}. This operation has an intuitive graphical representation: given $\Circ{c}$, take its mirror image $\CircOp[+]{c}$ and then its photographic negative $\CircOp[-]{c}$. For instance, the linear adjoint of $\boxCirc[+]{R}$ is $\boxOpCirc[-]{R}$.

\paragraph{First order bicategories} The final step is to characterise how cartesian, cocartesian and linear bicategories combine: %
\textit{(i)} white and black (co)monoids are linear adjoints that \textit{(ii)} satisfy a ``linear'' version of the Frobenius law. We dub the result \emph{first order bicategories}.
	We shall see that this is a complete axiomatisation for first order logic, yet all of the algebraic machinery is compactly summarised at the right of Fig.~\ref{fig:diagsyntax}.

\paragraph{Functorial semantics for first order theories} In the spirit of functorial semantics, we take the free first order bicategory $\LCB[\T{T}]$ generated by a theory $\T{T}$ and observe that models of $\T{T}$ in a first order bicategory $\Cat{C}$ are
morphisms %
$\mathcal{M}\colon \LCB[\T{T}] \to \Cat{C}$. Taking $\Cat{C}=\Rel$, the first order bicategory of sets and relations, these are models in the sense of $\FOL$ with one notable exception: in $\FOL$ models with the empty domain are forbidden. %
As we shall wee, theories with empty models are exactly the propositional theories. %

\paragraph{Completeness}  We prove that the laws of first order bicategories provide a complete axiomatisation for first order logic. The proof is a diagrammatic adaptation of Henkin's proof \cite{henkin_1949} of G\"odel's completeness theorem. However, in order to properly consider models with an empty domain, we make a slight additional step to go beyond G\"odel completeness.

\paragraph{A taste of diagrammatic logic} Before we introduce the calculus of neo-Peircean relations, we start with a short worked example to give the reader a taste of using the calculus to prove a non-trivial result of first order logic. Doing so lets us illustrate the methodology of proof within the calculus, which is sometimes referred to as diagrammatic reasoning or string diagram surgery.

\begin{wrapfigure}{R}{0.19\textwidth}
    \vspace{-12pt}
    $\!\!\!
    \InputIfFileExists{existsForallR.tikz}{}{\input{tikz/existsForallR.tikz}}
 \seminclusion 
    \InputIfFileExists{forallExistsR.tikz}{}{\input{tikz/forallExistsR.tikz}}
$
    \vspace{-12pt}
\end{wrapfigure}
Let $R$ be a symbol with arity $2$ and coarity $0$. %
The two diagrams on the right correspond to $\FOL$ formulas
$\exists x.\,\forall y.\,R(x,y)$ and
$\forall y.\,\exists x.\,R(x,y)$: see \S~\ref{sec:fol} for a dictionary of translating between $\FOL$ and diagrams. %
It is well-known that $\exists x.\,\forall y.\,R(x,y) \models \forall y.\,\exists x.\,R(x,y)$, i.e.\ in any model, if the first formula evaluates to true then so does the second. Within our calculus, this statement is expressed as the above inequality.
This can be proved by mean of the axiomatisation we introduce in this work: %
\input{tikz/proofs/forallExists.tex}
The central step relies on the particularly good behaviour of \emph{maps}, intuitively those relations that are functional. In particular $\scalebox{0.8}{\discardCirc[+]}$ is an  example. The details are not important at this stage.

\paragraph{Synopsis} We begin by recalling $\CR$ in \S~\ref{sec:calculusrelations}. The calculus of neo-Peircean relations is introduced in \S~\ref{sec:NPR}, together with the statement of our main result (Theorem~\ref{thm:completeness}). We recall (co)cartesian and linear bicategories in \S~\ref{sec:cartesianbi} and \S~\ref{sec:linbic}. The categorical structures most important for our work are first-order bicategories, introduced in \S~\ref{sec:fobic}. In \S~\ref{sec:theories} we consider first order theories, the diagrammatic version of the deduction theorem (Theorem~\ref{th:deduction}) and some subtle differences with $\FOL$ that play an important role on the proof of completeness in \S~\ref{sec:completeness}. Translations of  $\CR$ and $\FOL$ into the calculus of neo-Peircean relations are given in \S~\ref{ssec:CRrevisited} and \S~\ref{sec:fol}. The encoding of $\PF$ and additional material omitted due to space restrictions are in Appendix~\ref{app:additional}. All proofs are in the remaining appendices.

\section{The Calculus of Binary Relations}\label{sec:calculusrelations}
The calculus %
of binary relations, in an original presentation given by Peirce in~\cite{peirce1883_studies-in-logic.-by-members-of-the-johns-hopkins-university}, features two forms of relational compositions $\seq[+][][]$ and  $\seq[-][][]$, defined for all relations $R\subseteq X \times Y$ and $S \subseteq Y \times Z$ as
\begin{equation}\label{eq:seqRel}
   \begin{aligned}
      \seq[+][R][S] &\defeq \{(x,z) \mid \exists y\!\in\! Y \,.\, (x,y)\in R  \wedge (y,z)\in S\}\subseteq X\times Z \; \text{and} \\
      \seq[-][R][S] &\defeq \{(x,z) \mid \forall y\!\in\! Y \,.\, (x,y)\in R \vee (y,z)\in S\}\subseteq X\times Z   
   \end{aligned}
   \end{equation}
with units the equality and the difference relations respectively, defined for all sets $X$ as
\begin{equation}\label{eq:idRel}
      \id[+][X]\! \defeq \! \{(x,y) \!\mid\! x=y\}\!\subseteq\! X \times X \text{ and }
      \id[-][X] \!\defeq \! \{(x,y) \!\mid \! x \neq y\}\! \subseteq\! X \times X \text{.}   
\end{equation}
Beyond the usual %
union $\cup$, intersection $\cap$, and their units $\bot$ and $\top$, the calculus also features two unary operations $\op{(\cdot)}$ and $\nega{(\cdot)}$ denoting the opposite and the complement:
$\op{R} \defeq \{ (y,x) \mid (x,y)\in R \}$ and $\nega{R} \defeq \{ (x,y) \mid (x,y) \notin R \}$.
In summary, its syntax is given by the following context free grammar
\begin{equation}\tag{$\CRS$} \label{eq:calculusofrelation}
   \begin{array}{rc c@{\;\;\; \mid \;\;}c@{\;\; \mid \;\;\;}c@{\;\;\; \mid \;\;}c@{\;\; \mid \;\;\;}c@{\;\;\; \mid \;\;\;}c}
   E & \Coloneqq & R      & \id[+] & \seq[+][E][E] & \id[-] &  \seq[-][E][E] & \\
     &     & \op{E} & \top   & E \cap E      & \bot   &  E \cup E      & \nega{E}
   \end{array}
   \end{equation}
where $R$ is taken from a given set $\Sigma$ of generating symbols.
The semantics is defined wrt a  \emph{relational interpretation} $\interpretation$, that is, a set $X$ together with a binary relation $\rho(R)\subseteq X \times X$ for each $R\in \Sigma$.
\begin{equation}\label{eq:sematicsExpr}
   \hspace*{-1.6em}
   {\small
   \arraycolsep=3pt
   \begin{array}{ccc}
      \arraycolsep=1.5pt
      \begin{array}{rcl}
         \dsemRel{R} & \defeq & \rho(R) \\
         \dsemRel{\op{E}} & \defeq & \op{\dsemRel{E}} \\
         \dsemRel{\nega{E}} & \defeq & \nega{\dsemRel{E}} \\
         & &
      \end{array}
      &
      \arraycolsep=1.5pt
      \begin{array}{rcl}
          \dsemRel{\id[+]} & \defeq & \id[+][X] \\
          \dsemRel{\id[-]} & \defeq & \id[-][X] \\
          \dsemRel{\bot} & \defeq & \varnothing \\
          \dsemRel{\top} & \defeq & X \times X
      \end{array}
      &
      \arraycolsep=1.5pt
      \begin{array}{rcl}
          \dsemRel{E_1 \seq[+] E_2} & \defeq & \dsemRel{E_1} \seq[+] \dsemRel{E_2} \\
          \dsemRel{E_1 \seq[-] E_2} & \defeq & \dsemRel{E_1} \seq[-] \dsemRel{E_2} \\
          \dsemRel{E_1 \cup E_2} & \defeq & \dsemRel{E_1} \cup \dsemRel{E_2} \\
          \dsemRel{E_1 \cap E_2} & \defeq & \dsemRel{E_1} \cap \dsemRel{E_2}
      \end{array}
  \end{array}
   }
\end{equation}
Two expressions $E_1$, $E_2$ are said to be \emph{equivalent}, written $E_1 \equiv_{\CR}E_2$, if and only if $\dsemRel{E_1} = \dsemRel{E_2}$, for all interpretations $\interpretation$. Inclusion, denoted by $\minorExpression$, is defined analogously by replacing $=$ with $\subseteq$.
For instance, the following inclusions hold, witnessing the fact that $\seq[+][][]$ \emph{linearly distributes} over $\seq[-][][]$.
\begin{equation}\label{eq:distributivityExpres}
\seq[+][R][(\seq[-][S][T])] \minorExpression  (R\seq[+][][] S) \seq[-][][] T  \qquad \seq[+][(\seq[-][R][S])][T] \minorExpression  \seq[-][R][(\seq[+][S][T])]
\end{equation}
Along with the boolean laws, in `Note B' \cite{peirce1883_studies-in-logic.-by-members-of-the-johns-hopkins-university} Peirce states \eqref{eq:distributivityExpres} and stresses its importance. However, since $\seq[-][R][S] \equiv_{\CR} \nega{\seq[+][\nega{R}][\nega{S}]}$ and $\id[-] \equiv_{\CR} \nega{\id[+]}$, both $\seq[-]$ and $\id[-]$ are often considered redundant, for instance by Tarski~\cite{tarski1941calculus} and much of the later work. %

Tarski asked whether $\equiv_{\CR}$ can be axiomatised, i.e., is there a basic set of laws from which one can prove all the valid equivalences? Unfortunately, %
there is no finite complete axiomatisations for the whole calculus~\cite{monk} nor for several fragments, e.g.,~\cite{hodkinson2000axiomatizability,redko1964defining,freyd1990categories,doumane2020non,DBLP:conf/stacs/Pous18}. %

Our work returns to the same problem, but from a radically different perspective: we see the calculus of relations as a sub-calculus of a more general system for arbitrary (i.e.\ not merely binary) relations.
The latter is strictly more expressive than $\CRS$ -- actually it is as expressive as first order logic ($\FOL$)-- but allows for an elementary complete axiomatisation based on the interaction of two influential algebraic structures%
: that of linear bicategories and cartesian bicategories. %

\begin{table*}[ht]
	\caption{Typing rules (top); inductive definitions of syntactic sugar (middle); structural congruence (bottom)}
	\label{fig:typingrules}
	\label{fig:sugar}
	\label{fig:freestricmmoncatax}
	{\tiny
	\begin{tabular}{c}
		\toprule
		$
		\begin{array}{ll}
			\begin{array}{c}
				{\id[][0]\colon 0 \to 0} \qquad\quad {\id[][1] \colon 1 \to 1} \qquad\quad {\symm[][1][1] \colon 2 \to 2} \\
				\copier[][1]\colon 1 \to 2 \qquad \discard[][1]\colon 1 \to 0 \qquad \cocopier[][1]\colon 2 \to 1 \qquad \codiscard[][1]\colon 0 \to 1
			\end{array}
			&
			\inferrule{\ari(R) =n \and \coar(R)=n}{R^\circ \colon n \to m} \qquad \inferrule{\ari(R) =n \and \coar(R)=m}{R^\bullet \colon m \to n} \qquad
			\inferrule{c \colon n_1 \to m_1 \and d \colon n_2 \to m_2}{\tensor[][c][d] \colon n_1 + n_2 \to m_1 + m_2}  \qquad
			\inferrule{c \colon n \to m \and d \colon m \to o}{\seq[][c][d] \colon n \to o}
		\end{array}
		$
		\\
		\midrule
		$\arraycolsep=3pt
		\begin{array}{llll}
		\begin{array}{ll}
		\copier[][0]=\id[][0] &
		\copier[][n+1]=(\copier[][1]\tensor[]\copier[][n] )\seq[] (\id[][1]\tensor[] \symm[][1][n] \tensor[] \id[][n]) \\
		\cocopier[][0]=\id[][0] &
		\cocopier[][n+1]= (\id[][1]\tensor[] \symm[][1][n] \tensor[] \id[][n]) \seq[]
		(\cocopier[][1]\tensor[]\cocopier[][n] )
		\end{array}
		&
		\begin{array}{ll}
		\discard[][0]=\id[][0] &
		\discard[][n+1]=\discard[][1]\tensor[] \discard[][n] \\
		\codiscard[][0]=\id[][0] &
		\codiscard[][n+1]= \codiscard[][1]\tensor[] \codiscard[][n]
		\end{array}
		&
		\begin{array}{l}
		\id[][0]=\id[][0] \\
		\id[][n+1]=\id[][1] \tensor[] \id[][n] \\
		\end{array}
		&
		\begin{array}{ll}
			\symm[][0][0]=\id[][0] &  \symm[][1][0]=\symm[][0][1]=\id[][1] \\
			\symm[][1][n+1]=(\symm[][1][n] \tensor[] \id[][1]) \seq[](\id[][n] \tensor[] \symm[][1][1]) & \symm[][m+1][n]=(\id[][1] \tensor[] \symm[][m][n]) \seq[] (\symm[][1][n] \tensor[] \id[][m])
		\end{array}
		\end{array}$ \\
		\midrule
		\begin{tabular}{@{}c@{\;\;\;\;}c@{\;\;\;\;}c@{\;\;\;\;}c@{\;\;\;\;}c@{\;\;\;\;}c@{\;\;\;\;}c@{}}
		$\seq[][a][(\seq[][b][c])] = \seq[][(\seq[][a][b])][c]$
		&
		$\seq[][\id[][n]][c]=c=\seq[][c][\id[][m]]$
		&
		$\tensor[][(\tensor[][a][b])][c] = \tensor[][a][(\tensor[][b][c])]$
		&
		$\tensor[][\id[][0]][c] = c = \tensor[][\id[][0]][c] $
		&
		$\seq[][(\tensor[][a][b])][ (\tensor[][c][d])] = \tensor[][(\seq[][a][c] )][(\seq[][b][d])]$
		&
		$\symm[][1][1] \seq[][][] \symm[][1][1] = \id[][2]$
		&
		$(c \tensor[][][] \id[][o]) \seq[][][] \symm[][m][o]  = \symm[][n][o] \seq[][][] (\id[][o] \tensor[][][] c)$
		\end{tabular}
		\\
		\bottomrule
	\end{tabular}
	}
\end{table*}

\section{Neo-Peircean Relations}\label{sec:NPR} %
Here we introduce the calculus of \emph{neo-Peircean relations} ($\NPR$). %

The first step is to move from binary relations $R\subseteq X \times X$ to relations $R\subseteq X^n \times X^m$ where, for any $n \in \nat$, $X^n$ denotes the set of row vectors $(x_1, \dots, x_n)$ with all $x_i\in X$. In particular, $X^0$ is the one element set $\singleton\defeq\{\star\}$. Considering this kind of relations allows us to identify two novel fundamental constants: the \emph{copier} $\copier[+][X] \subseteq X \times X^2$ which is the diagonal function $\langle \id[+][X], \id[+][X]\rangle \colon X \to X \times X$ (considered as a relation) and the \emph{discharger} $\discard[+][X] \subseteq X \times \singleton$ which is, similarly, the unique function from $X$ to $\singleton$. By combining them with opposite and complement we obtain, in total, 8 basic relations.
\begin{equation}\label{eq:comonoidsREL}
	\hspace*{-1em}
		\begin{tabular}{rcl rcl}
			$\copier[+][X]$   & $\!\!\!\!\!\defeq\!\!\!\!\!$ & $\{(x, \; (y,z)) \mid x=y \wedge x=z\}$  	    &  $\discard[+][X]$   & $\!\!\!\!\!\defeq\!\!\!\!\!$ & $\{(x, \star) \mid x\in X\}$ \\
			$\cocopier[+][X]$ & $\!\!\!\!\!\defeq\!\!\!\!\!$ & $\{((y,z),\; x) \mid x=y \wedge x=z\}$         &  $\codiscard[+][X]$ & $\!\!\!\!\!\defeq\!\!\!\!\!$ &  $\{(\star,x) \mid x\in X\}$ \\
			\midrule
			$\copier[-][X]$   & $\!\!\!\!\!\defeq\!\!\!\!\!$ & $\{(x, \; (y,z)) \mid x\neq y \vee x \neq z\}$   &  $\discard[-][X]$   & $\!\!\!\!\!\defeq\!\!\!\!\!$ & $\varnothing$ \\
			$\cocopier[-][X]$ & $\!\!\!\!\!\defeq\!\!\!\!\!$ & $\{((y,z), \; x  ) \mid x\neq y \vee x \neq z\}$ &  $\codiscard[-][X]$ & $\!\!\!\!\!\defeq\!\!\!\!\!$ & $\varnothing$
		\end{tabular}
\end{equation}
Together with $\id[+][X]$ and $\id[-][X]$ and the compositions $\seq[+]$ and $\seq[-]$ from~\eqref{eq:idRel}, there are black and white \emph{symmetries}: $\symm[+][X][Y] \defeq \{(\;(x,y), (y,x)\;) \mid x\in X, y \in Y\} %
$ and $\symm[-][X][Y] \defeq \nega{\symm[+][X][Y]}$.  The calculus does \emph{not} feature the boolean operators nor the opposite and the complement: these can be derived using the above structure and two \emph{monoidal products} $\tensor[+]$ and $\tensor[-]$, defined for $R\subseteq X \times Y$ and $S \subseteq V \times W$  as
\begin{equation}\label{eq:tensorREL}
	\begin{tabular}{rcl}
	$R \tensor[+]S$ & $\defeq$ & $\{ ( \,(x,v), (y,w) \,) \mid (x,y)\in R \wedge (v,w) \in S \}$\\
	$R \tensor[-]S$ & $\defeq$ & $\{ ( \,(x,v), (y,w) \,) \mid (x,y)\in R \vee (v,w) \in S \}$ \text{.}
	\end{tabular}
\end{equation}

\noindent\paragraph{Syntax}
Terms are defined by the following context free grammar
\begin{equation}\tag{$\NPR$}\label{eq:syntax} \arraycolsep=1.1pt
	\begin{array}{rc c@{\; \mid \;}c@{\; \mid \;}c@{\; \mid \;}c@{\; \mid \;}c@{\; \mid \;}c@{\; \mid \;}c@{\; \mid \;}c@{\; \mid \;}c@{\; \mid \;}c}
		\!\!c  & \Coloneqq  & \copier[+][1] & \discard[+][1] & R^\circ & \codiscard[+][1] & \cocopier[+][1] & \id[+][0] & \id[+][1] & \symm[+][1][1] & c \seq[+] c & c \tensor[+] c \mid \\
		   &            & \copier[-][1] & \discard[-][1] & R^\bullet & \codiscard[-][1] & \cocopier[-][1] & \id[-][0] & \id[-][1] & \symm[-][1][1] & c \seq[-] c & \!\!\! c \tensor[-] c \\
	  \end{array}
  \end{equation}
where $R$, like in $\CRS$, belongs to a fixed set $\sign$ of \emph{generators}. Differently than in $\CRS$, each $R\in \sign$ comes with two natural numbers: arity $\ari(R)$ and coarity $\coar(R)$. The tuple $(\sign, \ari, \coar)$, usually simply $\sign$, is a \emph{monoidal signature}. Intuitively, every $R\in \sign$ represents some relation $R\subseteq X^{\ari(R)} \times X^{\coar(R)}$. %

In the first row of~\eqref{eq:syntax} there are eight constants and two operations: white composition ($\seq[+]$) and white monoidal product ($\tensor[+]$). These, together with identities ($\id[+][0]$ and $\id[+][1]$) and symmetry ($\symm[+][1][1]$) are typical of symmetric monoidal categories. Apart from $R^{\switchLabelS{+}}$, the constants are the copier ($\copier[+][1]$), discharger ($\discard[+][1]$) and their opposite cocopier ($\cocopier[+][1]$) and codischarger ($\codiscard[+][1]$). The second row contains the ``black'' versions of the same constants and operations. %
Note that our syntax does not have variables, no quantifiers, nor the usual associated meta-operations like capture-avoiding substitution.

We shall refer to the terms generated by the first row as the \emph{white fragment}, while to those of second row as the \emph{black fragment}. Sometimes, we use the gray colour to be agnostic wrt white or black. The rules in top of Table \ref{fig:typingrules} assigns to each term at most one type $n \to m$. We consider only those terms that can be typed. For all $n,m \in \nat$, $\id[][n]\colon n \to n$, $\symm[][n][m] \colon n+m \to m+n$, $\copier[][n]\colon n \to n+n$, $\cocopier[][n]\colon n+n \to n$, $\discard[][n]\colon n \to 0$ and $\codiscard[][n]\colon 0 \to n$ are as in middle of Table~\ref{fig:sugar}. %

\paragraph{Semantics}
As for $\CRS$, the semantics of $\NPR$ needs an interpretation $\interpretation=(X,\rho)$: a set $X$, the \emph{semantic domain}, and $\rho(R) \subseteq X^{\ari(R)} \times X^{\coar(R)}$ for each $R\in\Sigma$. The semantics of terms is: %
\begin{equation}\label{fig:semantics}\arraycolsep=1.4pt%
	{\footnotesize
	\begin{array}{l}
	\begin{array}{l@{\quad}l@{\quad}l@{\quad}l}
		\interpretationFunctor ( \copier[][1] ) \defeq \copier[][X]
		&
		\interpretationFunctor ( \discard[][1] ) \defeq \discard[][X]
		&
		\interpretationFunctor ( \cocopier[][1] ) \defeq \cocopier[][X]
		&
		\interpretationFunctor ( \codiscard[][1] ) \defeq \codiscard[][X]
		\\
		\interpretationFunctor ( \id[][0] ) \defeq \id[][\singleton]
		&
		\interpretationFunctor ( \id[][1] ) \defeq \id[][X]
		&
		\interpretationFunctor ( \symm[][1][1] ) \defeq \symm[][X][X]
		&
		\interpretationFunctor (R^{\switchLabelS{+}} ) \defeq \rho(R)
		\end{array}
		\\
		\begin{array}{l@{\;\;}l@{\;\;}l}
		\interpretationFunctor \!( c \!\seq[]\! d ) \!\defeq\! \interpretationFunctor ( c) \! \seq[] \! \interpretationFunctor ( d )
		&
		\interpretationFunctor \!( c \!\tensor[]\! d ) \!\defeq\! \interpretationFunctor ( c) \! \tensor[] \! \interpretationFunctor ( d )
		&
		\interpretationFunctor \!(R^{\switchLabelS{-}} ) \!\defeq\! \op{\nega{\rho(R)}} %
	\end{array}
	\end{array}
	}
\end{equation}

 The constants and operations appearing on the right-hand-side of the above equations are amongst those defined in \eqref{eq:seqRel}, \eqref{eq:idRel}, \eqref{eq:comonoidsREL},  \eqref{eq:tensorREL}. %
A simple inductive argument confirms that $\interpretationFunctor$ maps terms $c$ of type $n \to m$ to relations $R \subseteq X^n \times X^m$. In particular, $\id[][0] \colon 0 \to 0$ is sent to $\id[][\singleton]\subseteq \singleton \times \singleton$, since $X^0=\singleton$ independently of $X$. Note that there are only two relations on the singleton set $\singleton=\{\star\}$: the relation $\{(\star,\star)\} \subseteq \singleton \times \singleton$ and the empty relation $\varnothing\subseteq \singleton \times \singleton$. These are, by \eqref{eq:idRel}, $\id[+][\singleton]$ and $\id[-][\singleton]$, embodying \emph{truth} and \emph{falsity}. %
\vspace{-5pt}
\begin{example}\label{eq:intersectionandtop}
Take $\sign$ with two symbols $R$ and $S$ with arity and coarity $1$. From Table~\ref{fig:typingrules}, the two terms below have type $1 \to 1$.
\begin{equation}\label{eq:exampletopint}
\discard[+][1] \seq[+] \codiscard[+][1] \qquad   \copier[+][1] \seq[+] ( (R^{\switchLabelS{+}} \tensor[+] S^{\switchLabelS{+}})\seq[+] \cocopier[+][1])
\end{equation}
For any interpretation $\interpretation = (X, \rho)$, $\interpretationFunctor(\discard[+][1] \seq[+] \codiscard[+][1])$ is the top $X\times X$:
\begin{align*}
	\interpretationFunctor(\discard[+][1] \seq[+] \codiscard[+][1]) &= \discard[+][X] \seq[+] \codiscard[+][X] = \{(x, \star) \mid x\in X\} \seq[+] \{(\star, x) \mid x\in X\} \\
	&= \{(x,y) \mid x,y\in X\} = X \times X = \dsemRel{\top}\text{.}
\end{align*}
Similarly, %
$\interpretationFunctor( \copier[+][1] \seq[+] ( (R^{\switchLabelS{+}} \tensor[+] S^{\switchLabelS{+}})\seq[+] \cocopier[+][1]) = \rho(R) \cap \rho(S)=\dsemRel{R\cap S}$.
\end{example}

\vspace{-10pt}
\begin{remark} %
$\NPR$ is as expressive as $\FOL$. We draw the reader's attention to the simplicity of the inductive definition of semantics compared to the traditional $\FOL$ approach where variables and quantifiers make the definition more involved.
Moreover, in traditional accounts, the domain of an interpretation is required to be a non-empty set. In our calculus this is unnecessary and it is \emph{not} a mere technicality: in \S~\ref{sec:theories} we shall see that empty models capture the propositional calculus. %
\end{remark}

\vspace{-5pt}
Two terms $c,d\colon n \to m$ are \emph{semantically equivalent}, written $c \semequivalence d$, if and only if $\interpretationFunctor (c) = \interpretationFunctor(d)$, for all interpretations $\interpretation$. \emph{Semantic inclusion} ($\seminclusion$) is defined analogously replacing $=$ with $\subseteq$.

By definition $\semequivalence$ and $\seminclusion$ only relate terms of the same type. Throughout the paper, we will encounter several relations amongst terms of the same type. To avoid any confusion with the relations denoted by the terms, we call them \emph{well-typed relations} and use symbols $\wtrel$ rather than the usual $R, S, T$. In the following, we write $c \basicR d$ for $(c,d)\in \basicR$ and $\pcong{\wtrel}$ for the smallest precongruence (w.r.t. $\seq[+]$, $\seq[-]$, $\tensor[+]$ and $\tensor[-]$) generated by $\basicR$, i.e., the relation inductively generated as
\begin{equation}\label{eq:pc}\arraycolsep=2pt
{\scriptsize
\begin{array}{c@{}c@{}c}
\inferrule*[right=($id$)]{c \wtrel d}{c \,\pcong{\wtrel} \,d}
\quad\quad\quad
&
\inferrule*[right=($r$)]{-}{c\,  \pcong{\wtrel} \,c}
\quad\quad\quad
&
\inferrule*[right=($t$)]{a \, \pcong{\wtrel} \,b \quad b \, \pcong{\wtrel}\, c}{a  \,\pcong{\wtrel} \,c}
\\[5px]
\multicolumn{3}{c}{
\inferrule*[right=($   {\seq}   $)]{c_1  \,\pcong{\wtrel} \,c_2 \quad d_1 \, \pcong{\wtrel} \, d_2}{c_1\seq[] d_1\,  \pcong{\wtrel} \, c_2 \seq[] d_2}
\qquad
\inferrule*[right=($   {\tensor}  $)]{c_1 \, \pcong{\wtrel} \,c_2 \quad d_1 \, \pcong{\wtrel} \, d_2}{c_1\tensor[] d_1 \, \pcong{\wtrel} \, c_2 \tensor[] d_2}
}
\end{array}
}
\end{equation}

\paragraph{Axioms} %
Fig. \ref{fig:textual axioms} in App. \ref{app:additional} illustrates a complete system of axioms for $\seminclusion$.
Let $\mathbb{FOB}$ be the well-typed relation obtained by substituting $a,b,c,d$ in Fig. \ref{fig:textual axioms} with terms of the appropriate type and and call its precongruence closure \emph{syntactic inclusion}, written $\syninclusion$. In symbols $\syninclusion = \pcong{\mathbb{FOB}}$. We will also write $\synequivalence \defeq \syninclusion \cap \syninclusionop$. Our main result is:
\begin{theorem}\label{thm:completeness}
For all terms $c,d\colon n \to m$, $c\syninclusion d$ iff $c\seminclusion d$.
\end{theorem}
The axiomatisation is far from minimal and is redundant in several respects. We chose the more verbose presentation in order to emphasise both the underlying categorical structures and the various dualities that we will highlight in the next sections.

\begin{figure*}[t]
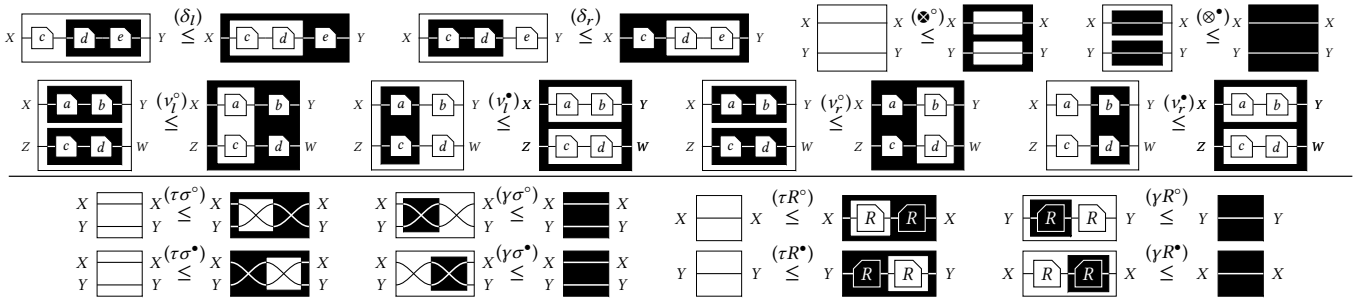

    \mylabel{ax:leftLinDistr}{$\delta_l$}
    \mylabel{ax:rightLinDistr}{$\delta_r$}
    \mylabel{ax:linStrn1}{$\nu^\circ_l$}
    \mylabel{ax:linStrn2}{$\nu^\circ_r$}
    \mylabel{ax:linStrn3}{$\nu^\bullet_l$}
    \mylabel{ax:linStrn4}{$\nu^\bullet_r$}
    \mylabel{ax:tensorPlusIdMinus}{$\tensor[+]^\bullet$}
    \mylabel{ax:tensorMinusIdPlus}{$\tensor[-]^\circ$}
    \mylabel{ax:tauSymmPlus}{$\tau\symm[+]$}
    \mylabel{ax:tauRPlus}{$\tau R^\circ$}
    \mylabel{ax:gammaSymmPlus}{$\gamma\symm[+]$}
    \mylabel{ax:gammaRPlus}{$\gamma R^\circ$}
    \mylabel{ax:tauSymmMinus}{$\tau\symm[-]$}
    \mylabel{ax:tauRMinus}{$\tau R^\bullet$}
    \mylabel{ax:gammaSymmMinus}{$\gamma\symm[-]$}
    \mylabel{ax:gammaRMinus}{$\gamma R^\bullet$}
    $
    \begin{array}{@{}c@{}}
        \begin{array}{@{}c@{\;}c@{\;}c@{\;}c@{}}
            \multicolumn{4}{@{}c@{}}{
                \begin{array}{@{\!\!\!\!}c@{\;}c@{\;}c@{\;}c@{}}
                    \scalebox{0.8}{
    \InputIfFileExists{axiomsNEW/leftLinDistr1.tikz}{}{\input{tikz/axiomsNEW/leftLinDistr1.tikz}}
}  \!\!\!\!\Lleq{\eqref*{ax:leftLinDistr}}\!\!\!\! \scalebox{0.8}{
    \InputIfFileExists{axiomsNEW/leftLinDistr2.tikz}{}{\input{tikz/axiomsNEW/leftLinDistr2.tikz}}
}
                    &
                    \scalebox{0.8}{
    \InputIfFileExists{axiomsNEW/rightLinDistr1.tikz}{}{\input{tikz/axiomsNEW/rightLinDistr1.tikz}}
} \!\!\!\!\Lleq{\eqref*{ax:rightLinDistr}}\!\!\!\! \scalebox{0.8}{
    \InputIfFileExists{axiomsNEW/rightLinDistr2.tikz}{}{\input{tikz/axiomsNEW/rightLinDistr2.tikz}}
}
                    &
                    \scalebox{0.8}{
    \InputIfFileExists{axiomsNEW/id2Pbig.tikz}{}{\input{tikz/axiomsNEW/id2Pbig.tikz}}
} \!\!\!\!\Lleq{\eqref*{ax:tensorMinusIdPlus}}\!\!\!\! \scalebox{0.8}{
    \InputIfFileExists{axiomsNEW/idPMinusidP.tikz}{}{\input{tikz/axiomsNEW/idPMinusidP.tikz}}
}
                    &
                    \scalebox{0.8}{
    \InputIfFileExists{axiomsNEW/idMPlusidM.tikz}{}{\input{tikz/axiomsNEW/idMPlusidM.tikz}}
} \!\!\!\!\Lleq{\eqref*{ax:tensorPlusIdMinus}}\!\!\!\! \scalebox{0.8}{
    \InputIfFileExists{axiomsNEW/id2Mbig.tikz}{}{\input{tikz/axiomsNEW/id2Mbig.tikz}}
}
                \end{array}
            }
            \\[10pt]
            \scalebox{0.8}{
    \InputIfFileExists{axiomsNEW/linStr1_1.tikz}{}{\input{tikz/axiomsNEW/linStr1_1.tikz}}
} \!\!\!\!\Lleq{\eqref*{ax:linStrn1}}\!\!\!\! \scalebox{0.8}{
    \InputIfFileExists{axiomsNEW/linStr1_2.tikz}{}{\input{tikz/axiomsNEW/linStr1_2.tikz}}
}
            &
            \scalebox{0.8}{
    \InputIfFileExists{axiomsNEW/linStr3_1.tikz}{}{\input{tikz/axiomsNEW/linStr3_1.tikz}}
} \!\!\!\!\Lleq{\eqref*{ax:linStrn3}}\!\!\!\! \scalebox{0.8}{
    \InputIfFileExists{axiomsNEW/linStr3_2.tikz}{}{\input{tikz/axiomsNEW/linStr3_2.tikz}}
}
            &
            \scalebox{0.8}{
    \InputIfFileExists{axiomsNEW/linStr1_1.tikz}{}{\input{tikz/axiomsNEW/linStr1_1.tikz}}
} \!\!\!\!\Lleq{\eqref*{ax:linStrn2}}\!\!\!\! \scalebox{0.8}{
    \InputIfFileExists{axiomsNEW/linStr2_2.tikz}{}{\input{tikz/axiomsNEW/linStr2_2.tikz}}
}
            & 
            \scalebox{0.8}{
    \InputIfFileExists{axiomsNEW/linStr4_1.tikz}{}{\input{tikz/axiomsNEW/linStr4_1.tikz}}
} \!\!\!\!\Lleq{\eqref*{ax:linStrn4}}\!\!\!\! \scalebox{0.8}{
    \InputIfFileExists{axiomsNEW/linStr3_2.tikz}{}{\input{tikz/axiomsNEW/linStr3_2.tikz}}
}
        \end{array}
        \\
        \midrule
        \begin{array}{cc}
            \begin{array}{@{}c@{}c@{}c c@{}c@{}c@{}}
                
    \InputIfFileExists{axiomsNEW/idXYP.tikz}{}{\input{tikz/axiomsNEW/idXYP.tikz}}
 & \!\!\Lleq{\eqref*{ax:tauSymmPlus}}\!\!   & 
    \InputIfFileExists{axiomsNEW/linadj/symMsym.tikz}{}{\input{tikz/axiomsNEW/linadj/symMsym.tikz}}
 & 
    \InputIfFileExists{axiomsNEW/linadj2/symPsym.tikz}{}{\input{tikz/axiomsNEW/linadj2/symPsym.tikz}}
     & \!\!\Lleq{\eqref*{ax:gammaSymmPlus}}\!\!   & 
    \InputIfFileExists{axiomsNEW/idXYM.tikz}{}{\input{tikz/axiomsNEW/idXYM.tikz}}
 \\
                
    \InputIfFileExists{axiomsNEW/idXYP.tikz}{}{\input{tikz/axiomsNEW/idXYP.tikz}}
 & \!\!\Lleq{\eqref*{ax:tauSymmMinus}}\!\!   & 
    \InputIfFileExists{axiomsNEW/linadj2/symMsym.tikz}{}{\input{tikz/axiomsNEW/linadj2/symMsym.tikz}}
 & 
    \InputIfFileExists{axiomsNEW/linadj/symPsym.tikz}{}{\input{tikz/axiomsNEW/linadj/symPsym.tikz}}
     & \!\!\Lleq{\eqref*{ax:gammaSymmMinus}}\!\!   & 
    \InputIfFileExists{axiomsNEW/idXYM.tikz}{}{\input{tikz/axiomsNEW/idXYM.tikz}}

            \end{array}
            &
            \begin{array}{@{}c@{}c@{}c c@{}c@{}c@{}}
                \idCirc[+][X] & \Lleq{\eqref*{ax:tauRPlus}}  & 
    \InputIfFileExists{axiomsNEW/linadj/rMrop.tikz}{}{\input{tikz/axiomsNEW/linadj/rMrop.tikz}}
 & 
    \InputIfFileExists{axiomsNEW/linadj/ropPr.tikz}{}{\input{tikz/axiomsNEW/linadj/ropPr.tikz}}
       & \Lleq{\eqref*{ax:gammaRPlus}}  & \idCirc[-][Y] \\
                \idCirc[+][Y] & \Lleq{\eqref*{ax:tauRMinus}}  & 
    \InputIfFileExists{axiomsNEW/linadj2/rMrop.tikz}{}{\input{tikz/axiomsNEW/linadj2/rMrop.tikz}}
 & 
    \InputIfFileExists{axiomsNEW/linadj2/ropPr.tikz}{}{\input{tikz/axiomsNEW/linadj2/ropPr.tikz}}
       &\Lleq{\eqref*{ax:gammaRMinus}}  & \idCirc[-][X]
            \end{array}
        \end{array}
    \end{array}
    $
    \caption{Axioms of closed symmetric monoidal linear bicategories}\label{fig:closed lin axioms}
\end{figure*}

\paragraph{Diagrams}
We confined the complete axiomatisation to the appendix because the axioms in Fig. \ref{fig:textual axioms} appear also in Figs.~\ref{fig:cb axioms}, \ref{fig:cocb axioms}, \ref{fig:closed lin axioms}, \ref{fig:fo bicat axioms} in diagrammatic form. This allows a more principled, staged presentation and place each axiom in its proper context, highlighting their provenance from one of the categorical structures involved.

Diagrams, inspired by
string diagrams \cite{joyal1991geometry,Selinger2009}, take centre stage in our presentation. %
A term $c\colon n \to m$ is drawn as a diagram with $n$ ports on the left and $m$ ports on the right; $\seq[]$ is depicted as horizontal composition while $\tensor[]$ by vertically ``stacking'' diagrams. The two compositions  $\seq[+]$ and $\seq[-]$ and two monoidal products $\tensor[+]$ and $\tensor[-]$ are distinguished with different colours. %
All constants in the white fragment have white background, mutatis mutandis for the black fragment: for instance $\id[+][1]$ and $\id[-][1]$ are drawn
$\idCirc[+]$ and $\idCirc[-]$. Indeed, the diagrammatic version of \eqref{eq:syntax} is the grammar in Fig.\ref{fig:diagsyntax}.

To better grasp the correspondence between terms and diagrams, the reader may compare the diagrammatic version of the axioms (Fig.s \ref{fig:cb axioms}, \ref{fig:cocb axioms}, \ref{fig:closed lin axioms}, \ref{fig:fo bicat axioms}) with the term-based one (in Figure \ref{fig:textual axioms}).

\begin{wrapfigure}{R}{0.06\textwidth}
    \vspace{-12pt}
    $\!\!\!\!\!\!\!\intersectionCirc{R}{S}$
    \vspace{-12pt}
\end{wrapfigure}
Note that one diagram may correspond to more than one term:  for instance the diagram on the right
does not only represent the  rightmost term in \eqref{eq:exampletopint}, namely $\copier[+][1] \seq[+] ( (R^{\switchLabelS{+}} \tensor[+] S^{\switchLabelS{+}})\seq[+] \cocopier[+][1])$,  but also  $(\copier[+][1] \seq[+]  (R^{\switchLabelS{+}} \tensor[+] S^{\switchLabelS{+}}))\seq[+] \cocopier[+][1]$. Indeed, it is clear that traditional term-based syntax carries more information than the diagrammatic one (e.g. associativity). From the point of view of the semantics, however, this bureaucracy is irrelevant and is conveniently discarded by the diagrammatic notation. To formally show this, we recall that diagrams capture only the axioms of symmetric monoidal categories \cite{joyal1991geometry,Selinger2009}, illustrated in Table \ref{fig:freestricmmoncatax}; we call
\emph{structural congruence}, written $\structuralcong$, the well-typed congruence generated by such axioms and we observe that $\structuralcong \subseteq \semequivalence$.

\paragraph{Proofs as diagrams rewrites}
Proofs in $\NPR$ are rather different from those of traditional proof systems: since the only inference rules are those in~\eqref{eq:pc}, any proof of $c\syninclusion d$ consists of a sequence of applications of axioms. %
As an example consider \eqref{eq:forallexists} from the Introduction
(see App. \ref{app:additional proofs} for a proof not using Prop. \ref{prop:maps}). Note that, when applying axioms, we are in fact performing diagram rewriting: an instance of the left hand side of an axiom is found within a larger diagram and replaced with the right hand side. Since such rewrites can happen anywhere, there is a close  connection between proofs in $\NPR$ and work on \emph{deep inference}~\cite{hughes2021combinatorial,DBLP:phd/de/Brunnler2003,guglielmi2007system} -- see Ex.~\ref{ex:propcalculus}. %

\begin{figure*}[t]
\mylabel{ax:comPlusAssoc}{$\copier[+]$-as}
\mylabel{ax:comPlusUnit}{$\copier[+]$-un}
\mylabel{ax:comPlusComm}{$\copier[+]$-co}
\mylabel{ax:monPlusAssoc}{$\cocopier[+]$-as}
\mylabel{ax:monPlusUnit}{$\cocopier[+]$-un}
\mylabel{ax:monPlusComm}{$\cocopier[+]$-co}
\mylabel{ax:plusSpecFrob}{S$^\circ$}
\mylabel{ax:plusFrob}{F$^\circ$}
\mylabel{ax:comPlusLaxNat}{$\copier[+]$-nat}
\mylabel{ax:discPlusLaxNat}{$\discard[+]$-nat}
\mylabel{ax:plusCodiscDisc}{$\epsilon\discard[+]$}
\mylabel{ax:plusDiscCodisc}{$\eta\discard[+]$}
\mylabel{ax:plusCocopyCopy}{$\epsilon\!\copier[+]$}
\mylabel{ax:plusCopyCocopy}{$\eta\!\copier[+]$}
\centering
$
\begin{array}{@{}c@{\!}c@{\!}c @{} c@{\!}c@{\!}c @{} c@{\!}c@{\!}c @{}|@{} c@{\!}c@{\!}c @{}|@{} c@{\!}c@{\!}c@{}}
        \scalebox{0.8}{
    \InputIfFileExists{axiomsNEW/cb/plus/comAssoc1.tikz}{}{\input{tikz/axiomsNEW/cb/plus/comAssoc1.tikz}}
}   & \Leq{\eqref*{ax:comPlusAssoc}} &  \scalebox{0.8}{
    \InputIfFileExists{axiomsNEW/cb/plus/comAssoc2.tikz}{}{\input{tikz/axiomsNEW/cb/plus/comAssoc2.tikz}}
}
        &
        \scalebox{0.8}{
    \InputIfFileExists{axiomsNEW/cb/plus/comUnit.tikz}{}{\input{tikz/axiomsNEW/cb/plus/comUnit.tikz}}
}     & \Leq{\eqref*{ax:comPlusUnit}}  &  \scalebox{0.8}{\idCirc[+][X]}
        &
        \scalebox{0.8}{
    \InputIfFileExists{axiomsNEW/cb/plus/comComm.tikz}{}{\input{tikz/axiomsNEW/cb/plus/comComm.tikz}}
}     & \Leq{\eqref*{ax:comPlusComm}}  &  \scalebox{0.8}{\copierCirc[+][X]}
        &
        \scalebox{0.8}{
    \InputIfFileExists{axiomsNEW/cb/plus/specFrob.tikz}{}{\input{tikz/axiomsNEW/cb/plus/specFrob.tikz}}
}    & \Leq{\eqref*{ax:plusSpecFrob}} &  \scalebox{0.8}{\idCirc[+][X]}
        &
        \scalebox{0.8}{
    \InputIfFileExists{axiomsNEW/cb/plus/copierLaxNat1.tikz}{}{\input{tikz/axiomsNEW/cb/plus/copierLaxNat1.tikz}}
} &\Lleq{\eqref*{ax:comPlusLaxNat}}& 
    \InputIfFileExists{axiomsNEW/cb/plus/copierLaxNat2.tikz}{}{\input{tikz/axiomsNEW/cb/plus/copierLaxNat2.tikz}}

        \\
        \scalebox{0.8}{
    \InputIfFileExists{axiomsNEW/cb/plus/monAssoc1.tikz}{}{\input{tikz/axiomsNEW/cb/plus/monAssoc1.tikz}}
}   & \Leq{\eqref*{ax:monPlusAssoc}} &  \scalebox{0.8}{
    \InputIfFileExists{axiomsNEW/cb/plus/monAssoc2.tikz}{}{\input{tikz/axiomsNEW/cb/plus/monAssoc2.tikz}}
}
        &
        \scalebox{0.8}{
    \InputIfFileExists{axiomsNEW/cb/plus/monUnit.tikz}{}{\input{tikz/axiomsNEW/cb/plus/monUnit.tikz}}
}     & \Leq{\eqref*{ax:monPlusUnit}}  &  \scalebox{0.8}{\idCirc[+][X]}
        &
        \scalebox{0.8}{
    \InputIfFileExists{axiomsNEW/cb/plus/monComm.tikz}{}{\input{tikz/axiomsNEW/cb/plus/monComm.tikz}}
}     & \Leq{\eqref*{ax:monPlusComm}}  &  \scalebox{0.8}{\cocopierCirc[+][X]}
        &
        \scalebox{0.8}{
    \InputIfFileExists{axiomsNEW/cb/plus/frob1.tikz}{}{\input{tikz/axiomsNEW/cb/plus/frob1.tikz}}
}       & \Leq{\eqref*{ax:plusFrob}}     &  \scalebox{0.8}{
    \InputIfFileExists{axiomsNEW/cb/plus/frob2.tikz}{}{\input{tikz/axiomsNEW/cb/plus/frob2.tikz}}
}
        &
        \scalebox{0.8}{
    \InputIfFileExists{axiomsNEW/cb/plus/discardLaxNat.tikz}{}{\input{tikz/axiomsNEW/cb/plus/discardLaxNat.tikz}}
} &\Lleq{\eqref*{ax:discPlusLaxNat}}& \scalebox{0.8}{\discardCirc[+][X][X]}
        \\
        \midrule
        \multicolumn{15}{c}{
                \begin{array}{c@{\;\;}c@{\;\;}c c@{}c@{}c c@{}c@{}c c@{}c@{}c}
                        \scalebox{0.8}{
    \InputIfFileExists{axiomsNEW/cb/plus/codiscDisc.tikz}{}{\input{tikz/axiomsNEW/cb/plus/codiscDisc.tikz}}
} & \Lleq{\eqref*{ax:plusCodiscDisc}}& \scalebox{0.8}{\emptyCirc[+]}
                        &
                        \scalebox{0.8}{\idCirc[+][X]}     &\Lleq{\eqref*{ax:plusDiscCodisc}}& \scalebox{0.8}{
    \InputIfFileExists{axiomsNEW/top.tikz}{}{\input{tikz/axiomsNEW/top.tikz}}
}
                        &
                        \scalebox{0.8}{
    \InputIfFileExists{axiomsNEW/cb/plus/cocopierCopier.tikz}{}{\input{tikz/axiomsNEW/cb/plus/cocopierCopier.tikz}}
} & \Lleq{\eqref*{ax:plusCocopyCopy}}& \scalebox{0.8}{
    \InputIfFileExists{axiomsNEW/id2P.tikz}{}{\input{tikz/axiomsNEW/id2P.tikz}}
}
                        &
                        \scalebox{0.8}{\idCirc[+][X]}     &\Lleq{\eqref*{ax:plusCopyCocopy}}& \scalebox{0.8}{
    \InputIfFileExists{axiomsNEW/cb/plus/specFrob.tikz}{}{\input{tikz/axiomsNEW/cb/plus/specFrob.tikz}}
}
                \end{array}
        }
\end{array}
$
\caption{Axioms of cartesian bicategories}\label{fig:cb axioms}
\end{figure*}

\begin{figure*}[t]
        \mylabel{ax:comMinusAssoc}{$\copier[-]$-as}
        \mylabel{ax:comMinusUnit}{$\copier[-]$-un}
        \mylabel{ax:comMinusComm}{$\copier[-]$-co}
        \mylabel{ax:monMinusAssoc}{$\cocopier[-]$-as}
        \mylabel{ax:monMinusUnit}{$\cocopier[-]$-un}
        \mylabel{ax:monMinusComm}{$\cocopier[-]$-co}
        \mylabel{ax:minusSpecFrob}{S$^\bullet$}
        \mylabel{ax:minusFrob}{F$^\bullet$}
        \mylabel{ax:comMinusLaxNat}{$\copier[-]$-nat}
        \mylabel{ax:discMinusLaxNat}{$\discard[-]$-nat}
        \mylabel{ax:minusCodiscDisc}{$\eta\codiscard[-]$}
        \mylabel{ax:minusDiscCodisc}{$\epsilon\codiscard[-]$}
        \mylabel{ax:minusCocopyCopy}{$\eta\!\cocopier[-]$}
        \mylabel{ax:minusCopyCocopy}{$\epsilon\!\cocopier[-]$}
        \centering
        $
        \begin{array}{@{}c@{\!}c@{\!}c @{} c@{\!}c@{\!}c @{} c@{\!}c@{\!}c @{}|@{} c@{\!}c@{\!}c @{}|@{} c@{\!}c@{\!}c@{}}
                \scalebox{0.8}{
    \InputIfFileExists{axiomsNEW/cb/minus/comAssoc1.tikz}{}{\input{tikz/axiomsNEW/cb/minus/comAssoc1.tikz}}
}   & \Leq{\eqref*{ax:comMinusAssoc}} &  \scalebox{0.8}{
    \InputIfFileExists{axiomsNEW/cb/minus/comAssoc2.tikz}{}{\input{tikz/axiomsNEW/cb/minus/comAssoc2.tikz}}
}
                &
                \scalebox{0.8}{
    \InputIfFileExists{axiomsNEW/cb/minus/comUnit.tikz}{}{\input{tikz/axiomsNEW/cb/minus/comUnit.tikz}}
}     & \Leq{\eqref*{ax:comMinusUnit}}  &  \scalebox{0.8}{\idCirc[-][X]}
                &
                \scalebox{0.8}{
    \InputIfFileExists{axiomsNEW/cb/minus/comComm.tikz}{}{\input{tikz/axiomsNEW/cb/minus/comComm.tikz}}
}     & \Leq{\eqref*{ax:comMinusComm}}  &  \scalebox{0.8}{\copierCirc[-][X]}
                &
                \scalebox{0.8}{
    \InputIfFileExists{axiomsNEW/cb/minus/specFrob.tikz}{}{\input{tikz/axiomsNEW/cb/minus/specFrob.tikz}}
}    & \Leq{\eqref*{ax:minusSpecFrob}} &  \scalebox{0.8}{\idCirc[-][X]}
                &
                
    \InputIfFileExists{axiomsNEW/cb/minus/copierLaxNat2.tikz}{}{\input{tikz/axiomsNEW/cb/minus/copierLaxNat2.tikz}}
 &\Lleq{\eqref*{ax:comMinusLaxNat}}& \scalebox{0.8}{
    \InputIfFileExists{axiomsNEW/cb/minus/copierLaxNat1.tikz}{}{\input{tikz/axiomsNEW/cb/minus/copierLaxNat1.tikz}}
}
                \\
                \scalebox{0.8}{
    \InputIfFileExists{axiomsNEW/cb/minus/monAssoc1.tikz}{}{\input{tikz/axiomsNEW/cb/minus/monAssoc1.tikz}}
}   & \Leq{\eqref*{ax:monMinusAssoc}} &  \scalebox{0.8}{
    \InputIfFileExists{axiomsNEW/cb/minus/monAssoc2.tikz}{}{\input{tikz/axiomsNEW/cb/minus/monAssoc2.tikz}}
}
                &
                \scalebox{0.8}{
    \InputIfFileExists{axiomsNEW/cb/minus/monUnit.tikz}{}{\input{tikz/axiomsNEW/cb/minus/monUnit.tikz}}
}     & \Leq{\eqref*{ax:monMinusUnit}}  &  \scalebox{0.8}{\idCirc[-][X]}
                &
                \scalebox{0.8}{
    \InputIfFileExists{axiomsNEW/cb/minus/monComm.tikz}{}{\input{tikz/axiomsNEW/cb/minus/monComm.tikz}}
}     & \Leq{\eqref*{ax:monMinusComm}}  &  \scalebox{0.8}{\cocopierCirc[-][X]}
                &
                \scalebox{0.8}{
    \InputIfFileExists{axiomsNEW/cb/minus/frob1.tikz}{}{\input{tikz/axiomsNEW/cb/minus/frob1.tikz}}
}       & \Leq{\eqref*{ax:minusFrob}}     &  \scalebox{0.8}{
    \InputIfFileExists{axiomsNEW/cb/minus/frob2.tikz}{}{\input{tikz/axiomsNEW/cb/minus/frob2.tikz}}
}
                &
                \scalebox{0.8}{\discardCirc[-][X][X]} &\Lleq{\eqref*{ax:discMinusLaxNat}}& \scalebox{0.8}{
    \InputIfFileExists{axiomsNEW/cb/minus/discardLaxNat.tikz}{}{\input{tikz/axiomsNEW/cb/minus/discardLaxNat.tikz}}
}
                \\
                \midrule
                \multicolumn{15}{c}{
                        \begin{array}{c@{}c@{}c c@{\;\;}c@{\;\;}c c@{}c@{}c c@{}c@{}c}
                                \scalebox{0.8}{
    \InputIfFileExists{axiomsNEW/bottom.tikz}{}{\input{tikz/axiomsNEW/bottom.tikz}}
}     &\Lleq{\eqref*{ax:minusDiscCodisc}}& \scalebox{0.8}{\idCirc[-][X]}
                                &
                                \scalebox{0.8}{\emptyCirc[-]} & \Lleq{\eqref*{ax:minusCodiscDisc}}& \scalebox{0.8}{
    \InputIfFileExists{axiomsNEW/cb/minus/codiscDisc.tikz}{}{\input{tikz/axiomsNEW/cb/minus/codiscDisc.tikz}}
}
                                &
                                \scalebox{0.8}{
    \InputIfFileExists{axiomsNEW/cb/minus/specFrob.tikz}{}{\input{tikz/axiomsNEW/cb/minus/specFrob.tikz}}
}     &\Lleq{\eqref*{ax:minusCopyCocopy}}& \scalebox{0.8}{\idCirc[-][X]}
                                &
                                \scalebox{0.8}{
    \InputIfFileExists{axiomsNEW/id2M.tikz}{}{\input{tikz/axiomsNEW/id2M.tikz}}
} & \Lleq{\eqref*{ax:minusCocopyCopy}}& \scalebox{0.8}{
    \InputIfFileExists{axiomsNEW/cb/minus/cocopierCopier.tikz}{}{\input{tikz/axiomsNEW/cb/minus/cocopierCopier.tikz}}
}
                        \end{array}
                }
        \end{array}
        $
        \caption{Axioms of cocartesian bicategories}\label{fig:cocb axioms}
        \end{figure*}

\section{(Co)Cartesian Bicategories}\label{sec:cartesianbi}

Although the term bicategory might seem ominous, the beasts considered in this paper are actually quite simple. %
We consider \emph{poset enriched symmetric monoidal categories}:  every homset carries a partial order $\leq$, and composition $\seq[]$ and monoidal product $\tensor$ are monotone. That is, if $a\leq b$ and $c\leq d$ then $a\seq[] c \leq b\seq[] d$ and $a\tensor[] c \leq b\tensor[] d$. A \emph{poset enriched symmetric monoidal functor} is a (strong, and usually strict) symmetric monoidal functor that preserves the order $\leq$.
The notion of \emph{adjoint arrows}, which will play a key role, amounts to the following: for $c \colon X \to Y$ and $d \colon Y \to X$,  $c$ is  \emph{left adjoint} to $d$, or $d$ is \emph{right adjoint} to $c$, written $d \vdash c$, if $\id[][X] \leq c \seq[] d$  and $d \seq[] c \leq \id[][Y]$.

For a symmetric monoidal bicategory $(\Cat{C}, \tensor[][][], \unittensor)$, we will write $\opposite{\Cat{C}}$ for the bicategory having the same objects as $\Cat{C}$ but homsets $\opposite{\Cat{C}}[X,Y] \defeq \Cat{C}[Y,X]$: ordering, identities and monoidal product are defined as in $\Cat{C}$, while composition $c \seq[] d$ in $\opposite{\Cat{C}}$ is $d \seq[] c$ in $\Cat{C}$.
Similarly, we will write  $\co{\Cat{C}}$ to denote the bicategory having the same objects and arrows of $\Cat{C}$ but equipped with the reversed ordering $\geq$. Composition, identities and monoidal product are defined as in $\Cat{C}$.
In this paper, we will often tacitly use the facts that, by definition, both $\opposite{(\opposite{\Cat{C}})}$ and $\co{(\co{\Cat{C}})}$ are $\Cat{C}$ and that $\opposite{(\co{\Cat{C}})}$ is $\co{(\opposite{\Cat{C}})}$. %

All monoidal categories considered throughout this paper are tacitly assumed to be strict \cite{mac_lane_categories_1978}, i.e.\ $(X\tensor[] Y)\tensor[] Z = X \tensor[] (Y \tensor[] Z)$ and $\unittensor \tensor[] X = X =X \tensor[] \unittensor$ for all objects $X,Y,Z$. This is harmless: strictification~\cite{mac_lane_categories_1978} allows to transform any monoidal category into a strict one, enabling the sound use of string diagrams. These will be exploited in this and the next two sections to describe the categorical structures of interest. In particular, in the following definition $\copier[+][X]\colon X \to X\tensor[+]X$,  $\discard[+][X] \colon X \to \unittensor$, $\cocopier[+][X]\colon X\tensor[+]X \to X$ and $\codiscard[+][X] \colon \unittensor \to X $ are drawn, respectively, as $\scalebox{0.8}{\copierCirc[+][X]}$, $\scalebox{0.8}{\discardCirc[+][X]}$,  $\scalebox{0.8}{\cocopierCirc[+][X]}$ and $\scalebox{0.8}{\codiscardCirc[+][X]}$.

\begin{definition}\label{def:cartesian bicategory}
A \emph{cartesian bicategory} $(\Cat{C}, \tensor[+], \unittensor, \copier[+], \discard[+], \cocopier[+], \codiscard[+])$, shorthand $(\Cat{C}, \copier[+], \cocopier[+])$,
is a poset enriched symmetric monoidal category $(\Cat{C}, \tensor[+], \unittensor)$ and, for every object $X$ in $\Cat{C}$,
arrows $\copier[+][X]\colon X \to X\tensor[+]X$, $\discard[+][X]\colon X \to \unittensor$, $\cocopier[+][X] \colon X \tensor[+]X \to X$,  $\codiscard[+][X]\colon \unittensor \to X$ s.t.

\noindent 1. $(\copier[+][X], \discard[+][X])$ is a comonoid and $(\cocopier[+][X], \codiscard[+][X])$ a monoid (i.e., \eqref{ax:comPlusAssoc}, \eqref{ax:comPlusUnit},  \eqref{ax:comPlusComm} and \eqref{ax:monPlusAssoc}, \eqref{ax:monPlusUnit}, \eqref{ax:monPlusComm} in Fig.~\ref{fig:cb axioms} hold);

\noindent 2. arrows $c \colon X \to Y$ are lax comonoid morphisms (\eqref{ax:comPlusLaxNat}, \eqref{ax:discPlusLaxNat});

\noindent 3.  $(\copier[+][X], \discard[+][X])$ are left adjoints to $(\cocopier[+][X], \codiscard[+][X])$ (\eqref{ax:plusCopyCocopy}, \eqref{ax:plusCocopyCopy}, \eqref{ax:plusDiscCodisc}, \eqref{ax:plusCodiscDisc});

\noindent 4. $(\copier[+][X], \discard[+][X])$ and  $(\cocopier[+][X], \codiscard[+][X])$ form special Frobenius algebras (\eqref{ax:plusFrob}, \eqref{ax:plusSpecFrob});

\noindent 5. $(\copier[+][X], \discard[+][X])$ and $(\cocopier[+][X], \codiscard[+][X])$ satisfy the coherence conditions:\footnote{Note that the coherence conditions are not in Fig.~\ref{fig:cb axioms} since they hold in $\NPR$, given the inductive definitions of Tab.~\ref{fig:sugar}.}
\[
\begin{array}{r@{\;}c@{\;}l @{\qquad} r@{\;}c@{\;}l}
\copier[+][\unittensor]    &=& \id[+][\unittensor] & \copier[+][X\tensor[+]Y]      &=& (\copier[+][X] \tensor[+] \copier[+][Y]) \seq[+]( \id[+][X] \tensor[+]\symm[+][X][Y] \tensor[+] \id[+][Y]) \\
\cocopier[+][\unittensor]  &=& \id[+][\unittensor] & \cocopier[+][X\tensor[+]Y]    &=& ( \id[+][X] \tensor[+]\symm[+][X][Y] \tensor[+] \id[+][Y]) \seq[+] (\cocopier[+][X] \tensor[+] \cocopier[+][Y]) \\
\multicolumn{6}{c}{
\;\,\discard[+][\unittensor]   = \id[+][\unittensor] \qquad\,\, \discard[+][X \tensor[+] Y]   = \discard[+][X] \tensor[+]\discard[+][Y]
\qquad
\codiscard[+][\unittensor] = \id[+][\unittensor] \qquad\;\,  \codiscard[+][X \tensor[+] Y] = \codiscard[+][X] \tensor[+]\codiscard[+][Y]
}
\end{array}
\]
$\Cat{C}$ is a \emph{cocartesian bicategory} if  $\co{\Cat{C}}$ is a cartesian bicategory. %
A \emph{morphism of (co)cartesian bicategories} is a poset enriched strong symmetric monoidal functor preserving monoids and comonoids.
\end{definition}

The archetypal example of a cartesian bicategory is  $(\Relp, \copier[+], \cocopier[+])$. $\Relp$ the bicategory of sets and relations ordered by inclusion $\subseteq$ with white composition $\seq[+]$ and identities $\id[+]$ defined as in \eqref{eq:seqRel} and \eqref{eq:idRel}. The monoidal product on objects is the cartesian product of sets with unit $\unittensor$ the singleton set $\singleton$. on arrows, $\tensor[+]$  is defined as in \eqref{eq:tensorREL}. It is immediate to check that, for every set $X$, the arrows $\copier[+][X]$, $\discard[+][X]$ defined in \eqref{eq:comonoidsREL} form a comonoid in $\Relp$, while $\cocopier[+][X]$, $\codiscard[+][X]$ a monoid. Simple computations also proves all the (in)equalities in Fig.~\ref{fig:cb axioms}. %

The fact that relations are lax comonoid homomorphisms is the most interesting to show: %
since $R \seq[+] \copier[+][Y] = \{(x,(y,y)) \mid (x,y)\in R\}$ is included in $\{(x,(y,z)) \mid (x,y)\in R \wedge (x,z)\in R\} = \copier[+][X] \seq[+] (R \tensor[+] R)$ and $R\seq[+] \discard[+][Y] = \{(x,\star) \mid \exists y\in X \;. \; (x,y)\in R\} $ in $\{(x,\star) \mid x\in X\} = \discard[+][X]$ for any relation $R\subseteq X \times Y$, \eqref{ax:comPlusLaxNat} and  \eqref{ax:discPlusLaxNat} hold.
The reversed inclusions are interesting to consider: $R \seq[+] \copier[+][Y] \supseteq \copier[+][X] \seq[+] (R \tensor[+] R)$ holds iff the relation $R$ is single valued, while $R\seq[+] \discard[+][Y] \supseteq \discard[+][X]$ iff $R$ a total. That is, the two inequalities in Definition \ref{def:cartesian bicategory}.(2) are equalities iff the relation $R$ is a function. This justifies the following:

\begin{definition}\label{def:maps} %
An arrow $c\colon X \to Y$ is a \emph{map} if  %
\[ 
    \InputIfFileExists{axiomsNEW/cb/plus/copierLaxNat1.tikz}{}{\input{tikz/axiomsNEW/cb/plus/copierLaxNat1.tikz}}
 \geq 
    \InputIfFileExists{axiomsNEW/cb/plus/copierLaxNat2.tikz}{}{\input{tikz/axiomsNEW/cb/plus/copierLaxNat2.tikz}}
 \qquad\quad 
    \InputIfFileExists{axiomsNEW/cb/plus/discardLaxNat.tikz}{}{\input{tikz/axiomsNEW/cb/plus/discardLaxNat.tikz}}
 \!\!\!\geq \discardCirc[+][X] \]
\end{definition}

It is easy to see that maps form a monoidal subcategory of $\Cat{C}$~\cite{carboni1987cartesian}, hereafter denoted by $\texttt{Map}(\Cat{C})$. In fact, it is cartesian.

Given a cartesian bicategory $(\Cat{C},\copier[+],\cocopier[+])$, one can take $\opposite{\Cat{C}}$, swap monoids and comonoids and thus, obtain a cartesian bicategory $(\opposite{\Cat{C}}, \cocopier[+],\copier[+])$. Most importantly, there is an  identity on objects isomorphism %
$\op{(\cdot)}\colon \Cat{C} \to \opposite{\Cat{C}}$ defined for all arrows $c \colon X \to Y$ as %
\begin{equation}\label{eqdagger}
\op{c} \defeq \daggerCirc[+]{c}[Y][X]
\end{equation}%

\begin{proposition}\label{prop:opcartesianfunctor}
$\op{(\cdot)}\colon \Cat{C} \to \opposite{\Cat{C}}$ is an isomorphism of cartesian bicategories, namely the laws in the first three rows of Table \ref{table:daggerproperties}.(a) hold.
\end{proposition}

Hereafter, we write $\CircOp[+]{c}$ for $\op{\Circ[+]{c}}$ and we call it the \emph{mirror image} of $\Circ[+]{c}$.
 Note that in \S~\ref{sec:calculusrelations}, we used the same symbol $\op{(\cdot)}$ to denote the converse relation. This is no accident:  in the cartesian bicategory $(\Relp, \copier[+], \cocopier[+])$, $\op{R}$ as in \eqref{eqdagger} is exactly $\{(y,x) \mid (x,y)\in R\}$. %

In a cartesian bicategory, one can also define, for all arrows $c,d\colon X \to Y$, $c \sqcap d$ and $\top$ as follows. %
\begin{equation}\label{eq:def:cap}c \sqcap d \defeq \intersectionCirc{c}{d}[X][Y] \qquad  \top \defeq \topCirc[X][Y]
\end{equation}
We have already seen in Example~\ref{eq:intersectionandtop} that these terms, when interpreted in $\Relp$, denote respectively intersection and top. It is easy to show that in any cartesian bicategory $\Cat{C}$, $\sqcap$ is associative, commutative, idempotent and has $\top$ as unit. Namely, $\Cat{C}[X,Y]$ is a meet-semilattice with top. However, $\Cat{C}$ is usually \emph{not} enriched over meet-semilattices since $\seq[+]$ distributes only laxly over $\sqcap$. Indeed, in $\Relp$,  $R\seq[+](S \cap T) \subseteq (R \seq[+] S)\cap(R\seq[+]T) $ holds but the reverse does not.

Let us now turn to \emph{co}cartesian bicategories. Our main example is $(\Relm, \copier[-], \cocopier[-])$. $\Relm$ is the bicategory of sets and relations ordered by $\subseteq$ with composition $\seq[-]$, identities $\id[-]$ and $\tensor[-]$ defined as in \eqref{eq:seqRel}, \eqref{eq:idRel} and \eqref{eq:tensorREL}. Comonoids $(\copier[-][X],\discard[-][X])$ and  monoids $(\cocopier[-][X] , \codiscard[-][X])$ are those of~\eqref{eq:comonoidsREL}. To see that $\Relm$ is a cocartesian bicategory, observe that the complement $\nega{(\cdot)}$ is a poset-enriched symmetric monoidal isomorphism $\nega{(\cdot)} \colon \co{(\Relp)} \to \Relm$ preserving (co)monoids.

We draw arrows of cocartesian bicategories in black: $\copier[-][X]$,$\discard[-][X]$, $\cocopier[-][X]$ and $\codiscard[-][X]$ are drawn  $\scalebox{0.8}{\copierCirc[-][X]}$, $\scalebox{0.8}{\discardCirc[-][X]}$, $\scalebox{0.8}{\cocopierCirc[-][X]}$ and $\scalebox{0.8}{\codiscardCirc[-][X]}$. Following this convention, the axioms of cocartesian bicategories are in Fig.~\ref{fig:cocb axioms}; they can also be obtained from Fig.~\ref{fig:cb axioms} by inverting both the colours and the order.

It is not surprising that in a cocartesian bicategory $\Cat{C}$, every homset $\Cat{C}[X,Y]$ carries a join semi-lattice with bottom, where $c\sqcup d$ and $\bot$ are defined for all arrows $c,d\colon X \to Y$ as follows.
\begin{equation}\label{eq:def:cup}
c \sqcup d \defeq \unionCirc{c}{d}[X][Y]
\qquad
\bot \defeq \bottomCirc[X][Y]
\end{equation}

\section{Linear Bicategories}\label{sec:linbic}
We have seen that $\Relp$ forms a cartesian, and $\Relm$ a cocartesian bicategory. Categorically, they are remarkably similar --- as evidenced by the isomorphism $\nega{(\cdot)}$ --- but from a logical viewpoint they represent two complimentary parts of $\FOL$: $\Relp$ the existential conjunctive fragment, and $\Relm$ the universal disjunctive fragment. To discover the full story, we must merge them into one entity and study the algebraic interactions between them. However, the coexistence of two different compositions $\seq[+]$ and $\seq[-]$ brings us out of the realm of ordinary categories. The solution is linear bicategories~\cite{cockett2000introduction}. %
Here $\seq[+]$ linearly distributes over $\seq[-]$, as in Pierce's calculus.
To keep our development easier, we stick to the poset enriched case and rely on diagrams, using white and black to distinguish $\seq[+]$ and $\seq[-]$.

\begin{definition}\label{def:linear bicategory}
A \emph{linear bicategory} $(\Cat{C}, \seq[+], \id[+], \seq[-], \id[-])$ consists of two poset enriched categories $(\Cat{C}, \seq[+], \id[+])$ and $(\Cat{C}, \seq[-], \id[-])$ with the same objects, arrows and orderings but possibly different identities and compositions such that  $\seq[+]$ linearly distributes over $\seq[-]$ (i.e., \eqref{ax:leftLinDistr} and \eqref{ax:rightLinDistr} in Fig.~\ref{fig:closed lin axioms} hold). %
 A \emph{symmetric monoidal linear bicategory} $(\Cat{C}, \seq[+], \id[+], \seq[-], \id[-], \tensor[+],\symm[+], \tensor[-], \symm[-], \unittensor)$, shortly  $(\Cat{C},\tensor[+], \tensor[-], \unittensor)$, consists of
a linear bicategory $(\Cat{C}, \seq[+], \id[+], \seq[-], \id[-])$ and  two poset enriched symmetric monoidal categories $(\Cat{C}, \tensor[+],  \unittensor)$ and $(\Cat{C}, \tensor[-], \unittensor)$ such that \  $\tensor[+]$ and $\tensor[-]$ agree on objects, i.e., $X \tensor[+]Y= X\tensor[-]Y$, share the same unit $\unittensor$ and

\noindent 1. there are linear strengths for $(\tensor[+],\tensor[-])$, (i.e., \eqref{ax:linStrn1}, \eqref{ax:linStrn2}, \eqref{ax:linStrn3}, \eqref{ax:linStrn4});

\noindent 2. $\tensor[-]$  preserves $\id[+]$ colaxly and $\tensor[+]$  preserves $\id[-]$ laxly
(\eqref{ax:tensorPlusIdMinus}, \eqref{ax:tensorMinusIdPlus}).

A \emph{morphism of symmetric monoidal linear bicategories} $\mathcal{F}\colon (\Cat{C_1}, \tensor[+], \tensor[-],  \unittensor) \to (\Cat{C_2},\tensor[+], \tensor[-], \unittensor)$ consists of two poset enriched symmetric monoidal  functors $\mathcal{F}^\circ \colon(\Cat{C_1},  \tensor[+], \unittensor) \to (\Cat{C_2}, \tensor[+],  \unittensor)$ and $\mathcal{F}^\bullet \colon(\Cat{C_1}, \ \tensor[-], \unittensor) \to (\Cat{C_2},  \tensor[-],  \unittensor)$ that agree on objects and arrows: $\mathcal{F}^{\circ} (X) = \mathcal{F}^{\bullet}(X)$ and $\mathcal{F}^{\circ} (c) = \mathcal{F}^{\bullet}(c)$. %
\end{definition}
\begin{remark}
In the literature %
 $\seq[+]$, $\id[+]$, $\seq[-]$ and $\id[-]$ are written with the linear logic notation $\otimes$, $\top$, $\oplus$ and $\bot$. %
Modulo this, %
the traditional notion of linear bicategory (Definition 2.1 in \cite{cockett2000introduction}) coincides with the one in Definition \ref{def:linear bicategory} whenever the 2-structure is collapsed to a poset. %

Monoidal products on linear bicategories are not much studied although %
the axioms in Definition~\ref{def:linear bicategory}.1 already appeared in \cite{naeimabadiconstructing}. They are the linear strengths of the pair $(\tensor[+], \tensor[-])$ seen as a linear functor (Definition 2.4 in \cite{cockett2000introduction}), a notion of morphism that crucially differs from ours on the fact that the $\mathcal{F}^\circ$ and $\mathcal{F}^\bullet$ may not coincide on arrows. Instead the inequalities  \eqref{ax:tensorPlusIdMinus} and \eqref{ax:tensorMinusIdPlus}  are, to the best of our knowledge, novel. Beyond being natural, they are crucial for Lemma \ref{lm:linearly distr cat} below.
\end{remark}

All linear bicategories in this paper are symmetric monoidal. We therefore omit the adjective \emph{symmetric monoidal} and refer to them simply as linear bicategories.
For a linear bicategory $(\Cat{C}, \tensor[+], \tensor[-], \unittensor)$, we will often refer to $(\Cat{C}, \tensor[+], \unittensor)$ as the \emph{white structure}, shorthand $\Cat{C}^\circ$, and to $(\Cat{C}, \tensor[-], \unittensor)$ as the \emph{black structure}, $\Cat{C}^\bullet$. Note that a morphism $\mathcal{F}$ is a mapping of objects and arrows that preserves the ordering, the white and black structures; thus we write $\mathcal{F}$ for both $\mathcal{F}^\circ$ and $\mathcal{F}^\bullet$.

If $(\Cat{C}, \tensor[+], \tensor[-], \unittensor)$ is  linear bicategory then  $(\opposite{\Cat{C}}, \tensor[+], \tensor[-], \unittensor)$ is  a linear bicategory. Similarly $(\co{\Cat{C}}, \tensor[-], \tensor[+], \unittensor)$, the bicategory obtained from $\Cat{C}$ by reversing the ordering and swapping the white and the black structure, is a linear bicategory. %

Our main example is the linear bicategory $\Rel$ of sets and relations ordered by $\subseteq$. The white structure is the symmetric monoidal category $(\Relp, \tensor[+], \singleton)$, introduced in the previous section and the black structure is $(\Relm, \tensor[-], \singleton)$. Observe that the two have the same objects, arrows and ordering. The white and black monoidal products $\tensor[+]$ and $\tensor[-]$ agree on objects and are the cartesian product of sets. As common unit object, they have the singleton set $\singleton$.
We already observed in \eqref{eq:distributivityExpres} that the white composition $\seq[+]$ distributes over $\seq[-]$ and thus \eqref{ax:leftLinDistr} and \eqref{ax:rightLinDistr} hold. By using the definitions in \eqref{eq:seqRel}, \eqref{eq:idRel} and \eqref{eq:tensorREL}, the reader can easily check also the inequalities in Definition \ref{def:linear bicategory}.1,2.

\begin{lemma}\label{lm:mix cat}\label{lm:linearly distr cat}%
	Let $(\Cat{C}, \tensor[+], \tensor[-], \unittensor)$ be a linear bicategory. For all arrows $a,b,c$ the following hold: %
	\[ (1) \; \id[-][\unittensor] \leq \id[+][\unittensor] \quad (2) \; a \tensor[+] b \leq a \tensor[-] b  \quad (3) \; (a \tensor[-] b) \tensor[+] c \leq a \tensor[-] (b \tensor[+] c) \]
\end{lemma}

\begin{remark}
As $\tensor[+]$ linearly distributes over $\tensor[-]$, it may seem that symmetric monoidal linear bicategories of Definition \ref{def:linear bicategory} are linearly distributive~\cite{de1991dialectica,cockett1997weakly}. Moreover (1), (2) of Lemma \ref{lm:mix cat} may suggest that they are mix categories \cite{cockett1997proof}. This is not the case: functoriality of $\tensor[+]$ over  $\seq[-]$ and of $\tensor[-]$ over $\seq[+]$ fails in general.
\end{remark}

\paragraph{Closed linear bicategories}
In \S~\ref{sec:cartesianbi}, we recalled adjoints of arrows in bicategories; in linear bicategories one can define \emph{linear} adjoints.
For $a \colon X \to Y$ and $b \colon Y \to X$, $a$ is  \emph{left linear adjoint} to $b$, or $b$ is \emph{right linear adjoint} to $a$, written $b \Vdash a$, if 
$\id[+][X] \leq a \seq[-] b$ and $b \seq[+] a \leq \id[-][Y]$.

Next we discuss some properties of right linear adjoints. Those of left adjoints are analogous but they do not feature in our exposition since in the categories of interest --- in next section --- left and right linear adjoint coincide. As expected, linear adjoints are unique.
\begin{lemma}\label{lemma:uniquenessla}
If $b \Vdash a$ and $c \Vdash a$, then $b=c$.
\end{lemma}

\begin{wrapfigure}{R}{0.10\textwidth}
    \vspace{-18pt}
    $\!\!\!\!\!\!\!\!\!\!\!\xymatrix{
    X \ar[r]^a & Y \\
    Z \ar[ru]_b \ar[u]|{b \seq[-]\rla{a}} \ar@(ul,dl)[u]^{c}
    }$
    \vspace{-15pt}
\end{wrapfigure}
By virtue of the above result we can write $\rla{a}\colon Y \to X$ for \emph{the} right linear adjoint of $a\colon X \to Y$. %
With this notation one can write the \emph{left residual} of $b\colon Z \to Y$ by $a\colon X \to Y$ as $b\seq[-]\rla{a} \colon Z \to X$. The left residual is the greatest arrow $Z \to X$ making the diagram on the right commute laxly in $\Cat{C}^\circ$, namely if $c\seq[+] a \leq b$ then $c\leq b\seq[-]\rla{a}$. This can be equivalently expressed as: 

\begin{lemma}[Residuation]\label{lm:residuation}
$a \leq b$ iff $\id[+][X] \leq b \seq[-] \rla{a} $. %
\end{lemma}

\begin{definition}
A linear bicategory $(\Cat{C},\tensor[+],\tensor[-],\unittensor)$ is said to be \emph{closed} if every $a\colon X \to Y$ has both a left and a right linear adjoint and the white symmetry is both left and right linear adjoint to the black symmetry, i.e.\ \eqref{ax:tauSymmPlus}, \eqref{ax:gammaSymmPlus}, \eqref{ax:tauSymmMinus} and \eqref{ax:gammaSymmMinus} in Fig.~\ref{fig:closed lin axioms} hold.
\end{definition}

$\Rel$ is a a closed linear bicategory: both left and right linear adjoints of a relation $R \subseteq X \times Y $ are given by $\op{\nega{R}}=\{(y,x) \mid (x,y) \notin R \} \subseteq Y \times X$. With this, it is easy to see that $\symm[-] \Vdash \symm[+] \Vdash \symm[-]$ in $\Rel$.

Observe that if a linear bicategory $(\Cat{C},\tensor[+],\tensor[-],\unittensor)$ is closed, then also $(\opposite{\Cat{C}},\tensor[+],\tensor[-],\unittensor)$ and $(\co{\Cat{C}}, \tensor[-], \tensor[+], \unittensor)$ are closed.
The assignment $a \mapsto \rla{a}$ gives rise to an identity on objects functor $\rla{(\cdot)} \colon \Cat{C} \to \opposite{(\co{\Cat{C}})}$.

\begin{proposition}\label{prop:rlamorphism}
$\rla{(\cdot)} \colon \Cat{C} \to \opposite{(\co{\Cat{C}})}$ is a morphism of linear bicategories, i.e., the laws in the first two columns of Table \ref{table:rlaproperties}.(b) hold.
\end{proposition}
Hereafter, the diagram obtained from $\Circ{c}$, by taking its mirror image $\CircOp[+]{c}$ and then its photographic negative $\CircOp[-]{c}$ will denote $\rla{\Circ{c}}$.

\begin{table*}[t]\caption{Properties of first order bicategories.}\label{table:daggerrlalattice}\label{table:daggerproperties}\label{table:rlaproperties}\label{table:enrichment}
    \[
        \tiny{
        \begin{array}{@{}c@{\;}|@{\;}c@{\;}|@{\;}c@{\,}|@{\,}c@{}}
            \toprule
            \begin{array}{@{}c@{\;}c@{\;}c@{\;}c@{}}
                \multicolumn{4}{c}{
                    \text{(a) Properties of $\op{(\cdot)} \colon (\Cat{C},\copier[+]\!, \cocopier[+]\!,  \copier[-]\!, \cocopier[-]) \to (\opposite{\Cat{C}},\cocopier[+]\!, \copier[+]\!,  \cocopier[-]\!, \copier[-])$}
                }
                \\
                \midrule
                \multicolumn{2}{c}{
                    \text{if }c\leq d\text{ then }\op{c} \leq \op{d}
                }
                &
                \multicolumn{2}{c}{
                    \op{(\op{c})}= c
                }
                \\
                \op{(c \seq[+] d)} = \op{d} \seq[+] \op{c}
                &\op{(\id[+][X])}=\id[+][X]
                &\op{(\cocopier[+][X])}= \copier[+][X]
                &\op{(\codiscard[+][X])}= \discard[+][X]
                \\
                \op{(c \tensor[+] d)} = \op{c} \tensor[+] \op{d}
                &\op{(\symm[+][X][Y])} = \symm[+][Y][X]
                &\op{(\copier[+][X])}= \cocopier[+][X]
                &\op{(\discard[+][X])}= \codiscard[+][X]
                \\
                \midrule
                \op{(c \seq[-] d)} = \op{d} \seq[-] \op{c}
                &\op{(\id[-][X])} = \id[-][X]
                &\op{(\cocopier[-][X])}= \copier[-][X]
                &\op{(\codiscard[-][X])}= \discard[-][X]
                \\
                \op{(c \tensor[-] d)} = \op{c} \tensor[-] \op{d}
                &\op{(\symm[-][X][Y])} = \symm[-][Y][X]
                &\op{(\copier[-][X])}= \cocopier[-][X]
                &\op{(\discard[-][X])}= \codiscard[-][X]
            \end{array}
            &
            \begin{array}{@{}c@{\;}c@{\,}|@{\,}c@{\;}c@{}}
                \multicolumn{4}{c}{
                    \text{(b) Properties of $\rla{(\cdot)} \colon (\Cat{C},\copier[+]\!, \cocopier[+]\!,  \copier[-]\!, \cocopier[-]) \to (\opposite{(\co{\Cat{C}})},\cocopier[-]\!, \copier[-]\!,  \cocopier[+]\!, \copier[+])$}
                }\\
                \midrule
                \multicolumn{2}{c}{
                    \text{if }c\leq d\text{ then }\rla{c} \geq \rla{d} \vphantom{\op{(c)}}
                }
                &
                \multicolumn{2}{c}{
                    \rla{(\rla{c})}= c \vphantom{\op{(c)}}
                }
                \\
                \rla{(c \seq[+] d)} = \rla{d} \seq[-] \rla{c}   \vphantom{\op{(\copier[+][X])}}
                &\rla{(\id[+][X])}=\id[-][X]  \vphantom{\op{(\copier[+][X])}}
                &\rla{(\cocopier[+][X])}= \copier[-][X] \vphantom{\op{(\copier[+][X])}}
                &\rla{(\codiscard[+][X])}= \discard[-][X] \vphantom{\op{(\copier[+][X])}}
                \\
                \rla{(c \tensor[+] d)} = \rla{c} \tensor[-] \rla{d}  \vphantom{\op{(\copier[+][X])}}
                &\rla{(\symm[+][X][Y])} = \symm[-][Y][X]  \vphantom{\op{(\copier[+][X])}}
                &\rla{(\copier[+][X])}= \cocopier[-][X] \vphantom{\op{(\copier[+][X])}}
                &\rla{(\discard[+][X])}= \codiscard[-][X] \vphantom{\op{(\copier[+][X])}}
                \\[1.2em]
                \rla{(c \seq[-] d)} = \rla{d} \seq[+] \rla{c} \vphantom{\op{(\copier[+][X])}}
                &\rla{(\id[-][X])} = \id[+][X] \vphantom{\op{(\copier[+][X])}}
                &\rla{(\cocopier[-][X])}= \copier[+][X] \vphantom{\op{(\copier[+][X])}}
                &\rla{(\codiscard[-][X])}= \discard[+][X] \vphantom{\op{(\copier[+][X])}}
                \\
                \rla{(c \tensor[-] d)} = \rla{c} \tensor[+] \rla{d}  \vphantom{\op{(\copier[+][X])}}
                &\rla{(\symm[-][X][Y])} = \symm[+][Y][X]  \vphantom{\op{(\copier[+][X])}}
                &\rla{(\copier[-][X])}= \cocopier[+][X] \vphantom{\op{(\copier[+][X])}}
                &\rla{(\discard[-][X])}= \codiscard[+][X] \vphantom{\op{(\copier[+][X])}}
            \end{array}
            &
            \begin{array}{c@{\;}c}
                \multicolumn{2}{c}{\text{(c) Interaction of $\op{\cdot}$ and $\rla{\cdot}$ with $\sqcap$ and $\sqcup$}}
                \\ \midrule
                \addlinespace[3pt]
                \op{(c \sqcap d)} = \op{c} \sqcap \op{d}          
                & 
                \op{\top} = \top       
                \\[2pt]
                \op{(c \sqcup d)} = \op{c} \sqcup \op{d}          
                & 
                \op{\bot} = \bot 
                \\[2pt]
                \rla{(c \sqcap d)} = \rla{c} \sqcup \rla{d}       
                &
                \rla{(\top)} = \bot 
                \\[2pt]
                \rla{(c \sqcup d)} = \rla{c} \sqcap \rla{d}
                &
                \rla{(\bot)} = \top  
                \\[2pt]
                \multicolumn{2}{c}{   
                    \rla{(\op{c})} = \op{(\rla{c})} 
                }
                \\
                \addlinespace[3pt]
            \end{array}
            &
            \begin{array}{@{}cccc@{}}
                \multicolumn{4}{@{}c@{}}{
                    \text{(d) Laws of Boolean algebras}
                }  
                \\
                \midrule
                \addlinespace[5pt]
                \multicolumn{4}{@{}c@{}}{
                    \begin{array}{cc@{}}
                        \multicolumn{2}{c}{
                            c \sqcap (d \sqcup e) = (c \sqcap d) \sqcup (c \sqcap e)     
                        }
                        \\[2pt]
                        \multicolumn{2}{c}{
                            c \sqcup (d \sqcap e) = (c \sqcup d) \sqcap (c \sqcup e) 
                        }
                        \\[2pt]
                        \nega{(c \sqcap d)} = \nega{c} \sqcup \nega{d} 
                        &  
                        \nega{\top} = \bot  
                        \\[2pt]
                        \nega{(c \sqcup d)} = \nega{c} \sqcap \nega{d} 
                        & 
                        \nega{\bot} = \top 
                        \\[2pt]
                        c \sqcap \nega{c} = \bot
                        &
                        c \sqcup \nega{c} = \top
                        \\[3pt]
                    \end{array}
                }
            \end{array}
            \\
            \midrule
            \multicolumn{4}{@{}c@{}}{
                \begin{array}{@{}l|cccccc@{}}
                    \multirow{2}{*}{
                        \makecell[l]{
                            \text{(e) Enrichment over} \\ \text{join-meet semilattices}
                        }
                    }
                    &
                    c \seq[+] (d \sqcup e) = (c \seq[+] d) \sqcup (c \seq[+] e)
                    &
                    (d \sqcup e) \seq[+] c = (d \seq[+] c) \sqcup (e \seq[+] c)
                    &
                    c \seq[+] \bot = \bot = \bot \seq[+] c 
                    &
                    c \tensor[+] (d \sqcup e) = (c \tensor[+] d) \sqcup (c \tensor[+] e)
                    &
                    (d \sqcup e) \tensor[+] c = (d \tensor[+] c) \sqcup (e \tensor[+] c)
                    &
                    c \tensor[+] \bot = \bot = \bot \tensor[+] c
                    \\
                    &
                    c \seq[-] (d \sqcap e) = (c \seq[-] d) \sqcap (c \seq[-] e)
                    &
                    (d \sqcap e) \seq[-] c = (d \seq[-] c) \sqcap (e \seq[-] c)
                    &
                    c \seq[-] \top = \top = \top \seq[-] c
                    &
                    c \tensor[-] (d \sqcap e) = (c \tensor[-] d) \sqcap (c \tensor[-] e)
                    &
                    (d \sqcap e) \tensor[-] c = (d \tensor[-] c) \sqcap (e \tensor[-] c)
                    &
                    c \tensor[-] \top = \top = \top \tensor[-] c
                \end{array}
            }
            \\
            \bottomrule
        \end{array}
        }
    \]
\end{table*}

\section{First Order Bicategories}\label{sec:fobic}

\begin{figure*}[t]
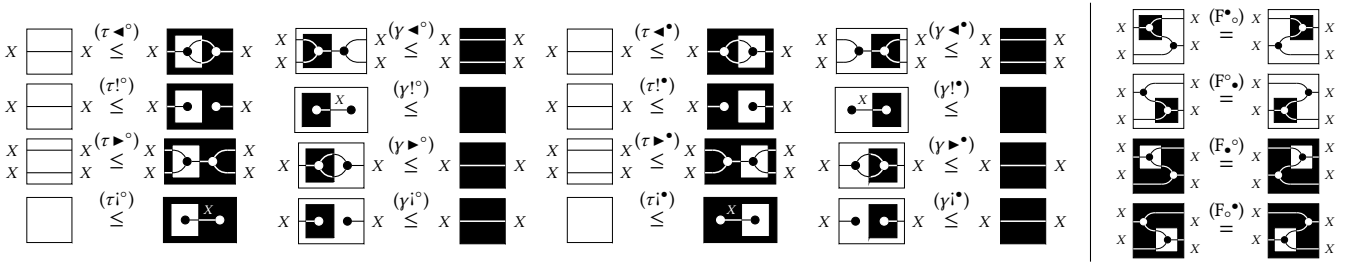

    \mylabel{ax:tauCopierPlus}{$\tau\!\copier[+]$}
    \mylabel{ax:tauDiscardPlus}{$\tau\discard[+]$}
    \mylabel{ax:tauCocopierPlus}{$\tau\!\cocopier[+]$}
    \mylabel{ax:tauCodiscardPlus}{$\tau\codiscard[+]$}

    \mylabel{ax:gammaCopierPlus}{$\gamma\!\copier[+]$}
    \mylabel{ax:gammaDiscardPlus}{$\gamma\discard[+]$}
    \mylabel{ax:gammaCocopierPlus}{$\gamma\!\cocopier[+]$}
    \mylabel{ax:gammaCodiscardPlus}{$\gamma\codiscard[+]$}

    \mylabel{ax:tauCopierMinus}{$\tau\!\copier[-]$}
    \mylabel{ax:tauDiscardMinus}{$\tau\discard[-]$}
    \mylabel{ax:tauCocopierMinus}{$\tau\!\cocopier[-]$}
    \mylabel{ax:tauCodiscardMinus}{$\tau\codiscard[-]$}

    \mylabel{ax:gammaCopierMinus}{$\gamma\!\copier[-]$}
    \mylabel{ax:gammaDiscardMinus}{$\gamma\discard[-]$}
    \mylabel{ax:gammaCocopierMinus}{$\gamma\!\cocopier[-]$}
    \mylabel{ax:gammaCodiscardMinus}{$\gamma\codiscard[-]$}

    \mylabel{ax:bwFrob}{F$\tiny{\begin{array}{@{}c@{}c@{}} \bullet & \\[-3pt] & \circ \end{array}}$}
    \mylabel{ax:bwFrob2}{F$\tiny{\begin{array}{@{}c@{}c@{}} \circ & \\[-3pt] & \bullet \end{array}}$}
    \mylabel{ax:wbFrob}{F$\tiny{\begin{array}{@{}c@{}c@{}}  & \circ  \\[-3pt] \bullet & \end{array}}$}
    \mylabel{ax:wbFrob2}{F$\tiny{\begin{array}{@{}c@{}c@{}}  & \bullet  \\[-3pt] \circ & \end{array}}$}
    $
        \begin{array}{@{}c@{}c@{}c@{}c|c@{}}
            \begin{array}{@{}c@{}c@{}c@{}}
                \idCirc[+][X] & \!\!\Lleq{\eqref*{ax:tauCopierPlus}}\!\! & 
    \InputIfFileExists{axiomsNEW/linadj/comMmon.tikz}{}{\input{tikz/axiomsNEW/linadj/comMmon.tikz}}
 \\
                \idCirc[+][X] & \!\!\Lleq{\eqref*{ax:tauDiscardPlus}}\!\! & 
    \InputIfFileExists{axiomsNEW/linadj/bangMcobang.tikz}{}{\input{tikz/axiomsNEW/linadj/bangMcobang.tikz}}
 \\
                
    \InputIfFileExists{axiomsNEW/id2P.tikz}{}{\input{tikz/axiomsNEW/id2P.tikz}}
 & \!\!\Lleq{\eqref*{ax:tauCocopierPlus}}\!\! &  
    \InputIfFileExists{axiomsNEW/linadj/monMcom.tikz}{}{\input{tikz/axiomsNEW/linadj/monMcom.tikz}}
 \\
                \emptyCirc[+]         & \!\!\Lleq{\eqref*{ax:tauCodiscardPlus}}\!\! & 
    \InputIfFileExists{axiomsNEW/linadj/cobangMbang.tikz}{}{\input{tikz/axiomsNEW/linadj/cobangMbang.tikz}}

            \end{array} &
            \begin{array}{@{}c@{}c@{}c@{}}
                
    \InputIfFileExists{axiomsNEW/linadj2/monPcom.tikz}{}{\input{tikz/axiomsNEW/linadj2/monPcom.tikz}}
     & \!\!\Lleq{\eqref*{ax:gammaCopierPlus}}\!\! & 
    \InputIfFileExists{axiomsNEW/id2M.tikz}{}{\input{tikz/axiomsNEW/id2M.tikz}}
 \\
                
    \InputIfFileExists{axiomsNEW/linadj2/cobangPbang.tikz}{}{\input{tikz/axiomsNEW/linadj2/cobangPbang.tikz}}
 & \!\!\Lleq{\eqref*{ax:gammaDiscardPlus}}\!\! & \emptyCirc[-]  \\
                
    \InputIfFileExists{axiomsNEW/linadj/comPmon.tikz}{}{\input{tikz/axiomsNEW/linadj/comPmon.tikz}}
     & \!\!\Lleq{\eqref*{ax:gammaCocopierPlus}}\!\! & \idCirc[-][X] \\
                
    \InputIfFileExists{axiomsNEW/linadj/bangPcobang.tikz}{}{\input{tikz/axiomsNEW/linadj/bangPcobang.tikz}}
 & \!\!\Lleq{\eqref*{ax:gammaCodiscardPlus}}\!\! & \idCirc[-][X]
            \end{array} &
            \begin{array}{@{}c@{}c@{}c@{}}
                \idCirc[+][X] & \!\!\Lleq{\eqref*{ax:tauCopierMinus}}\!\! & 
    \InputIfFileExists{axiomsNEW/linadj2/comMmon.tikz}{}{\input{tikz/axiomsNEW/linadj2/comMmon.tikz}}
 \\
                \idCirc[+][X] & \!\!\Lleq{\eqref*{ax:tauDiscardMinus}}\!\! & 
    \InputIfFileExists{axiomsNEW/linadj2/bangMcobang.tikz}{}{\input{tikz/axiomsNEW/linadj2/bangMcobang.tikz}}
 \\
                
    \InputIfFileExists{axiomsNEW/id2P.tikz}{}{\input{tikz/axiomsNEW/id2P.tikz}}
 & \!\!\Lleq{\eqref*{ax:tauCocopierMinus}}\!\! &  
    \InputIfFileExists{axiomsNEW/linadj2/monMcom.tikz}{}{\input{tikz/axiomsNEW/linadj2/monMcom.tikz}}
 \\
                \emptyCirc[+]         & \!\!\Lleq{\eqref*{ax:tauCodiscardMinus}}\!\! & 
    \InputIfFileExists{axiomsNEW/linadj2/cobangMbang.tikz}{}{\input{tikz/axiomsNEW/linadj2/cobangMbang.tikz}}

            \end{array} &
            \begin{array}{@{}c@{}c@{}c@{}}
                
    \InputIfFileExists{axiomsNEW/linadj/monPcom.tikz}{}{\input{tikz/axiomsNEW/linadj/monPcom.tikz}}
     & \!\!\Lleq{\eqref*{ax:gammaCopierMinus}}\!\! & 
    \InputIfFileExists{axiomsNEW/id2M.tikz}{}{\input{tikz/axiomsNEW/id2M.tikz}}
 \\
                
    \InputIfFileExists{axiomsNEW/linadj/cobangPbang.tikz}{}{\input{tikz/axiomsNEW/linadj/cobangPbang.tikz}}
 & \!\!\Lleq{\eqref*{ax:gammaDiscardMinus}}\!\! & \emptyCirc[-]  \\
                
    \InputIfFileExists{axiomsNEW/linadj2/comPmon.tikz}{}{\input{tikz/axiomsNEW/linadj2/comPmon.tikz}}
     & \!\!\Lleq{\eqref*{ax:gammaCocopierMinus}}\!\! & \idCirc[-][X] \\
                
    \InputIfFileExists{axiomsNEW/linadj2/bangPcobang.tikz}{}{\input{tikz/axiomsNEW/linadj2/bangPcobang.tikz}}
 & \!\!\Lleq{\eqref*{ax:gammaCodiscardMinus}}\!\! & \idCirc[-][X]
            \end{array}
            &
            \begin{array}{@{}c@{}c@{}c@{}}
                \scalebox{0.8}{
    \InputIfFileExists{axiomsNEW/bwS2.tikz}{}{\input{tikz/axiomsNEW/bwS2.tikz}}
} & \!\!\Leq{\eqref*{ax:bwFrob}}\!\! & \scalebox{0.8}{
    \InputIfFileExists{axiomsNEW/bwZ2.tikz}{}{\input{tikz/axiomsNEW/bwZ2.tikz}}
} \\
                \scalebox{0.8}{
    \InputIfFileExists{axiomsNEW/bwS.tikz}{}{\input{tikz/axiomsNEW/bwS.tikz}}
} & \!\!\Leq{\eqref*{ax:bwFrob2}}\!\! & \scalebox{0.8}{
    \InputIfFileExists{axiomsNEW/bwZ.tikz}{}{\input{tikz/axiomsNEW/bwZ.tikz}}
} \\
                \scalebox{0.8}{
    \InputIfFileExists{axiomsNEW/wbS2.tikz}{}{\input{tikz/axiomsNEW/wbS2.tikz}}
} & \!\!\Leq{\eqref*{ax:wbFrob}}\!\! & \scalebox{0.8}{
    \InputIfFileExists{axiomsNEW/wbZ2.tikz}{}{\input{tikz/axiomsNEW/wbZ2.tikz}}
} \\
                \scalebox{0.8}{
    \InputIfFileExists{axiomsNEW/wbS.tikz}{}{\input{tikz/axiomsNEW/wbS.tikz}}
} & \!\!\Leq{\eqref*{ax:wbFrob2}}\!\! & \scalebox{0.8}{
    \InputIfFileExists{axiomsNEW/wbZ.tikz}{}{\input{tikz/axiomsNEW/wbZ.tikz}}
}
            \end{array}
        \end{array}
    $
    \caption{Additional axioms for fo-bicategories}\label{fig:fo bicat axioms}
\end{figure*}

Here we focus on the most important and novel part of the axiomatisation.
Indeed, having introduced the two main ingredients, cartesian and linear bicategories, it is time to fire up the Bunsen burner.
The remit of this section is to
 understand how the cartesian and the linear bicategory structures interact in the context of relations.
We introduce \emph{first order bicategories} that make these interactions precise. The resulting axioms echo those of cartesian bicategories but in the linear bicategory setting: recall that in a cartesian bicategory the monoid and comonoids are adjoint and satisfy the Frobenius law. Here, the white and black (co)monoids are again related, but by \emph{linear} adjunctions; moreover, they also satisfy appropriate ``linear'' counterparts of the Frobenius equations.

\begin{definition}\label{def:fobicategory}
    A \emph{first order bicategory} $(\Cat{C},\tensor[+],\tensor[-], \unittensor, \copier[+], \discard[+], \cocopier[+], \codiscard[+], \copier[-], \discard[-], \cocopier[-], \codiscard[-])$, shorthand \emph{fo-bicategory} $(\Cat{C},\copier[+], \cocopier[+],  \copier[-], \cocopier[-])$, consists of
    
    \noindent 1. a closed linear bicategory $(\Cat{C}, \tensor[+], \tensor[-],\unittensor)$,

    \noindent 2. a cartesian bicategory $(\Cat{C},\tensor[+], \unittensor, \copier[+], \discard[+], \cocopier[+], \codiscard[+])$ and

    \noindent 3. a cocartesian bicategory $(\Cat{C},\tensor[-], \unittensor, \copier[-], \discard[-], \cocopier[-], \codiscard[-])$, such that

    \noindent 4.  the white comonoid  $(\copier[+], \discard[+])$ is left and right linear adjoint to black monoid $(\cocopier[-], \codiscard[-])$ and $(\cocopier[+], \codiscard[+])$ is left and right linear adjoint to $(\copier[-], \discard[-])$, %
    i.e. the inequalities on the left of Figure \ref{fig:fo bicat axioms} hold;%

    \noindent 5. white and black (co)monoids satisfy the linear Frobenius laws, i.e. the equalities on the right of Fig.~\ref{fig:fo bicat axioms} hold.

A \emph{morphism of fo-bicategories} is a morphism of linear bicategories \emph{and} of (co)cartesian bicategories. %
\end{definition}
We have seen that $\Rel$ is a closed linear bicategory, $\Relp$ a cartesian bicategory and $\Relm$ a cocartesian bicategory.
Given~\eqref{eq:comonoidsREL}, it is easy to confirm linear adjointness  and linear Frobenius. %

Now if $(\Cat{C},\copier[+], \cocopier[+],  \copier[-], \cocopier[-])$ is a fo-bicategory then $(\opposite{\Cat{C}},\cocopier[+], \copier[+],  \cocopier[-], \copier[-])$  and $(\co{\Cat{C}},\copier[-], \cocopier[-],  \copier[+], \cocopier[+])$ are fo-bicategories: the laws of Fig.~\ref{fig:fo bicat axioms} are closed under mirror-reflection and photographic negative. %
The fourth condition in Definition \ref{def:fobicategory} entails that the linear bicategory morphism $\rla{(\cdot)}\colon \Cat{C} \to \opposite{(\co{\Cat{C}})}$ (see Prop.~\ref{prop:rlamorphism}) is a morphism of fo-bicategories and, similarly, the fifth condition that
also
$\op{(\cdot)} \colon \Cat{C} \to \opposite{\Cat{C}}$ (Prop.~\ref{prop:opcartesianfunctor}) is a morphism of fo-bicategories.

\begin{proposition}\label{prop:opfunctor}\label{prop:rlafunctor} Let $(\Cat{C},\copier[+], \cocopier[+],  \copier[-], \cocopier[-])$ be a fo-bicategory. Then
$\op{(\cdot)} \colon\! \Cat{C} \to \opposite{\Cat{C}}$ and $\rla{(\cdot)}\colon\! \Cat{C} \to \opposite{(\co{\Cat{C}})}$ are isomorphisms of fo-bicategories, namely the laws in Table~\ref{table:daggerproperties}.(a) and (b) hold.
\end{proposition}

\begin{corollary}\label{cor:oplinearadjoint}\label{cor:interactiondaggerrlalattice}
    The laws in Table~\ref{table:daggerrlalattice}.(c) hold.
\end{corollary}

The corollary follows from \eqref{eq:def:cap} and \eqref{eq:def:cup} and the laws in Tables \ref{table:daggerproperties}.(a) and (b). For instance, $\rla{(a \sqcap b)} = \rla{a} \sqcup \rla{b}$ is proved as follows.
\input{tikz/decomposedCapNeg.tex}

The next result about maps (Definition \ref{def:maps}) plays a crucial role. %
\begin{proposition}\label{prop:maps}%
For all maps $f\colon X \to Y$ and arrows $c\colon Y\to Z$, $f \seq[+] c = \rla{(\op{f})} \seq[-] c$ and thus
\[
    \begin{array}{cccc}
        \input{tikz/mapsCopier} &  \!\input{tikz/mapsDiscard.tex} &
        \!\input{tikz/mapsCocopier} & \input{tikz/mapsCodiscard.tex}
    \end{array}
\]
\end{proposition}

\noindent\begin{minipage}{0.65\columnwidth}
    \hspace*{0.3cm} For fo-bicategory $\Cat{C}$, we have the four isomorphisms in the diagram on the right, which commutes
     by Corollary \ref{cor:oplinearadjoint}.
    We can thus define the complement as the diagonal of the square, namely
    $\nega{(\cdot)} \defeq \op{(\rla{(\cdot)})}$. 
    \end{minipage}
    \begin{minipage}{0.28\columnwidth}\vspace{-0.3cm}
    \[\xymatrix{\Cat{C} \ar[r]^{\op{(\cdot)}} \ar[d]_{\rla{(\cdot)}} & \opposite{\Cat{C}} \ar[d]^{\rla{(\cdot)}} \\
    \opposite{(\co{\Cat{C}})} \ar[r]_{\op{(\cdot)}} & \co{\Cat{C}}}\]
    \end{minipage}

    In diagrams, given $\Circ[+]{c}$, its negation is $\op{(\rla{\Circ[+]{c}})} = \op{\CircOp[-]{c}} = \Circ[-]{c}$.

    Clearly $\nega{(\cdot)} \colon \Cat{C} \to \co{\Cat{C}}$ is an isomorphism of fo-bicategories.
    Moreover, it induces a Boolean algebra on each homset of $\Cat{C}$.

\begin{proposition}\label{prop:enrichment}
    Let $(\Cat{C},\copier[+], \cocopier[+],  \copier[-], \cocopier[-])$ be a fo-bicategory. Then every homset of $\Cat{C}$ is a Boolean algebra: the laws in Tab.~\ref{table:daggerrlalattice}.(d) hold. Further, $(\Cat{C}, \tensor[+], \unittensor)$ is monoidally enriched over $\sqcup$-semilattices with $\bot$, while $(\Cat{C}, \tensor[-], \unittensor)$ over $\sqcap$-semilattices with $\top$: the laws in Tab.~\ref{table:enrichment}.(e) hold.
\end{proposition}
The monoidal enrichment is interesting: as we mentioned in \S~\ref{sec:cartesianbi}, the white structure is not enriched over $\sqcap$, but it \emph{is} enriched over $\sqcup$. In $\Rel$, this is the fact that $R \seq[+](S \cup T)= (R\seq[+]S)\cup (R\seq[+]T)$. %

We conclude with a result that extends Lemma \ref{lm:residuation} with five different possibilities to express the concept of logical entailment. 
\begin{lemma}\label{lm:implications}%
In a fo-bicategory, the following are equivalent:
\[
    \begin{array}{l@{\;}l}
        (1) \boxCirc[+]{a}[X][Y] \!\!\leq\!\! \boxCirc[+]{b}[X][Y] & (2) \idCirc[+][X] \!\!\leq\!\! \seqCirc[-]{b}{a}[X][X][w][b] \\[8pt]
        (3) \idCirc[+][Y] \!\!\leq\!\! \seqCirc[-]{a}{b}[Y][Y][b] & (4) \topCirc[X][Y] \!\!\leq\!\! \unionCirc{a}{b}[X][Y][b] \\[10pt]
        \multicolumn{2}{c}{
            (5)\; \emptyCirc[+] \leq \circleCirc{a}{b}[b]
        }
    \end{array}
\]
\end{lemma}

\subsection{The calculus of neo-Peircean relations as a freely generated first order bicategory}\label{sec:freely}

We now return to $\NPR$. Recall that $\syninclusion$ is the precongruence  obtained from the axioms in Fig.s~\ref{fig:cb axioms}, \ref{fig:cocb axioms}, \ref{fig:closed lin axioms} and \ref{fig:fo bicat axioms}. Its soundness (half of Theorem~\ref{thm:completeness}) is immediate since $\Rel$ is a fo-bicategory.
\begin{proposition}\label{prop:soundness}
For all terms $c,d\colon n \to m$, if $c\syninclusion d$ then $c\seminclusion d$.
\end{proposition}

Next, we show how $\NPR$ gives rise to a fo-bicategory $\LCB$. Objects are natural numbers and monoidal products $\tensor[][][]$ are defined as addition with unit object $0$.
Arrows from $n$ to $m$ are terms $c\colon n \to m$ modulo syntactic equivalence $\synequivalence$, namely $\LCB{}[n,m] \defeq \{[c]_{\synequivalence} \mid c\colon n \to m\}$. Observe that this is well defined since $\synequivalence$ is well-typed. Since $\synequivalence$ is a congruence, the operations  $\seq[]$ and  $\tensor[]$ on terms are well defined on equivalence classes: $[t_1]_{\synequivalence}\seq[][][] [t_2]_{\synequivalence} \defeq [t_1\seq[][][] t_2]_{\synequivalence}$ and $[t_1]_{\synequivalence}\tensor[][][] [t_2]_{\synequivalence} \defeq [t_1\tensor[][][] t_2]_{\synequivalence}$. By fixing as partial order the syntactic inclusion $\syninclusion$, one can easily prove the following.

\begin{proposition}\label{prop:LCBfo} 
$\LCB$ is a first order bicategory.
\end{proposition}
A  useful consequence is that, for any interpretation $\interpretation=(X,\rho)$, the semantics $\interpretationFunctor$ gives rise to a morphism $\interpretationFunctor \colon \LCB \to \Rel$ of fo-bicategories: it is defined on objects as $n \mapsto X^n$ and on arrows by the inductive definition in \eqref{fig:semantics}. To see that it is a morphism, note that, by \eqref{fig:semantics}, all the structure of (co)cartesian bicategories and of linear bicategories is preserved (e.g. $\interpretationFunctor (\copier[+][1])=\copier[+][X]$). Moreover, the ordering is preserved by Prop.~\ref{prop:soundness}. Note that, by construction,
\begin{equation}\label{eq:free}
\interpretationFunctor(1)= X \text{ and }\interpretationFunctor(R^\circ)= \rho (R) \text{ for all }R\in \sign\text{.}
\end{equation}

Actually,  $\interpretationFunctor$ is the unique such morphism of fo-bicategories.
This is a consequence of a more general universal property: $\Rel$ can be replaced with an arbitrary fo-bicategory $\Cat{C}$.
To see this, we first need to generalise the notion of interpretation.

\begin{definition}\label{def:intCAT}
Let $\sign$ be a monoidal signature and $\Cat{C}$ a first order bicategory. An \emph{interpretation} $\interpretation=(X,\rho)$ of $\sign$ in $\Cat{C}$ consists of an object $X$ of $\Cat{C}$ and an arrow $\rho(R)\colon X^{n} \to X^{m}$ for each $R \in \sign[n,m]$. %
\end{definition}

With this definition, we can state that $\LCB$ is the fo-bicategory freely generated by $\sign$. %

\begin{proposition}\label{prop:free}
Let $\sign$ be a monoidal signature, $\Cat{C}$ a first order bicategory and $\interpretation=(X,\rho)$  an interpretation of $\sign$ in $\Cat{C}$. There exists a unique morphism of fo-bicategories $\interpretationFunctor\colon \LCB \to \Cat{C}$ such that  $\interpretationFunctor(1)= X$ and $\interpretationFunctor(R^\circ)= \rho (R)$ for all $R\in \sign$.
\end{proposition}

\begin{figure*}
$
    \arraycolsep=2pt
    \begin{array}{ccccccc}
        \scalebox{0.8}{\propCirc[+]{c}} \!\!\Lleq{\eqref*{ax:comPlusLaxNat}}\!\! \scalebox{0.8}{\propTensorCirc[+]{c}{c}},
        &
        \scalebox{0.8}{\propTensorCirc[-]{c}{c}} \!\!\Lleq{\eqref*{ax:comMinusLaxNat}}\!\! \scalebox{0.8}{\propCirc[-]{c}},
        &
        \scalebox{0.8}{\propCirc[+]{c}} \!\!\Lleq{\eqref*{ax:discPlusLaxNat}}\!\! \scalebox{0.8}{\emptyCirc[+]},
        &
        \scalebox{0.8}{\emptyCirc[-]} \!\!\Lleq{\eqref*{ax:discMinusLaxNat}}\!\! \scalebox{0.8}{\propCirc[-]{c}},
        &
        \scalebox{0.8}{
    \InputIfFileExists{axiomsNEW/propositional/distributivity1.tikz}{}{\input{tikz/axiomsNEW/propositional/distributivity1.tikz}}
} \!\!\stackrel{\footnotesize{\stackanchor{\eqref*{ax:leftLinDistr}}{\eqref*{ax:rightLinDistr}}}}{\leq}\!\! \scalebox{0.8}{
    \InputIfFileExists{axiomsNEW/propositional/distributivity2.tikz}{}{\input{tikz/axiomsNEW/propositional/distributivity2.tikz}}
},
        &
        \scalebox{0.8}{\emptyCirc[+]} \!\!\stackrel{\footnotesize{\stackanchor{\eqref*{ax:tauRPlus}}{\eqref*{ax:tauRMinus}}}}{\leq}\!\!  \scalebox{0.8}{
    \InputIfFileExists{axiomsNEW/propositional/tauR.tikz}{}{\input{tikz/axiomsNEW/propositional/tauR.tikz}}
},
        &
        \scalebox{0.8}{
    \InputIfFileExists{axiomsNEW/propositional/gammaR.tikz}{}{\input{tikz/axiomsNEW/propositional/gammaR.tikz}}
} \!\!\stackrel{\footnotesize{\stackanchor{\eqref*{ax:gammaRPlus}}{\eqref*{ax:gammaRMinus}}}}{\leq}\!\! \scalebox{0.8}{\emptyCirc[-]}
    \end{array}
$
\caption{The axioms in Figures \ref{fig:cb axioms}, \ref{fig:cocb axioms} and \ref{fig:closed lin axioms} reduce to those above for diagrams of type $\unittensor \to \unittensor$}
\label{fig:propositionalcalculus}
\end{figure*}

\section{Diagrammatic first order theories}\label{sec:theories}%
Here we take the first steps towards completeness and show that for first order theories, fo-bicategories play an analogous role to cartesian categories in Lawvere's functorial semantics~\cite{LawvereOriginalPaper}. %

A \emph{first order theory} $\T{T}$ is a pair $(\sign, \T{I})$ where $\sign$ is a signature and $\T{I}$ is a set of \emph{axioms}: pairs $(c,d)$ for  $c,d \colon n \to m$ in $\LCB$.
 A \emph{model} of $\T{T}$ is an interpretation $\interpretation$ of $\sign$ where if $(c,d) \in \T{I}$, then $\interpretationFunctor(c) \subseteq \interpretationFunctor(d)$.

\begin{example}\label{ex:lin ord}%
The simplest case is $\sign=\T{I}=\varnothing$.
An interpretation is a set: all sets, including the empty set $\varnothing$, are models.

Next take  $\sign = \varnothing$ and $\T{I}= \{ (\, \scalebox{0.8}{\emptyCirc[+]} , \scalebox{0.8}{
    \InputIfFileExists{axioms/cb/plus/codiscDisc.tikz}{}{\input{tikz/axioms/cb/plus/codiscDisc.tikz}}
} \,) \}$. An interpretation $\interpretation$ is a set $X$. By~\eqref{fig:semantics}, $\interpretationFunctor{( \scalebox{0.8}{
    \InputIfFileExists{axioms/cb/plus/codiscDisc.tikz}{}{\input{tikz/axioms/cb/plus/codiscDisc.tikz}}
})} = \{(\star,x)\mid x\in X\} \seq[+] \{(x,\star) \mid x\in X\}$, so  $\interpretationFunctor{( \scalebox{0.8}{
    \InputIfFileExists{axioms/cb/plus/codiscDisc.tikz}{}{\input{tikz/axioms/cb/plus/codiscDisc.tikz}}
})} = \{(\star,\star)\}$ if $X \neq \varnothing$, but
$\varnothing$ if $X = \varnothing$.
Instead, $\interpretationFunctor{\scalebox{0.8}{(\emptyCirc[+]} )} = \{(\star,\star)\}$ always, since $X^0$ is always $\singleton$.
 Succinctly, $\interpretationFunctor{\scalebox{0.8}{(\emptyCirc[+]} )} \subseteq \interpretationFunctor{( \scalebox{0.8}{
    \InputIfFileExists{axioms/cb/plus/codiscDisc.tikz}{}{\input{tikz/axioms/cb/plus/codiscDisc.tikz}}
})}$ iff $X \neq \varnothing$: models are \emph{non-empty sets}.

Finally, take $\sign = \{ R \colon 1 \to 1 \}$ and let $\mathbb{I}$ be as follows:

\noindent      $ \{\, ( \scalebox{0.7}{\idCirc[+]} , \scalebox{0.7}{
    \InputIfFileExists{linOrdR.tikz}{}{\input{tikz/linOrdR.tikz}}
}) ,\;
          (\scalebox{0.7}{
    \InputIfFileExists{linOrdSeq.tikz}{}{\input{tikz/linOrdSeq.tikz}}
}\!, \scalebox{0.7}{
    \InputIfFileExists{linOrdR.tikz}{}{\input{tikz/linOrdR.tikz}}
}\!) ,\;
          (\scalebox{0.7}{
    \InputIfFileExists{linearOrdIntersection.tikz}{}{\input{tikz/linearOrdIntersection.tikz}}
}\!, \scalebox{0.7}{\idCirc[+]}) ,\;
          (\scalebox{0.7}{
    \InputIfFileExists{linOrdTop.tikz}{}{\input{tikz/linOrdTop.tikz}}
}\!, \scalebox{0.7}{
    \InputIfFileExists{linearOrdUnion.tikz}{}{\input{tikz/linearOrdUnion.tikz}}
}\!)\,\} $.
An interpretation is a set $X$ and a relation $R \subseteq X \times X$. It is a model iff $R$ is an order, i.e., reflexive, transitive, antisymmetric and total. %
\end{example}

Monoidal signatures $\sign$, differently from usual $\FOL$ alphabets,
do not have function symbols. The reason is that, by adding the axioms below to $\mathbb{I}$, one forces a symbol $f\colon n \to 1\in\sign$ to be a function.%
\begin{equation}\label{TMAP}\tag{$\TMAP{f}$} 
    \InputIfFileExists{copierFunc1.tikz}{}{\input{tikz/copierFunc1.tikz}}
 \leq 
    \InputIfFileExists{copierFunc2.tikz}{}{\input{tikz/copierFunc2.tikz}}
 \qquad \qquad \discardCirc[+][n] \leq 
    \InputIfFileExists{discardFunc.tikz}{}{\input{tikz/discardFunc.tikz}}
 \end{equation} %
Indeed, as we remarked in \S~\ref{sec:cartesianbi}, $f\subseteq X^n \times X$ satisfies $\TMAP{f}$ if and only if it is single valued and total, i.e. a function. %
We depict functions as \!\!$\scalebox{0.8}{\funcCirc[+]{f}[n]}$ and constants, being $0\to 1$ functions, as $\scalebox{0.8}{\constCirc[+]{k}}$.

The axioms of a theory together with $\syninclusion$ form a deduction system. Formally, the \emph{deduction relation} induced by $\T{T}=(\Sigma, \T{I})$ is the closure (see \eqref{eq:pc}) of $\syninclusion \cup \;\wtrel$, i.e.\ $\syninclusionT{\T{T}} \defeq \pcong{\syninclusion \cup \;\wtrel}$. We write $\synequivalenceT{\T{T}}$ for $\syninclusionT{\T{T}} \cap \synreverdedinclusionT{\T{T}}$.
\begin{proposition}\label{prop:soundnessoftheories}
Let $\T{T}=(\sign, \T{I})$ be a theory. If $c \syninclusionT{\T{T}} d$, then $\interpretationFunctor(c) \subseteq \interpretationFunctor(d)$ for all models $\interpretation$.
\end{proposition}

\begin{example}
Consider the theory $\T{T}$ with $\sign=\{k\colon 0 \to 1\}$ and axioms $\TMAP{k}$.
By the definitions of $\copier[+][0]$ and $\discard[+][0]$ in
Tab.~\ref{fig:sugar}, these are:
\begin{equation}\label{TMAPk}\tag{$\TMAP{k}$}
    
    \InputIfFileExists{copierConst1.tikz}{}{\input{tikz/copierConst1.tikz}}
 \leq 
    \InputIfFileExists{copierConst2.tikz}{}{\input{tikz/copierConst2.tikz}}
 \qquad \qquad \emptyCirc[+] \leq 
    \InputIfFileExists{discardConst.tikz}{}{\input{tikz/discardConst.tikz}}

\end{equation}
An interpretation $\interpretation$ of $\sign$ consists of a set $X$ and a relation $k\subseteq \singleton \times X$. An interpretation is a model iff $k$ is a function of type $\singleton \to X$. %
One can easily prove that in all models the domain is non-empty:
\input{tikz/proofs/nonEmptiness.tex}
\end{example}

\paragraph{Contradictory vs trivial theories.}
The distinction between contradictory and trivial theories is so subtle that, as shown in Remark \ref{rmk:ambiguity}, it is invisible in $\FOL$. Let us start with the definition.
\begin{definition} %
A theory $\T{T}$ is \emph{contradictory} if $\scalebox{0.8}{\emptyCirc[+]} \syninclusionT{\T{T}} \scalebox{0.8}{\emptyCirc[-]}$. It is \emph{trivial} if $\scalebox{0.8}{\codiscardCirc[+]} \syninclusionT{\T{T}} \scalebox{0.8}{\codiscardCirc[-]}$.
\end{definition}
Triviality implies all models have domain $\varnothing$: $\interpretationFunctor{(\scalebox{0.8}{\codiscardCirc[+]})}=  \{(\star,x)\mid x\in X\}$ is included in $\varnothing =\interpretationFunctor{(\scalebox{0.8}{\codiscardCirc[-]})}$ iff $X=\varnothing$. On the other hand, contradictory theories cannot have a model, not even when $X=\varnothing$: since $\interpretationFunctor{(\scalebox{0.8}{\emptyCirc[+]})}=  \{(\star,\star)\}$ and $\interpretationFunctor{(\scalebox{0.8}{\emptyCirc[-]})}=  \varnothing$ independently of $X$.
Every contradictory theory is trivial (see Prop.~\ref{lemma:contraddictoryimpliestrivial} in App.~\ref{app:theories}).

In trivial theories diagrams of type $0 \to 0$ can be quite interesting (see Example \ref{ex:propcalculus}),
while those with a different type collapse: %
\begin{lemma}\label{lemma:trivalallequal}
Let $\T{T}$ be a trivial theory and $c\colon n \to m+1, d\colon m+1 \to n$ be arrows in $\LCB$.
Then $\top \syninclusionT{\T{T}} c \syninclusionT{\T{T}} \bot$ and $\top \syninclusionT{\T{T}} d \syninclusionT{\T{T}} \bot$.

\end{lemma}
\begin{example}[The trivial theory of propositional calculus]\label{ex:propcalculus}%
Let $\T{T}=(\sign, \T{I})$ be the theory where $\sign$ contains only symbols $P,Q,R\dots$ of type $0 \to 0$ and $\T{I}=\{  (\scalebox{0.8}{\codiscardCirc[+]} , \scalebox{0.8}{\codiscardCirc[-]}) \}$. In any model of $\T{T}$, the domain $X$ must be $\varnothing$, because of the only axiom in $\T{I}$. A model is a mapping of each of the symbols in $\sign$ to either $\{(\star,\star)\}$ or $\varnothing$. In other words, $P,Q,R,\dots$ act as propositional variables and any model is just an assignment of boolean values. By Lemma \ref{lemma:trivalallequal} all arrows collapse, with the exception of those of type $0\to 0$, that are exactly propositional formulas (see Prop. \ref{prop:propcalc} in App. \ref{app:propo calc}).
Our axiomatisation reduces to the one in Fig.~\ref{fig:propositionalcalculus}. %
The reader can check %
App. \ref{app:propo calc} to see that this is the deep inference system $\mathsf{SKSg}$ in~\cite{DBLP:phd/de/Brunnler2003}. %
\end{example}

Diagrams $c\colon 0\to 0$, which can be thought of as closed formulas of $\FOL$, also play an important role in the following result: a diagrammatic analogue of the deduction theorem (the reader may check App.~\ref{app:closedtheories} for a detailed comparison with theories in $\FOL$).

\begin{theorem}[Deduction theorem]\label{th:deduction}%
    Let $\T{T} = (\sign, \T{I})$ be a theory and $c \colon 0 \to 0$ in $\LCB$. Let $\T{I'} = \T{I} \cup \{(\id[+][0], c)\}$ and let $\T{T'}$ denote the theory $(\sign, \T{I'})$. Then, for every $a,b \colon n \to m$ arrows of $\LCB$,
    \begin{center}if $\;\;  \boxCirc[+]{a} \;\;\precongR{\T{T'}}\;\; \boxCirc[+]{b} \;\;$ then $\;\; 
    \InputIfFileExists{cPerId.tikz}{}{\input{tikz/cPerId.tikz}}
 \;\;\precongR{\T{T}}\;\!\!\!\! \seqCirc[-]{b}{a}[][][w][b]$\!\!\!\!.\end{center}

\end{theorem}
\begin{proof}%

By induction on the rules of~\eqref{eq:pc}.  We show only the case for $(\!\seq[+]\!)$. The remaining ones are  in App.~\ref{app:theories}.

Assume $a = a_1 \seq[+] a_2$ and $b = b_1 \seq[+] b_2$ for some $a_1, b_1 \colon n \to l, a_2, b_2 \colon l \to m$ such that $a_1 \;\precongR{\T{T'}}\; b_1$ and $a_2 \;\precongR{\T{T'}}\; b_2$. By induction hypothesis $c \tensor[+] \id[+][n] \;\precongR{\T{T}}\; b_1 \seq[-] \rla{a_1}$ and $c \tensor[+] \id[+][n] \;\precongR{\T{T}}\; b_2 \seq[-] \rla{a_2}$. Thus:
\input{tikz/proofs/deductionTheorem/seqPlusMain.tex}
\end{proof}
\begin{corollary}\label{cor:deduction}
Let $\T{T}=(\sign, \T{I})$ be a theory, $c\colon 0 \to 0$ in $\LCB$ and $\T{T}'=(\sign, \T{I}\cup \{(\id[+][0], \nega{c} ) \})$. Then $\id[+][0]\syninclusionT{\T{T}}c$ iff \ $\T{T'}$ is contradictory.
\end{corollary}

\subsection{Functorial semantics for first order theories}\label{sec:funsem}

Recall that the notion of interpretation of a signature $\sign$ in $\Rel$ has been generalised in Definition \ref{def:intCAT} to an arbitrary fo-bicategory. As expected, the same is possible also with the notion of model.
\begin{definition}
Let $\T{T}=(\sign, \T{I})$ be a theory and $\Cat{C}$ a first order bicategory. An interpretation $\interpretation$ of $\sign$ in $\Cat{C}$ is a model iff, for all $(c,d)\in \T{I}$,
$\interpretationFunctor(c) \leq \interpretationFunctor(d)$.
\end{definition}

For any theory $\T{T}=(\sign, \T{I})$, one can build a fo-bicategory $\LCB[\T{T}]$: this is like $\LCB$, but homsets are now $\LCB[\T{T}][n,m] =\{ [d]_{\synequivalenceT{\T{T}}} \mid d\in \LCB[\sign][n,m]\}$ ordered by $\syninclusionT{\T{T}}$. Since, by definition, $\syninclusion \subseteq \syninclusionT{\T{T}}$, $\LCB[\T{T}]$ is a fo-bicategory.
Thus, one can take an interpretation $\mathcal{Q}_{\T{T}}$ of $\Sigma$ in $\LCB[\T{T}]$: the domain $X$ is $1$ and $\rho (R) = [R^\circ]_{\synequivalenceT{\T{T}}}$ for all $R \in \Sigma$. By Prop.~\ref{prop:free}, $\mathcal{Q}_{\T{T}}$ induces a fo-bicategory morphism $\mathcal{Q}_{\T{T}}^\sharp \colon \LCB \to \LCB[\T{T}]$.

\begin{proposition}\label{prop:modelfactor}
Let $\T{T}=(\sign, \T{I})$ be a theory, $\Cat{C}$ a fo-bicategory and
$\interpretation$ an interpretation of $\Sigma$ in $\Cat{C}$. $\interpretation$ is a model of $\T{T}$ in $\Cat{C}$ iff $\interpretationFunctor \colon \LCB \to \Cat{C}$ factors uniquely through $\mathcal{Q}_{\T{T}}^\sharp \colon \LCB \to \LCB[\T{T}]$.
\end{proposition}
\begin{wrapfigure}{R}{0.2\linewidth}\vspace{-22pt}
    $\!\!\!\!\!\!\!\!\!\!\scalebox{0.8}{\xymatrix{\LCB \ar[rd]|{\interpretationFunctor} \ar[r]^{\mathcal{Q}_{\T{T}}^\sharp}& \LCB[\T{T}] \ar@{.>}[d]^{\interpretationFunctor_{\T{T}}}\\
    & \Cat{C}}}$
    \vspace{-22pt}
\end{wrapfigure}
\noindent In other words, there is a unique fo-bicategory morphism $\interpretationFunctor_{\T{T}} \colon \LCB[\T{T}] \to \Cat{C}$ s.t. the diagram on the right commutes. The assignment $\interpretation \mapsto \interpretationFunctor_{\T{T}}$ yields a 1-to-1 correspondence between models and morphisms.

\begin{corollary}\label{corollarymodelfunctor}
To give a model of \,$\T{T}$ in $\Cat{C}$ is to give a fo-bicategory morphism $\LCB[\T{T}] \to \Cat{C}$.
\end{corollary}

By virtue of the above, we can tacitly identify models and morphisms.
Proposition \ref{prop:modelfactor} can also be used to obtain the following result, useful for showing completeness in the next section.

\begin{lemma}\label{lemma:largertheories}
Let $\T{T}=(\sign, \T{I})$ and $\T{T}'=(\sign ', \T{I}')$ be theories s.t.\ $\sign \subseteq \sign'$ and $\T{I} \subseteq \T{I}'$. Then there exists an identity on objects fo-bicategory morphism $\mathcal{F}\colon \LCB[\T{T}] \to \LCB[\T{T}']$ mapping each $d$ of $\LCB[\T{T}]$ to $[d]_{\synequivalenceT{\T{T'}}}$.
\end{lemma}

\section{Beyond G\"odel's completeness}\label{sec:completeness}\label{ssec:henkinmodel}\label{sec:Beyond Godel}
Let $\T{T}= (\sign, \T{I})$  be a theory. %
First, we prove G\"odel completeness %
\begin{equation}\label{thm:Godel}
\text{if $\T{T}$ is non-trivial, then $\T{T}$ has a model}\tag{G\"odel}
\end{equation}
by adapting Henkin's~\cite{henkin_1949} proof to $\NPR$. We begin with two additional definitions. Note that when referring to arrows in the context of $\T{T}$, we mean arrows of $\LCB[\T{T}]$ (or of $\LCB$, it is immaterial).

\begin{definition}\label{def:henkin wit}
$\T{T}$ is \emph{syntactically complete} if for all $c \colon 0 \to 0$ either $\id[+][0] \;\precongR{\T{T}}\; c$ or $\id[+][0] \;\precongR{\T{T}}\; \overline{c}$.
    $\T{T}$ has \emph{Henkin witnesses} if for all $c \colon 1 \to 0$ there is a map $k \colon 0 \to 1$ s.t.
    $\scalebox{0.8}{
    \InputIfFileExists{henkinWit1.tikz}{}{\input{tikz/henkinWit1.tikz}}
} \!\!\syninclusionT{\T{T}}\!\! \scalebox{0.8}{
    \InputIfFileExists{henkinWit2.tikz}{}{\input{tikz/henkinWit2.tikz}}
}$\!\!.
\end{definition}
These properties do not hold for the theories we have considered so far. In terms of $\FOL$, syntactic completeness means that closed formulas either hold in all models of the theory or in none. A Henkin witness is a term $k$ such that $c(k)$ holds: a theory has Henkin witnesses if for every true formula $\exists x.c(x)$, there exists such a $k$.
We shall see in Theorem~\ref{thm:nontrivialALL} that non-trivial theories can be expanded to have Henkin witnesses, be non-contradictory and syntactically complete. The key idea of Henkin's proof, Theorem~\ref{thm:Hekinismodel}, is that these three properties yield a model. %

\smallskip

\noindent\begin{minipage}{0.58\columnwidth}
\hspace{0.3cm} To add a witness for $c\colon 1 \to 0$, one adds a constant $k\colon 0\to 1$ and the axiom $\mathbb{W}_k^c$, asserting that $k$ is a witness. This preserves non-triviality.
\end{minipage}
\;
\begin{minipage}{.4\columnwidth}\vspace{-0.5cm}
\[\mathbb{W}_k^c \defeq \{(\emptyCirc[+] \, , \scalebox{0.7}{
    \InputIfFileExists{proofs/nonTrivialCompleteness/newAxiom.tikz}{}{\input{tikz/proofs/nonTrivialCompleteness/newAxiom.tikz}}
})\}  \]
\end{minipage}
\begin{lemma}[Witness Addition]\label{lemma:addingHenkin}%
Let $\T{T}=(\Sigma, \mathbb{I})$ be a theory and consider an arbitrary $c\colon 1\to 0$.
Let $\T{T'} = ( \Sigma \cup \{k\colon 0 \to 1\}, \mathbb{I} \cup \TMAP{k} \cup\mathbb{W}_k^c )$. %
If \,$\T{T}$ is non-trivial then  $\T{T'}$ is non-trivial.
\end{lemma}

\begin{remark}%
Observe that the distinction between trivial and contradictory theories is essential for the above development. Indeed, under the conditions of Lemma \ref{lemma:addingHenkin}, it does \emph{not} hold that
\begin{center}if $\T{T}$ is non-contradictory, then  $\T{T'}$ is non-contradictory.\end{center}
As  counter-example, take as $\T{T}$ the theory consisting only of the trivialising axiom $(tr) \defeq (\scalebox{0.8}{\codiscardCirc[+]} , \scalebox{0.8}{\codiscardCirc[-]})$. By definition $\T{T}$ is trivial but non-contradictory. Instead $\T{T}'$ is contradictory: %
\begin{equation}\label{eq:remcounter}
    \emptyCirc[+] \stackrel{\eqref{eq:bone}}{\syninclusionT{\T{T}}} 
    \InputIfFileExists{axioms/cb/plus/codiscDisc.tikz}{}{\input{tikz/axioms/cb/plus/codiscDisc.tikz}}
 \stackrel{(tr)}{\syninclusionT{\T{T}}} 
    \InputIfFileExists{axioms/linadj/cobangPbang.tikz}{}{\input{tikz/axioms/linadj/cobangPbang.tikz}}
 \stackrel{\eqref{ax:gammaDiscardPlus}}{\syninclusionT{\T{T}}} \emptyCirc[-]
\end{equation}
This shows that adding Henkin witnesses to a non-contradictory theory may end up in a contradictory theory. Therefore, the usual Henkin proof for $\FOL$ works just for our \emph{non-trivial} theories. %
\end{remark}

By iteratively using Lemma \ref{lemma:addingHenkin}, one can transform a non-trivial theory into a non-trivial theory with Henkin witnesses. To obtain a syntactically complete theory, we use the standard argument featuring Zorn's Lemma (see Prop.~\ref{prop:syntacticallycomplete} in App.~\ref{app:proofCompl}). In summary:

\begin{theorem}\label{thm:nontrivialALL}
Let $\T{T}=(\Sigma, \mathbb{I})$ be a non-trivial theory. There exists a theory $\T{T'}=(\Sigma', \mathbb{I'})$ such that $\Sigma \subseteq \Sigma'$ and $\mathbb{I}\subseteq \mathbb{I'}$; $\T{T'}$ has Henkin witnesses; $\T{T'}$ is syntactically complete; $\T{T'}$ is non-contradictory.
\end{theorem}

Before introducing Henkin's interpretation, %
observe that any map $c\colon 0 \to n$ can be decomposed as $k_1 \tensor[+] \dots \tensor[+] k_n$ where each $k_i\colon 0 \to 1$ is a map (see Prop.~\ref{prop:nary maps} in App.~\ref{app:proofCompl}). %
 We thus write such $c$ as $\nterm{k}$, depicted as $\constCirc{\nterm{k}}[n]\!\!$, to make explicit its status as a vector.

\begin{definition}\label{def:henkin struct}
Let $\T{T}=(\Sigma,\mathbb{I})$ be a theory.
The \emph{Henkin interpretation} $\Hen$ of $\sign$, consists of a set $X \defeq \texttt{Map}(\LCB[\T{T}])[0,1]$ and a function $\rho$, defined for all $R  \colon n \to m \in \sign$ as:
    \[
        \rho(R) \defeq \{ (\nterm{k},\nterm{l}) \in X^n \times X^m \mid \scalebox{0.8}{\emptyCirc[+]} \syninclusionT{\T{T}} \scalebox{0.8}{\closedFormulaCirc{\nterm{k}}{R}{\nterm{l}}} \}
    \]
\end{definition}

 The domain is the set of constants of the theory. Then $R\colon n \to m$ is mapped to all pairs $(\nterm{k},\nterm{l})$ of vectors that make $R$ true in $\T{T}$. The following characterisation of $\HenFunctor \colon \LCB \to \Rel$ is crucial.

\begin{proposition}\label{prop:henkin}
    Let $\T{T}=(\Sigma,\mathbb{I})$ be a non-contradictory, syntactically complete theory with Henkin witnesses. Then, for any $c \colon n \to m$,
    $\HenFunctor(c) = \{ (\nterm{k},\nterm{l}) \in X^n \times X^m \mid \scalebox{0.8}{\emptyCirc[+]} \syninclusionT{\T{T}} \scalebox{0.8}{\closedFormulaCirc{\nterm{k}}{c}{\nterm{l}}} \}$.
\end{proposition}

\begin{theorem}\label{thm:Hekinismodel}
    \!\!If $\T{T}$ is non-contradictory, syntactically complete with Henkin witnesses, then $\Hen$ is a model.
\end{theorem}
\begin{proof} %
    We show that $c \syninclusionT{\T{T}} d$ gives $\HenFunctor(c) \subseteq \HenFunctor(d)$.
    If $(\nterm{k}, \nterm{l}) \in \HenFunctor(c)$ then $\scalebox{0.8}{\emptyCirc[+]} \!\!\!\syninclusionT{\T{T}}\!\!\! \scalebox{0.8}{\closedFormulaCirc{\nterm{k}}{c}{\nterm{l}}}$ by Prop.~\ref{prop:henkin}. Since $c \syninclusionT{\T{T}} d$,  $\scalebox{0.8}{\emptyCirc[+]} \!\!\!\syninclusionT{\T{T}}\!\!\! \scalebox{0.8}{\closedFormulaCirc{\nterm{k}}{c}{\nterm{l}}} \!\!\!\syninclusionT{\T{T}}\!\!\! \scalebox{0.8}{\closedFormulaCirc{\nterm{k}}{d}{\nterm{l}}}$ and by Prop.~\ref{prop:henkin}, $(\nterm{k}, \nterm{l}) \in \HenFunctor(d)$.
\end{proof}

Theorems \ref{thm:nontrivialALL} and \ref{thm:Hekinismodel} give us a proof for \eqref{thm:Godel}. %
\begin{proof}[Proof of \eqref{thm:Godel}]
Let $\T{T'}=(\Sigma', \mathbb{I'})$ be obtained via Theorem \ref{thm:nontrivialALL}.
Since $\Sigma \subseteq \Sigma'$ and $\mathbb{I}\subseteq \mathbb{I'}$, by Lemma \ref{lemma:largertheories}, we have $\mathcal{F}\colon \LCB[\T{T}] \to \LCB[\T{T}']$.
Since $\T{T'}$ has Henkin witnesses, is syntactically complete and non-contradictory, Theorem \ref{thm:Hekinismodel} gives   $\Hen_{\T{T'}}^\sharp \colon \LCB[\T{T}'] \to \Rel$. We thus have a morphism $ \LCB[\T{T}] \to \Rel$.
\end{proof}

Now, we would like to conclude Theorem \ref{thm:completeness} by means of \eqref{thm:Godel}, but this is not possible since, for the former one needs a model for all non-contradictory theories, while \eqref{thm:Godel} provides it only for non-trivial ones. %
Thankfully, the Henkin interpretation $\Hen$ gives us, once more, a model (see Prop.~\label{prop:booleanisamodel} in App.~\ref{app:proofCompl}) that allows us to prove
\begin{equation}\label{thm:trivialcom}
\text{\!\!if $\T{T}$ is trivial and non-contradictory,  then $\T{T}$ has a model.} \tag{Prop}
\end{equation}

\noindent From \eqref{thm:trivialcom} and \eqref{thm:Godel} we can prove general completeness
\begin{equation}\label{cor:gencompleteness}
\text{if $\T{T}$ is  non-contradictory, then  $\T{T}$ has a model}\tag{General}
\end{equation}
and thus deduce our main result.
\begin{proof}[Proof of \eqref{cor:gencompleteness} and  Theorem \ref{thm:completeness}] To prove \eqref{cor:gencompleteness} take $\T{T}$ to be a non-contradictory theory. If $\T{T}$ is trivial, then it has a model by  \eqref{thm:trivialcom}. Otherwise, it has a model by \eqref{thm:Godel}. Now, by means of traditional $\FOL $ arguments exploiting Corollary \ref{cor:deduction}, one can show that \eqref{cor:gencompleteness} entails Theorem \ref{thm:completeness} (see Prop.\ref{prop:entailment} in App.~\ref{app:proofCompl}).
\end{proof}

\begin{table*}[!htb]
  \caption{The encoding $\enc{\cdot} \colon \CRS \to \NPR $}\label{table:cr encoding}
  \centering
  \footnotesize{
  $
      \begin{array}{llllll}
          \toprule
            \enc{R} \defeq R^\circ
          & \enc{\id[+]} \defeq \id[+][1]
          & \enc{E_1 \seq[+] E_2} \defeq \enc{E_1} \seq[+] \enc{E_2}
          & \enc{\top} \defeq \discard[+][1] \seq[+] \codiscard[+][1]
          & \enc{E_1 \cap E_2} \defeq \copier[+][1] \seq[+] (\enc{E_1} \tensor[+] \enc{E_2}) \seq[+] \cocopier[+][1] 
          & \enc{\nega{E}} \defeq \nega{\enc{E}} \\
            \enc{\op{E}} \defeq \op{\enc{E}}
          & \enc{\id[-]} \defeq \id[-][1]
          & \enc{E_1 \seq[-] E_2} \defeq \enc{E_1} \seq[-] \enc{E_2}
          & \enc{\bot} \defeq \discard[-][1] \seq[-] \codiscard[-][1]
          & \enc{E_1 \cup E_2} \defeq \copier[-][1] \seq[-] (\enc{E_1} \tensor[-] \enc{E_2}) \seq[-] \cocopier[-][1] 
          & \\
          \bottomrule
      \end{array}
  $
  }
\end{table*}

\subsection{The Calculus of Binary Relations (revisited)}\label{ssec:CRrevisited}
The map $\enc{\cdot}$ defined in Table~\ref{table:cr encoding} is an econding of the calculus of relations into $\NPR$.
Since $\enc{\cdot}$ preserves the semantics (see Prop. \ref{prop:cr interpretation} in App.\ref{app:encoding}), from Theorem~\ref{thm:completeness} follows that one can prove inclusions of expressions of $\CRS$ by translating them into $\NPR$ via $\enc{\cdot}$ and then using the axioms in Fig.s \ref{fig:cb axioms}, \ref{fig:cocb axioms}, \ref{fig:closed lin axioms} and \ref{fig:fo bicat axioms}.

\begin{corollary}\label{cor:cr completeness}
    For all $E_1, E_2$, $E_1 \minorExpression E_2$ iff $\enc{E_1} \syninclusion \enc{E_2}$.
\end{corollary}

\newcommand{\varcontext}{\vdash}
\newcommand{\varlist}{\mathbf{x}_n}
\newcommand{\varlistred}{\mathbf{x}_{n-1}}
\newcommand{\divider}{\ \rightsquigarrow \ }
\newcommand{\from}{\mathrel{:}}

\section{First Order Logic with Equality}
\label{sec:fol}

\begin{figure*}[!htb]
  \centering
  {\small
  $
  \begin{array}{@{}c@{\!\!\!\!}|@{\,}c}
    \begin{array}{@{}r@{\;}c@{\;}l}
      \folEnc{n}{x_i} &\defeq& 
    \InputIfFileExists{encodingFOL2/xi.tikz}{}{\input{tikz/encodingFOL2/xi.tikz}}
 
      \\[18pt]
      \!\!\folEnc{n}{f(t_1,..,t_m)} &\defeq& \scalebox{0.8}{
    \InputIfFileExists{encodingFOL2/t.tikz}{}{\input{tikz/encodingFOL2/t.tikz}}
}   
    \end{array}
      &
    \begin{array}{@{}r@{\;}c@{\,\;}c @{\;} r@{\;}c@{\,\;}c @{} r@{\;}c@{\,\;}c}
      \folEnc{n}{\varphi_1 \wedge \varphi_2} &\defeq& \!\!
    \InputIfFileExists{encodingFOL2/and.tikz}{}{\input{tikz/encodingFOL2/and.tikz}}
  
      &
      \folEnc{n}{\varphi_1 \vee \varphi_2 } &\defeq& \!\!
    \InputIfFileExists{encodingFOL2/or.tikz}{}{\input{tikz/encodingFOL2/or.tikz}}
 
      &
      \folEnc{n}{R(t_1,..,t_m)}& \defeq & \!\!\!\scalebox{0.8}{
    \InputIfFileExists{encodingFOL2/R.tikz}{}{\input{tikz/encodingFOL2/R.tikz}}
}  
      \\
      \folEnc{n}{\top} &\defeq& \!\!\!\!\!\!\!\!\!\!\!\!\!\discardCirc[+][n]  
      &
      \folEnc{n}{\bot} &\defeq& \!\!\!\!\!\!\!\!\!\!\!\!\!\discardCirc[-][n]   
       &
      \folEnc{n}{t_1 = t_2} &\defeq& \!\!\!\scalebox{0.8}{
    \InputIfFileExists{encodingFOL2/eq.tikz}{}{\input{tikz/encodingFOL2/eq.tikz}}
} 
      \\
      \folEnc{n\!-\!1}{\exists x_n.\, \varphi} &\defeq& \!\!\!
    \InputIfFileExists{encodingFOL2/exists.tikz}{}{\input{tikz/encodingFOL2/exists.tikz}}
 
            &
      \folEnc{n\!-\!1}{\forall x_n.\, \varphi} &\defeq& \!\!
    \InputIfFileExists{encodingFOL2/forall.tikz}{}{\input{tikz/encodingFOL2/forall.tikz}}

&
      \folEnc{n}{\neg \varphi} &\defeq& \!\!\!\!\!\!\!\!\!\!\!\!\!\!\!\!
    \InputIfFileExists{encodingFOL2/neg.tikz}{}{\input{tikz/encodingFOL2/neg.tikz}}
           
    \end{array}
  \end{array}
  $}
  \caption{$\FOL$ encoding in $\NPR$.}
  \label{fig:fol encoding}
\end{figure*}

\hspace{-0.02cm}
\begin{figure*}
  \centering
    \begin{subfigure}{0.33\textwidth}
      \centering
      $
      
    \InputIfFileExists{eg2.tikz}{}{\input{tikz/eg2.tikz}}

      \;\; \leftrightsquigarrow \;\;
      
    \InputIfFileExists{eg.tikz}{}{\input{tikz/eg.tikz}}

      $
    \end{subfigure}
     \vline
     \begin{subfigure}{0.33\textwidth}
        \centering
         $
    \InputIfFileExists{peirce/itLHS.tikz}{}{\input{tikz/peirce/itLHS.tikz}}
 \synequivalence 
    \InputIfFileExists{peirce/itRHS.tikz}{}{\input{tikz/peirce/itRHS.tikz}}
$
     \end{subfigure}
     \vline
     \begin{subfigure}{0.33\textwidth}
         \centering
         \raisebox{-0.45\height}{\includegraphics*[scale=0.28]{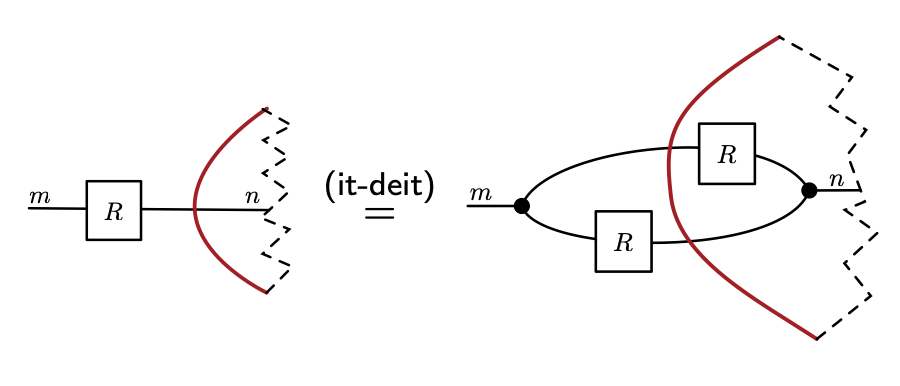}}
     \end{subfigure}
  \caption{An $\EG$ and its encoding in $\NPR$ (left); Peirce's (de)iteration rule in $\NPR$ (middle) and in~\cite{Haydon2020} (right).}
  \label{fig:itdeit}
\end{figure*}

As we already mentioned in the introduction the white fragment of $\NPR$ is as expressive as the existential-conjunctive fragment of first order logic with equality $(\FOL)$. The semantic preserving encodings between the two fragments are illustrated in \cite{GCQ}. From the fact that the full $\NPR$ can express negation, we get immediately semantic preserving encodings between $\NPR$ and the full $\FOL$. In this section we illustrate anyway a translation $\enc{\cdot} \colon \FOL \to \NPR$ to emphasise the subtle differences between the two. To go in the other way, the reader is referred to App.~\ref{app:fol}.

To ease the presentation, we consider $\FOL$ formulas $\varphi$ to be typed in the context of a list of variables that are allowed (but not required) to appear in $\varphi$. Fixing $\varlist \defeq \{x_1,\dots,x_{n}\}$ we write $\folFormula{n}{\varphi}$ if all free variables of $\varphi$ are contained in $\varlist$.
It is standard to present $\FOL$ in two steps: first terms and then formulas. For every function symbol $f$ of arity $m$ in $\FOL$, we have a symbol $f\colon m \to 1$ in the signature $\sign$ together with the equations $\TMAP{f}$ forcing $f$ to be interpreted as a function. The translation of $\folFormula{n}{t}$ to an $\NPR$ diagram $n\to 1$ is given inductively in the left part of Fig.~\ref{fig:fol encoding}.

Formulas $\folFormula{n}{\varphi}$ translate to $\NPR$ diagrams $n\to 0$. For every $n$-ary predicate symbol $R$ in $\FOL$ there is a symbol $R\colon n \to 0\in\sign$. In order not to over-complicate the presentation with bureaucratic details, we assume that it is always the last variable that is quantified over. Additional variable manipulation can be introduced: see App.~\ref{sec:quine} for an encoding of Quine's predicate functor logic.

The full encoding in Fig.~\ref{fig:fol encoding} should give the reader the spirit of the correspondence between $\NPR$ and traditional syntax. There is one aspect of the above translation that merits additional attention.
\begin{remark}\label{rmk:ambiguity}
By the definition of $\discard[][n]$ in Table \ref{fig:sugar}, we have that:
\[
  \folEnc{0}{\top} \defeq \emptyCirc[+]  \qquad \qquad \folEnc{0}{\bot} \defeq \emptyCirc[-]
\]
Thus $\top$ and $\bot$ translate to, respectively $\id[+][0]$, $\id[-][0]$ in the absence of free variables or to $\discard[+][n]$, $\discard[-][n]$, respectively, when $n>0$. This can be seen as an ambiguity in the traditional $\FOL$ syntax, which obscures the distinction between inconsistent and trivial theories in traditional accounts, and as a side effect requires the assumption on non-empty models in formal statements of G\"odel completeness. Instead, the syntax of $\NPR$ ensures that this pitfall is side-stepped.
\end{remark}

\section{Concluding Remarks}

The diagrammatic notation of $\NPR$ is closely related to system $\beta$ of Peirce's $\EG$s~\cite{peirce2020_the-logic-of-the-future:-history-and-applications,peirce2021_the-logic-of-the-future:-the-1903-lowell-lectures,peirce2021_the-logic-of-the-future:-the-logical-tracts, roberts1973_the-existential-graphs-of-charles-s.-peirce}. Consider the two diagrams on the left of Fig.~\ref{fig:itdeit} corresponding to the closed $\FOL$ formula $\exists x.\, p(x) \wedge \forall y.\, p(y) \to q(y)$.
In existential graph notation the circle enclosure (dubbed `cut' by Peirce) signifies negation. To move from $\EG$s to diagrams of $\NPR$ it suffices to treat lines and predicate symbols in the obvious way and each cut as a color switch.

A string diagrammatic approach to existential graphs appeared in~\cite{Haydon2020}. This exploits the white fragment of $\NPR$ with a primitive negation operator rendered as Peirce's cut, namely a circle around diagrams. However, this inhibits a fully compositional treatment since, for instance, negation is not functorial. %
As an example consider Peirce's (de)iteration rule in Fig.~\ref{fig:itdeit}: in $\NPR$ on the center, and in~\cite{Haydon2020} on the right. Note that the diagrams on the right require open cuts, a notational trick, allowing to express the rule for arbitrary contexts, i.e. any diagram eventually appearing inside the cut. In $\NPR$ this ad-hoc treatment of contexts is not needed as negation is not a primitive operation, but a derived one. %
A proof of the law in the middle of Fig.~\ref{fig:itdeit} can be found in App.~\ref{app:additional proofs}.

Other diagrammatic calculi of Peirce's $\EG$s appear in~\cite{mellies2016bifibrational} and~\cite{brady2000string}. The categorical treatment goes, respectively, through the notions of chiralities and doctrines. The formers consider a pair of categories $(\Rel_{\bullet}, \Rel_{\circ})$ that are significantly different from our $\Relp$ and $\Relm$: to establish a formal correspondence, it might be convenient to first focus on doctrines. To this aim, we plan to exploit the equivalence in~\cite{bonchi2021doctrines} between cartesian bicategories and certain doctrines (elementary existential with comprehensive diagonals and unique choice~\cite{maietti2013quotient}). Preliminary attempts suggests the same equivalence restrict to fo-bicategories and boolean hyperdoctrines but many details have to be carefully checked. The connection with allegories~\cite{freyd1990categories} is also worth to be explored: since cartesian bicategories are equivalent to unitary pretabular allegories, Prop.~\ref{prop:enrichment} suggests that fo-bicategories are closely related to Peirce allegories~\cite{olivier1997peirce}.

Through the Introduction, we have already emphasized the key features of the calculus of neo-Peircean relations. We hope that the reader has also appreciated its beauty. Quoting Dijkstra \cite{dijkstra1980some}:
\begin{center}
  ``When we recognize the battle against chaos, mess and unmastered complexity as one of computing science's major challenges, we must admit that Beauty is our Business.''  
\end{center}

\bibliographystyle{ACM-Reference-Format}
\bibliography{main}

\appendix

\section{A tribute to Charles S. Peirce}\label{app:peirce}

We have chosen the name ``Neo-Peircean Relations'' to emphasize several connections with the work of Charles S. Peirce. First of all, $\NPR$ and  the calculus of relations in `Note B' \cite{peirce1883_studies-in-logic.-by-members-of-the-johns-hopkins-university} share the same underlying philosophy: they both propose a relational analogue to Boole's algebra of classes.

Second, Peirce's presentation in `Note B' contains already several key ingredients of $\NPR$. As we have stressed, it singles out the two forms of composition ($\seq[+]$ and $\seq[-]$), presents linear distributivity (\eqref{ax:leftLinDistr} and \eqref{ax:rightLinDistr}) and linear adjunctions (\ \eqref{ax:tauSymmPlus}, \eqref{ax:tauSymmMinus}, \eqref{ax:gammaSymmPlus}, and \eqref{ax:gammaSymmMinus}), and even the cyclic conditions of Lemma \ref{lm:implications}.(2)-(3). With respect to the rules for linear distributivity and linear adjunction, Peirce states that the latter are ``highly important" and that the former are ``so constantly used that hardly anything can be done without them" (p. 192 \& 190). %

At around the same time as `Note B' Pierce gave a systematic study of residuation \cite[see ``On the Logic of Relatives"]{peirce1989_writings-of-charles-s.-peirce:-a-chronological-edition-volume-4:-18791884} and listed a set of equivalent expressions that includes the discussion given after Lemma \ref{lemma:uniquenessla}, where $c\seq[+] a \leq b$ iff $c\leq b\seq[-]\rla{a}$. 
In Peirce's words:
\begin{quote}
Hence the rule is that having a formula of the form $[c \seq[+]a \leq b]$, the three letters may be cyclically advanced one place in the order of writing, those which are carried from one side of the copula to the other being both negatived and converted. \cite[p. 341]{peirce1989_writings-of-charles-s.-peirce:-a-chronological-edition-volume-4:-18791884}
\end{quote}

Peirce took the principal defect of the presentation in ‘Note B’ to be its focus on binary relations \cite[8:831]{peirce1931_collected-papers-of-charles-sanders}. He went on to emphasize the \emph{teri-} or \emph{tri-}identity relation, arising from adding a `branch’ to the identity relation, as the key to moving from binary to arbitrary relations. Having the advantage now of ``treating triadic and higher relations as easily as dyadic relations... it's superiority to the human mind as an instrument of logic", he writes, ``is overwhelming" \cite[p. 173]{peirce2022_the-logic-of-the-future:-pragmaticism}. %

By moving from binary to arbitrary relations, Peirce felt the importance of a graphical syntax and developed the existential graphs.
\begin{quote}
``One of my earliest works was an enlargement of Boole’s idea so as to take into account ideas of relation, --- or at least of all ideas of existential relation… \emph{I was finally led to prefer what I call a diagrammatic syntax}. It is a way of setting down on paper any assertion, however intricate… “ [MS 515, emphasis in original, 1911]
\end{quote}

We refer the reader to \cite{Haydon2020} for a detailed explanation of Peirce's topological intuitions behind the Frobenius equations and the correspondence of some inference rules for 
$\EG$s with those of (co)cartesian bicategories.
Moreover, we now know that Peirce continued to study and draw graphs of residuation \cite{haydon2021_residuation-in-existential-graphs} and --- as affirmed in Fig.~\ref{fig:propositionalcalculus} --- we know the rules for propositional $\EG$s comprise a deep inference system \cite{ma2017_proof-analysis-of-peirces-alpha-system-of-graphs}.

In short, Peirce's development of $\EG$s shares many of the features that $\NPR$ has over other approaches, such as Tarski's presentation of relation algebra. We like to think that if Peirce had known category theory then he would have presented $\NPR$. %

\begin{figure*}[t!]
  \vspace{1cm}
  \centering
  $
      \begin{array}{@{}r@{\,\,}c@{\,\,}l @{\qquad} r@{\,\,}c@{\,\,}l@{}}
          \copier[+][1] \seq[+] (\id[+][1] \tensor[+] \copier[+][1]) &\stackrel{(\copier[+]\text{-as})}{=}& \copier[+][1] \seq[+] (\copier[+][1] \tensor[+] \id[+][1]) & (\id[+][1] \tensor[+] \cocopier[+][1]) \seq[+] \cocopier[+][1] &\stackrel{(\cocopier[+]\text{-as})}{=}& (\cocopier[+][1] \tensor[+] \id[+][1]) \seq[+] \cocopier[+][1] \\
          \copier[+][1] \seq[+] (\id[+][1] \tensor[+] \discard[+][1]) &\stackrel{(\copier[+]\text{-un})}{=}& \id[+][1]  & (\id[+][1] \tensor[+] \codiscard[+][1]) \seq[+] \cocopier[+][1] &\stackrel{(\cocopier[+]\text{-un})}{=}& \id[+][1] \\
          \copier[+][1] \seq[+] \symm[+][1][1] &\stackrel{(\copier[+]\text{-co})}{=}& \copier[+][1] & \symm[+][1][1] \seq[+] \cocopier[+][1] &\stackrel{(\cocopier[+]\text{-co})}{=}& \cocopier[+][1] \\
          \addlinespace[0.6em]
          \midrule
          \addlinespace[0.6em]
          (\copier[+][1] \tensor[+] \id[+][1]) \seq[+] (\id[+][1] \tensor[+] \cocopier[+][1]) &\stackrel{(\text{F}^{\switchLabelS{+}})}{=}& (\id[+][1] \tensor[+] \copier[+][1]) \seq[+] (\cocopier[+][1] \tensor[+] \id[+][1]) & \copier[+][1] \seq[+] \cocopier[+][1] &\stackrel{(\text{S}^{\switchLabelS{+}})}{=}& \id[+][1] \\
          \addlinespace[0.6em]
          \midrule
          \addlinespace[0.6em]
          \multicolumn{6}{c}{
              \begin{array}{c@{\qquad}c}
                  \begin{array}{r@{\,\,}c@{\,\,}l @{\qquad} r@{\,\,}c@{\,\,}l}
                      \codiscard[+][1] \seq[+] \discard[+][1] 
                      &\stackrel{\eqref*{ax:plusCodiscDisc}}{\leq}& 
                      \id[+][0] 
                      &
                      \cocopier[+][1] \seq[+] \copier[+][1] 
                      &\stackrel{\eqref*{ax:plusCocopyCopy}}{\leq}&
                      (\id[+][1] \tensor[+] \id[+][1])
                      \\
                      \id[+][1] 
                      &\stackrel{\eqref*{ax:plusDiscCodisc}}{\leq}&
                      \discard[+][1] \seq[+] \codiscard[+][1] 
                      & 
                      \id[+][1] 
                      &\stackrel{\eqref*{ax:plusCopyCocopy}}{\leq}& 
                      \copier[+][1] \seq[+] \cocopier[+][1] \\           
                  \end{array}
                  &
                  \begin{array}{r@{\,\,}c@{\,\,}l r@{\,\,}c@{\,\,}l}
                  c \seq[+] \copier[+][m] &\stackrel{\eqref*{ax:comPlusLaxNat}}{\leq}& \copier[+][n] \seq[+] (c \tensor[+] c) 
                  \\
                  c \seq[+] \discard[+][m] &\stackrel{\eqref*{ax:discPlusLaxNat}}{\leq}& \discard[+][n]
                  \end{array}
              \end{array}
          }
          \\
          \addlinespace[0.6em]
          \toprule
          \addlinespace[0.6em]
          \copier[-][1] \seq[-] (\id[-][1] \tensor[-] \copier[-][1]) &\stackrel{(\copier[-]\text{-as})}{=}& \copier[-][1] \seq[-] (\copier[-][1] \tensor[-] \id[-][1]) & (\id[-][1] \tensor[-] \cocopier[-][1]) \seq[-] \cocopier[-][1] &\stackrel{(\cocopier[-]\text{-as})}{=}& (\cocopier[-][1] \tensor[-] \id[-][1]) \seq[-] \cocopier[-][1] \\
          \copier[-][1] \seq[-] (\id[-][1] \tensor[-] \discard[-][1]) &\stackrel{(\copier[-]\text{-un})}{=}& \id[-][1]  & (\id[-][1] \tensor[-] \codiscard[-][1]) \seq[-] \cocopier[-][1] &\stackrel{(\cocopier[-]\text{-un})}{=}& \id[-][1] \\
          \copier[-][1] \seq[-] \symm[-][1][1] &\stackrel{(\copier[-]\text{-co})}{=}& \copier[-][1] & \symm[-][1][1] \seq[-] \cocopier[-][1] &\stackrel{(\cocopier[-]\text{-co})}{=}& \cocopier[-][1] \\
          \addlinespace[0.6em]
          \midrule
          \addlinespace[0.6em]
          (\copier[-][1] \tensor[-] \id[-][1]) \seq[-] (\id[-][1] \tensor[-] \cocopier[-][1]) &\stackrel{(\text{F}^{\switchLabelS{-}})}{=}& (\id[-][1] \tensor[-] \copier[-][1]) \seq[-] (\cocopier[-][1] \tensor[-] \id[-][1]) & \copier[-][1] \seq[-] \cocopier[-][1] &\stackrel{(\text{S}^{\switchLabelS{-}})}{=}& \id[-][1] \\
          \addlinespace[0.6em]
          \midrule
          \addlinespace[0.6em]
           \multicolumn{6}{c}{
            \begin{array}{c@{\qquad}c}
              \begin{array}{r@{\,\,}c@{\,\,}l @{\qquad} r@{\,\,}c@{\,\,}l}
                \discard[-][1] \seq[-] \codiscard[-][1] 
                &\stackrel{\eqref*{ax:minusDiscCodisc}}{\leq}&
                \id[-][1] 
                &
                \copier[-][1] \seq[-] \cocopier[-][1]
                &\stackrel{\eqref*{ax:minusCopyCocopy}}{\leq}& 
                \id[-][1]
                \\
                \id[-][0] 
                &\stackrel{\eqref*{ax:minusCodiscDisc}}{\leq}& 
                \codiscard[-][1] \seq[-] \discard[-][1] 
                & 
                \cocopier[-][1] \seq[-] \copier[-][1] 
                &\stackrel{\eqref*{ax:minusCocopyCopy}}{\leq}&
                (\id[-][1] \tensor[-] \id[-][1])
            \end{array}
            &
              \begin{array}{r@{\,\,}c@{\,\,}l r@{\,\,}c@{\,\,}l}
                \copier[-][n] \seq[-] (c \tensor[-] c) &\stackrel{\eqref*{ax:comMinusLaxNat}}{\leq}& c \seq[-] \copier[-][m]  
                \\
                \discard[-][n] &\stackrel{\eqref*{ax:discMinusLaxNat}}{\leq}& c \seq[-] \discard[-][m]
              \end{array}
            \end{array}
           }\\
          \addlinespace[0.6em]
          \toprule
          \addlinespace[0.6em]
          a \seq[+] (b \seq[-] c) &\stackrel{\eqref*{ax:leftLinDistr}}{\leq}& (a \seq[+] b) \seq[-] c & (a \seq[-] b) \seq[+] c &\stackrel{\eqref*{ax:rightLinDistr}}{\leq}& a \seq[-] (b \seq[+] c) 
          \\
          \addlinespace[0.6em]
          \midrule
          \addlinespace[0.6em]
          \multicolumn{6}{c}{
              \begin{array}{cc}
                  \begin{array}{r@{\,\,}c@{\,\,}l r@{\,\,}c@{\,\,}l}
                      \id[+][n+m] &\stackrel{\eqref*{ax:tauSymmPlus}}{\leq}& \symm[+][n][m] \seq[-] \symm[-][m][n]
                      &        
                      \symm[-][n][m] \seq[+] \symm[+][m][n] &\stackrel{\eqref*{ax:gammaSymmPlus}}{\leq}& \id[-][n+m]
                      \\
                      \id[+][n+m] &\stackrel{\eqref*{ax:tauSymmMinus}}{\leq}& \symm[-][n][m] \seq[-] \symm[+][m][n]
                      &
                      \symm[+][n][m] \seq[+] \symm[-][m][n] &\stackrel{\eqref*{ax:gammaSymmMinus}}{\leq}& \id[-][n+m]
                  \end{array}
                  &
                  \begin{array}{r@{\,\,}c@{\,\,}l r@{\,\,}c@{\,\,}l}
                      \id[+][n] &\stackrel{\eqref*{ax:tauRPlus}}{\leq}& R^\circ \seq[-] R^\bullet
                      &
                      R^\bullet \seq[+] R^\circ &\stackrel{\eqref*{ax:gammaRPlus}}{\leq}& \id[-][m]
                      \\
                      \id[+][m] &\stackrel{\eqref*{ax:tauRMinus}}{\leq}& R^\bullet \seq[-] R^\circ
                      &
                      R^\circ \seq[+] R^\bullet &\stackrel{\eqref*{ax:gammaRMinus}}{\leq}& \id[-][n]
                  \end{array}
              \end{array}
          }
          \\
          \addlinespace[0.6em]
          \midrule
          \addlinespace[0.6em]
          \id[+][n+m] &\stackrel{\eqref*{ax:tensorMinusIdPlus}}{\leq}& \id[+][n] \tensor[-] \id[+][m] 
          & 
          \id[-][n] \tensor[+] \id[-][m] &\stackrel{\eqref*{ax:tensorPlusIdMinus}}{\leq}& \id[-][n+m] 
          \\
          (a \seq[-] b) \tensor[+] (c \seq[-] d) &\stackrel{\eqref*{ax:linStrn1}}{\leq}& (a \tensor[+] c) \seq[-] (b \tensor[-] d)
          &
          (a \tensor[-] c) \seq[+] (b \tensor[+] d) &\stackrel{\eqref*{ax:linStrn3}}{\leq}& (a \seq[+] b) \tensor[-] (c \seq[+] d)
          \\
          (a \seq[-] b) \tensor[+] (c \seq[-] d) &\stackrel{\eqref*{ax:linStrn2}}{\leq}& (a \tensor[-] c) \seq[-] (b \tensor[+] d)
          &
          (a \tensor[+] c) \seq[+] (b \tensor[-] d) &\stackrel{\eqref*{ax:linStrn4}}{\leq}& (a \seq[+] b) \tensor[-] (c \seq[+] d)
          \\
          \addlinespace[0.6em]
          \toprule
          \addlinespace[0.6em]
          \multicolumn{6}{c}{
              \begin{array}{cc}
                  \begin{array}{r@{\,\,}c@{\,\,}l r@{\,\,}c@{\,\,}l}
                      \id[+][n] &\stackrel{\eqref*{ax:tauCopierPlus}}{\leq}& \copier[+][n] \seq[-] \cocopier[-][n] 
                      &
                      \cocopier[-][n] \seq[+] \copier[+][n] &\stackrel{\eqref*{ax:gammaCopierPlus}}{\leq}& \id[-][n+n] 
                      \\
                      \id[+][n] &\stackrel{\eqref*{ax:tauDiscardPlus}}{\leq}& \discard[+][n] \seq[-] \codiscard[-][n] 
                      &
                      \codiscard[-][n] \seq[+] \discard[+][n] &\stackrel{\eqref*{ax:gammaDiscardPlus}}{\leq}& \id[-][0] 
                      \\
                      \id[+][n] &\stackrel{\eqref*{ax:tauCopierMinus}}{\leq}& \copier[-][n] \seq[-] \cocopier[+][n]
                      &
                      \cocopier[+][n] \seq[+] \copier[-][n] &\stackrel{\eqref*{ax:gammaCopierMinus}}{\leq}& \id[-][n+n] 
                      \\
                      \id[+][n] &\stackrel{\eqref*{ax:tauDiscardMinus}}{\leq}& \discard[-][n] \seq[-] \codiscard[+][n]
                      &
                      \codiscard[+][n] \seq[+] \discard[-][n] &\stackrel{\eqref*{ax:gammaDiscardMinus}}{\leq}& \id[-][0] 
                  \end{array}
                  &
                  \begin{array}{r@{\,\,}c@{\,\,}l r@{\,\,}c@{\,\,}l}
                      \id[+][n+n] &\stackrel{\eqref*{ax:tauCocopierPlus}}{\leq}& \cocopier[+][n] \seq[-] \copier[-][n] 
                      &
                      \copier[-][n] \seq[+] \cocopier[+][n] &\stackrel{\eqref*{ax:gammaCocopierPlus}}{\leq}& \id[-][n] 
                      \\
                      \id[+][0] &\stackrel{\eqref*{ax:tauCodiscardPlus}}{\leq}& \codiscard[+][n] \seq[-] \discard[-][n] 
                      &
                      \discard[-][n] \seq[+] \codiscard[+][n] &\stackrel{\eqref*{ax:gammaCodiscardPlus}}{\leq}& \id[-][n]
                      \\
                      \id[+][n+n] &\stackrel{\eqref*{ax:tauCocopierMinus}}{\leq}& \cocopier[-][n] \seq[-] \copier[+][n]
                      &
                      \copier[+][n] \seq[+] \cocopier[-][n] &\stackrel{\eqref*{ax:gammaCocopierMinus}}{\leq}& \id[-][n]  
                      \\
                      \id[+][0] &\stackrel{\eqref*{ax:tauCodiscardMinus}}{\leq}& \codiscard[-][n] \seq[-] \discard[+][n]
                      &
                      \discard[+][n] \seq[+] \codiscard[-][n] &\stackrel{\eqref*{ax:gammaCodiscardMinus}}{\leq}& \id[-][0]
                  \end{array} 
              \end{array}
          }
          \\
          \addlinespace[0.6em]
          \midrule
          \addlinespace[0.6em]
          \multicolumn{6}{c}{
              \begin{array}{r@{\,\,}c@{\,\,}l r@{\,\,}c@{\,\,}l}
                  (\copier[-][n] \tensor[+] \id[+][n]) \seq[+] (\id[+][n] \tensor[+] \cocopier[+][n]) 
                  &\stackrel{\eqref*{ax:bwFrob}}{=}&
                  (\id[+][n] \tensor[+] \copier[+][n]) \seq[+] (\cocopier[-][n] \tensor[+] \id[+][n])
                  &
                  (\copier[+][n] \tensor[+] \id[+][n]) \seq[+] (\id[+][n] \tensor[+] \cocopier[-][n]) 
                  &\stackrel{\eqref*{ax:bwFrob2}}{=}& 
                  (\id[+][n] \tensor[+] \copier[-][n]) \seq[+] (\cocopier[+][n] \tensor[+] \id[+][n])
                  \\
                  (\copier[+][n] \tensor[-] \id[-][n]) \seq[-] (\id[-][n] \tensor[-] \cocopier[-][n]) 
                  &\stackrel{\eqref*{ax:wbFrob}}{=}& 
                  (\id[-][n] \tensor[-] \copier[-][n]) \seq[-] (\cocopier[+][n] \tensor[-] \id[-][n])
                  &
                  (\copier[-][n] \tensor[-] \id[-][n]) \seq[-] (\id[-][n] \tensor[-] \cocopier[+][n]) 
                  &\stackrel{\eqref*{ax:wbFrob2}}{=}& 
                  (\id[-][n] \tensor[-] \copier[+][n]) \seq[-] (\cocopier[-][n] \tensor[-] \id[-][n])       
              \end{array}
          }
      \end{array}
  $
  \caption{Axioms for $\NPR$. Here $a,b,c,d$ are diagrams of the appropriate type.}
  \label{fig:textual axioms}
  \vspace{1cm}
\end{figure*}

\section{Additional material}\label{app:additional}
In Figure~\ref{fig:textual axioms} we give a term-based version of the axioms of $\NPR$. In the rest of this appendix we give some additional diagrammatic proofs; some more details on the trivial theory of Propositional Calculus (Example~\ref{ex:propcalculus}); an encoding of Quine's $\PFL$ in $\NPR$; and a translation of $\NPR$ diagrams into (typed) $\FOL$ formulas.

\subsection{Additional proofs}\label{app:additional proofs}

In Figure~\ref{fig:forallExists full} we give a completely axiomatic proof of the inclusion in~\eqref{eq:forallexists}. In Figure~\ref{fig:itdeitproof} we prove Peirce's (de)iteration rule (Figure~\ref{fig:itdeit}), showing the two inclusions separately.

\input{tikz/proofs/forallExistsFull.tex}

\input{tikz/peirce/itproof.tex}

\subsection{The trivial theory of Propositional Calculus (Example~\ref{ex:propcalculus})}\label{app:propo calc}

\begin{figure}[!htb]
  \centering
  \[
      \begin{array}{c@{}c}
            \begin{array}{cc}
              \scalebox{0.8}{\propCirc[+]{c}} \!\!\Lleq{\eqref*{ax:comPlusLaxNat}}\!\! \scalebox{0.8}{\propTensorCirc[+]{c}{c}}  & \raisebox{-0.8em}{\ensuremath{\inferrule*[left=(c$\uparrow$)]{c}{c \wedge c}}}
            \end{array}
          & \begin{array}{cc}
              \scalebox{0.8}{\propCirc[+]{c}} \!\!\Lleq{\eqref*{ax:discPlusLaxNat}}\!\! \scalebox{0.8}{\emptyCirc[+]}            & \raisebox{-0.8em}{\ensuremath{\inferrule*[left=(w$\uparrow$)]{c}{\top}}}
            \end{array}
          \\[20pt] 
              \begin{array}{cc}
              \scalebox{0.8}{\propTensorCirc[-]{c}{c}} \!\!\Lleq{\eqref*{ax:comMinusLaxNat}}\!\! \scalebox{0.8}{\propCirc[-]{c}} & \raisebox{-0.8em}{\ensuremath{\inferrule*[left=(c$\downarrow$)]{c \vee c}{c}}}
             \end{array}
          & \begin{array}{cc}
              \scalebox{0.8}{\emptyCirc[-]} \!\!\Lleq{\eqref*{ax:discMinusLaxNat}}\!\! \scalebox{0.8}{\propCirc[-]{c}}             & \raisebox{-0.8em}{\ensuremath{\inferrule*[left=(w$\downarrow$)]{\bot}{c}}}
             \end{array} 
          \\[20pt]
          \begin{array}{cc}
            \scalebox{0.8}{\emptyCirc[+]} \!\!\stackrel{\footnotesize{\stackanchor{\eqref*{ax:tauRPlus}}{\eqref*{ax:tauRMinus}}}}{\leq}\!\!  \scalebox{0.8}{
    \InputIfFileExists{axiomsNEW/propositional/tauR.tikz}{}{\input{tikz/axiomsNEW/propositional/tauR.tikz}}
}
          & \raisebox{-0.8em}{\ensuremath{\inferrule*[left=(i$\downarrow$)]{\top}{c \vee \nega{c}}}}
          \end{array}
          & 
          \begin{array}{cc}
                  \scalebox{0.8}{
    \InputIfFileExists{axiomsNEW/propositional/gammaR.tikz}{}{\input{tikz/axiomsNEW/propositional/gammaR.tikz}}
}\!\!\stackrel{\footnotesize{\stackanchor{\eqref*{ax:gammaRPlus}}{\eqref*{ax:gammaRMinus}}}}{\leq}\!\! \scalebox{0.8}{\emptyCirc[-]}
                & \raisebox{-0.8em}{\ensuremath{\inferrule*[left=(i$\uparrow$)]{c \wedge \nega{c}}{\bot}}}
          \end{array} \\[20pt]
          \multicolumn{2}{c}{
              \begin{array}{cc}
                  \scalebox{0.8}{
    \InputIfFileExists{axiomsNEW/propositional/distributivity1.tikz}{}{\input{tikz/axiomsNEW/propositional/distributivity1.tikz}}
} \!\!\stackrel{\footnotesize{\stackanchor{\eqref*{ax:leftLinDistr}}{\eqref*{ax:rightLinDistr}}}}{\leq}\!\! \scalebox{0.8}{
    \InputIfFileExists{axiomsNEW/propositional/distributivity2.tikz}{}{\input{tikz/axiomsNEW/propositional/distributivity2.tikz}}
} 
              & \raisebox{-0.8em}{\ensuremath{\inferrule*[left=(s)]{a \wedge (b \vee c)}{(a \wedge b) \vee c}}}
              \end{array}
          }
      \end{array}
  \]
  \caption{Correspondence between axioms in Figure~\ref{fig:propositionalcalculus} and rules of $\mathsf{SKSg}$ in~\cite{DBLP:phd/de/Brunnler2003}. By the laws of symmetric monoidal categories $\seq[+]$ and $\tensor[+]$ coincide: they both correspond to $\wedge$. Moreover they are associative, commutative and with unit $\id[+][\unittensor]$, corresponding to $\top$. Symmetrically $\seq[-]$ and $\tensor[-]$ coincide and correspond to $\vee$.}
  \label{fig:correspondencepropositionalcalculus}
\end{figure}

In this appendix we revisit the propositional case shortly illustrated in Example \ref{ex:propcalculus}. 

First, we better details why the axioms of fo-bicategories (in Fig.s \ref{fig:cb axioms}, \ref{fig:cocb axioms}, \ref{fig:closed lin axioms}, \ref{fig:fo bicat axioms}) collapse to those in  Fig.~\ref{fig:propositionalcalculus} for arrows of type $0 \to 0$. Consider for instance \eqref{ax:comPlusLaxNat}: by definition of $\copier[+][0]$ in Tab.~\ref{fig:sugar}, the two diagrams of  \eqref{ax:comPlusLaxNat} in Fig.~\ref{fig:cb axioms} reduce to those in Fig.~\ref{fig:propositionalcalculus}. The rules \eqref{ax:linStrn1}, \eqref{ax:linStrn2}, \eqref{ax:linStrn3} and \eqref{ax:linStrn4} become redundant since, by the axioms of symmetric monoidal categories,  $\seq[]$ and $\tensor[]$ coincide on diagrams $0 \to 0$ and are associative, commutative and with unit $\id[][0]$. 

Then, we draw reader attention toward  the correspondence  with~\cite{DBLP:phd/de/Brunnler2003}: this is illustrated in  Figure \ref{fig:correspondencepropositionalcalculus}.
We expect that there exists also a strong connection with Peirce's system $\alpha$ and its categorical treatment given in \cite{brady2000_a-categorical-interpretation-of-c.s.-peirces-propositional-logic-alpha} by means of *-autonomy. 

We conclude with the following proposition ensuring that diagrams $0\to 0$ are exactly propositional formulas.

\begin{proposition}\label{prop:propcalc} %
Let $\T{T}=(\sign, \T{I})$ be the theory of Example \ref{ex:propcalculus}. For every diagram $a\colon 0 \to 0$ in $\LCB$ there exists a $\synequivalenceT{\T{T}}$-equivalent diagram generated by the following grammar where $R\in \sign$.%
\[ \Circ{c} ::= \propVar[+]{} \; \mid \; \propVar[-]{} \; \mid \; \propVar[+]{R} \; \mid \; \propOpVar[-]{R} \; \mid \; \propSeqCirc[+]{c}{c} \; \mid \; \propSeqCirc[-]{c}{c} \] 
\end{proposition}
\begin{proof}
By induction on $a \colon 0 \to 0$.
Observe that there are only four base cases:  $\id[+][0]$, $\id[-][0]$, $R^\circ$ and $R^\bullet$. These already appear in the grammar above.
We have the usual four inductive cases:
\begin{enumerate}
\item $a= c\seq[+] d$. There are two sub-cases: either $c,d\colon 0 \to 0$ or $c\colon 0 \to n+1$ and $d\colon n+1 \to 0$. In the former we can use the inductive hypothesis to get $c'$ and $d'$ generated by the above grammar such that $c'\synequivalenceT{\T{T}} c$ and $d' \synequivalenceT{\T{T}}d$. Thus $a$ is $\synequivalenceT{\T{T}}$-equivalent to $c' \seq[+] d'$ that is generated by the above grammar. 

Consider now the case where $c\colon 0 \to n+1$ and $d\colon n+1 \to 0$. By Lemma \ref{lemma:trivalallequal}, $c\synequivalenceT{\T{T}} \codiscard[+][n+1]$ and $d\synequivalenceT{\T{T}} \discard[-][n+1]$. By axiom \eqref{ax:gammaDiscardMinus}, $\codiscard[+][n+1] \seq[+] \discard[-][n+1] \synequivalence \id[-][0]$. Thus $a \synequivalence \id[-][0]$.
\item $a=c \tensor[+] d$. Note that, in this case both $c$ and $d$ must have type $0 \to 0$. Thus we can use the inductive hypothesis to get $c'$ and $d'$ generated by the above grammar such that $c'\synequivalenceT{\T{T}} c$ and $d' \synequivalenceT{\T{T}}d$.  Thus $a \synequivalenceT{\T{T}} c' \tensor[+]d' \structuralcong c' \seq[+] d'$. Note that $c' \seq[+] d'$ is generated by the above grammar.
\item $a=c \seq[-] d$. The proof follows symmetrical arguments to the case $c\seq[+] d$.
\item $a=c \tensor[-] d$. The proof follows symmetrical arguments to the case $c\tensor[+] d$.\qedhere
\end{enumerate}
\end{proof}

\subsection{Quine's predicate functor logic}\label{sec:quine}

\begin{table*}[!htb]
  \vspace{0.5cm}
  \caption{$\PFL$: (top) syntax; (mid) typing rules; (bottom) semantics w.r.t. an interpretation $\interpretation =(X,\rho)$.}
  \label{fig:typingrulesQuine}
\tiny{
\begin{center}
  \begin{tabular}{c}
    \toprule
 $
 \begin{array}{rcl}
P & ::=&\; R  \; \mid \; I \; \mid  \;  \mathbf{p}P \mid \; \mathbf{P}P \; \mid  \;  [P \; \mid \; ]P \; \mid \;  P \cap P  \; \mid \;  \neg P, \;\;\; \text{ where }\; R \in \sign
\end{array}
$ \\
 \midrule
 $   \inferrule{-}{I\colon 2} \quad
       \inferrule{\ari(R)=n}{R \colon n} \quad\inferrule{P\colon n \; n\geq 2}{  \mathbf{p}P \colon n } \quad\inferrule{P\colon 1}{  \mathbf{p}P \colon 2 } \quad\inferrule{P\colon 0}{  \mathbf{p}P \colon 2 } \quad
        \inferrule{P \colon n}{ \mathbf{P}P \colon n}  \quad
        \inferrule{P_1 \colon n \;\; P_2\colon m \;\; n\geq m}{P_1 \cap P_2 \colon n} \; \inferrule{P_1 \colon n \;\; P_2\colon m\;\; n<m}{P_1 \cap P_2 \colon m} \quad
        \inferrule{P \colon n}{ \neg P \colon n}  \quad
 \inferrule{P \colon n}{ [ P \colon n+1}  \; \inferrule{P \colon n+1}{ ] P \colon n} \; \inferrule{P \colon 0}{ ] P \colon 0}
$\\
\midrule
$\begin{array}{@{}l@{}l}
  \begin{array}{@{}l@{}l}
    \dsemRel{R} \defeq \{\tau \mid (\tau_1, \dots, \tau_{n}) \in \rho(R)\} &\;\; \dsemRel{I} \defeq \{\tau \mid \tau_1=\tau_2\} \\
    \dsemRel{ P_1 \cap P_2} \defeq \dsemRel{ P_1} \cap \dsemRel{ P_2} &\;\; \dsemRel{ \neg P} \defeq \{\tau \mid \tau \notin \dsemRel{P}\}
  \end{array}
  & \;\;
  \begin{array}{@{}l}
    \dsemRel{]P} \defeq \{\tau \mid \tau_2 \cdot \tau_3 \cdots \in \dsemRel{P})\}\\
    \dsemRel{[P} \defeq \{x_0 \cdot \tau_1 \cdot \tau_2 \cdots \mid x_0\in X, \tau_1 \cdot \tau_2 \dots \in \dsemRel{P}\}
  \end{array} \\
  \dsemRel{\mathbf{P}P} \defeq \{ \tau \mid \tau_n \cdot \tau_2\cdots \tau_{n-1} \cdot \tau_1 \cdot \tau_{n+1} \cdots \in \dsemRel{P} \} & \;\;\dsemRel{ \mathbf{p}P} \defeq \{\tau  \mid \tau_2 \cdot \tau_1\cdots \in \dsemRel{P}\}
\end{array}
$
\\
\bottomrule
\end{tabular}
\end{center}
}
\vspace{0.5cm}
\end{table*}

\begin{table*}[!htb]
  \vspace{0.5cm}
  \caption{The encoding $\enc{\cdot} \colon \PFL \to \NPR $}\label{table:quine encoding}
  $\tiny{
  \begin{array}{c@{\qquad\qquad}c@{\qquad\qquad}c@{\qquad\qquad}c@{\qquad\qquad}c}
    \toprule
    \inferrule{-}{\enc{I} = \capCirc[+]} &
    \inferrule{\ari(R)=n}{\enc{R} \defeq 
    \InputIfFileExists{quine/R.tikz}{}{\input{tikz/quine/R.tikz}}
} &
    \inferrule{P\colon n \; n\geq 2}{  \enc{\mathbf{p}P} \defeq 
    \InputIfFileExists{quine/symm2.tikz}{}{\input{tikz/quine/symm2.tikz}}
 } &
    \inferrule{P\colon 1}{  \enc{\mathbf{p}P} \defeq 
    \InputIfFileExists{quine/symm1.tikz}{}{\input{tikz/quine/symm1.tikz}}
 } &
    \inferrule{P\colon 0}{  \enc{\mathbf{p}P} \defeq 
    \InputIfFileExists{quine/symm0.tikz}{}{\input{tikz/quine/symm0.tikz}}
 } 
    \\[25pt]
    \multicolumn{5}{c}{
      \begin{array}{c@{\qquad\qquad}c@{\qquad\qquad}c}
        \inferrule{P \colon n}{ \enc{\mathbf{P}P} \defeq 
    \InputIfFileExists{quine/symmn.tikz}{}{\input{tikz/quine/symmn.tikz}}
}  &
        \inferrule{P_1 \colon n \;\; P_2\colon m \;\; n\geq m}{\enc{P_1 \cap P_2} \defeq 
    \InputIfFileExists{quine/cap1.tikz}{}{\input{tikz/quine/cap1.tikz}}
} &
        \inferrule{P_1 \colon n \;\; P_2\colon m\;\; n<m}{\enc{P_1 \cap P_2} \defeq 
    \InputIfFileExists{quine/cap2.tikz}{}{\input{tikz/quine/cap2.tikz}}
}
      \end{array}
    } \\[25pt]
    \multicolumn{5}{c}{
      \begin{array}{c@{\qquad\qquad}c@{\qquad\qquad}c@{\qquad\qquad}c}
        \inferrule{P \colon n}{ \enc{\neg P} \defeq 
    \InputIfFileExists{quine/nega.tikz}{}{\input{tikz/quine/nega.tikz}}
}  &
        \inferrule{P \colon n}{ \enc{[ P} \defeq 
    \InputIfFileExists{quine/inj.tikz}{}{\input{tikz/quine/inj.tikz}}
}  &
        \inferrule{P \colon n+1}{ \enc{] P} \defeq 
    \InputIfFileExists{quine/proj.tikz}{}{\input{tikz/quine/proj.tikz}}
} &
        \inferrule{P \colon 0}{ \enc{] P} \defeq 
    \InputIfFileExists{quine/proj0.tikz}{}{\input{tikz/quine/proj0.tikz}}
}
      \end{array}
    }
    \\[8pt]
    \bottomrule
  \end{array}
  }$
\end{table*}

Inspired by combinatory logic, Quine~\cite{QUINE1971309} introduced \emph{predicate functor logic}, $\PFL$ for short, as a quantifier-free treatment of first order logic with equality. Several flavours of the logic have been proposed by Quine and others, here we focus on the treatment by Kuhn~\cite{kuhn1983}. Using the terminology of that thread of research, for
each $n\geq 0$ there is a collection of atomic $n$-ary predicates, corresponding to traditional $\FOL$ predicate symbols together with an additional binary predicate $I$ (identity). The term (predicate) constructors are called \emph{functors} -- here the terminology is unrelated to the notion of functor in category theory. These are divided into unary operations $\mathbf{p},\mathbf{P},[,]$  called \emph{combinatory functors} that, in the absence of explicit variables, capture the combinatorial aspects of handling variable lists as well as (existential) quantification. To get full expressivity of FOL, there are two additional \emph{alethic functors}: a binary conjunction and unary negation.

The syntax is reported on the top of Table \ref{fig:typingrulesQuine} where $R$ belong to $\sign$, a set of symbols with an associated arity. Similarly to $\NPR$, only the predicates that can be typed according to the rules in Table \ref{fig:typingrulesQuine} are considered. The semantics, on the bottom, is defined w.r.t. an interpretation $\interpretation$ consisting of a \emph{non-empty} set $X$ and a set $\rho(R)\subseteq X^n$ for all $R\in \sign$ of arity $n$. For all predicates $P$, $\dsemRel{P}$ is a subset of $X^\omega \defeq \{\tau_1 \cdot \tau_2 \cdots \mid \tau_i \in X \text{ for all }i\in\nat^+\}$. From $\interpretation =(X,\rho)$, one can define an interpretation of $\NPR$ $\mathcal{I}_p\defeq(X,\rho_p)$  where $\rho_p(R)\defeq \{(x,\star) \mid x\in \rho(R)\} \subseteq X^n\times \singleton$ for all $R \in \Sigma$ of arity $n$. The encoding of $\PFL$ into $\NPR$ is given in Table~\ref{table:quine encoding} where $
    \InputIfFileExists{quineSymm.tikz}{}{\input{tikz/quineSymm.tikz}}
$ is a suggestive representation for the permutation formally defined as $\symm[+][1][n-1]\seq[+] (\symm[+][n-2][1]\tensor[+] \id[+][1])$ for $n\geq 2$, $\id[+][n]$ for $n<2$.

\begin{proposition}\label{prop:Quine}
Let $P\colon n$ be a predicate of $\PFL$. Then
$\dsemRel{P} =\{\tau \mid ((\tau_1, \dots, \tau_n),\star)\in\interpretationFunctorP(\enc{P}) \}$.
\end{proposition}
\begin{proof}
  The proof goes by induction on the typing rules:
  
  Base cases:
  \begin{itemize}
  \item $I\colon 2$.  By definition $\dsemRel{I}=\{\tau \mid \tau_1=\tau_2\}$ and $\interpretationFunctorP (\enc{I})=\{((x_1,x_2),\star)\mid x_1=x_2\}$. Thus  $\dsemRel{I} =\{ \tau \mid ((\tau_1,\tau_2),\star)\in \interpretationFunctorP(\enc{I})\}$.
  \item $R\colon n$. Assume $\ari(R)=n$. By definition $\dsemRel{R} = \{\tau \mid (\tau_1, \dots, \tau_{n}) \in \rho(R)\}$ and $\interpretationFunctorP (\enc{R})=\{((x_1,\dots, x_n),\star)\mid (x_1, \dots x_n) \in \rho(R)\}$. Thus $\dsemRel{R} = \{ \tau \mid ((\tau_1,\dots, \tau_n),\star)\in \interpretationFunctorP(\enc{R})\}$.
  \end{itemize}
  The inductive cases follow always the same argument. We report below only the most interesting ones.
  \begin{itemize}
  \item $P_1 \cap P_2$. Assume $P_1\colon n$, $P_2 \colon m$ and $n\geq m$. %
  \begin{align*}
    &\dsemRel{P_1 \cap P_2} \\ 
  =\;\; &\dsemRel{P_1} \cap \dsemRel{P_2} \tag{def. $\dsemRel{\cdot}$}\\
  =\;\; &\begin{aligned}
      &\{ \tau \mid ((\tau_1, \dots, \tau_n),\star)\in\interpretationFunctorP(\enc{P_1}) \} \\[-5px]
      & \qquad \cap  \{\tau \mid ((\tau_1, \dots, \tau_m),\star)\in\interpretationFunctorP(\enc{P_2})\}
    \end{aligned} \tag{ind. hyp.}\\
  =\;\; &\begin{aligned}
    &\{ \tau \mid ((\tau_1, \dots, \tau_n),\star)\in\interpretationFunctorP(\enc{P_1}) \\[-5px]
    &\qquad\wedge ((\tau_1, \dots, \tau_m),\star)\in\interpretationFunctorP(\enc{P_2})\}
  \end{aligned} \\
  =\;\; &\{ \tau \mid ((\tau_1, \dots, \tau_n),\star)\in \interpretationFunctorP(\enc{P_1 \cap P_2} \} \tag{def. $\enc{\cdot}$ and $\interpretationFunctorP(\cdot)$}
  \end{align*}
  \item $\mathbf{p}P\colon 2$. Assume $P\colon 1$.
  \begin{align*}
  \dsemRel{\mathbf{p}P} &=  \{\tau  \mid \tau_2,\tau_1,\tau_3, \tau_4 \dots \in \dsemRel{P}\} \tag{def. $\dsemRel{\cdot}$}\\
  &=  \{\tau  \mid \tau_2,\tau_1,\dots \in \{ \tau \mid (\tau_1,\star)\in \interpretationFunctorP(\enc{P})\} \; \; \}  \tag{ind. hyp.}\\
  &=  \{\tau  \mid  (\tau_2,\star) \in \interpretationFunctorP(\enc{P})\}   \\
  & =  \{\tau  \mid ((\tau_1,\tau_2),\star) \in \interpretationFunctorP(\enc{\mathbf{p}P}) \} \tag{def. $\enc{\cdot}$ and $\interpretationFunctorP(\cdot)$}
  \end{align*}
  \item $]P\colon 0$. Assume $P\colon 0$. 
  \begin{align*}
  \dsemRel{]P} &=  \{\tau  \mid \tau_2,\tau_3,\dots \in \dsemRel{P}\} \tag{def. $\dsemRel{\cdot}$}\\
  &=  \{\tau  \mid \tau_2,\tau_3,\dots \in \{ \tau \mid (\star,\star)\in \interpretationFunctorP(\enc{P})\} \; \; \}  \tag{ind. hyp.}\\
  &=  \{\tau  \mid  (\star,\star) \in \interpretationFunctorP(\enc{P})\}   \\
  & =  \{\tau  \mid (\star,\star) \in \interpretationFunctorP(\enc{]P}) \} \tag{def. $\enc{\cdot}$ and $\interpretationFunctorP(\cdot)$}
  \end{align*}
  \end{itemize}
  
  \end{proof}

\subsection{Translation from $\NPR$ to $\FOL$}\label{app:fol}
In \S~\ref{sec:fol} we show how to translate typed formulas of $\FOL$ into diagrams of $\NPR$. Here we show the translation in the other direction.

\begin{figure*}
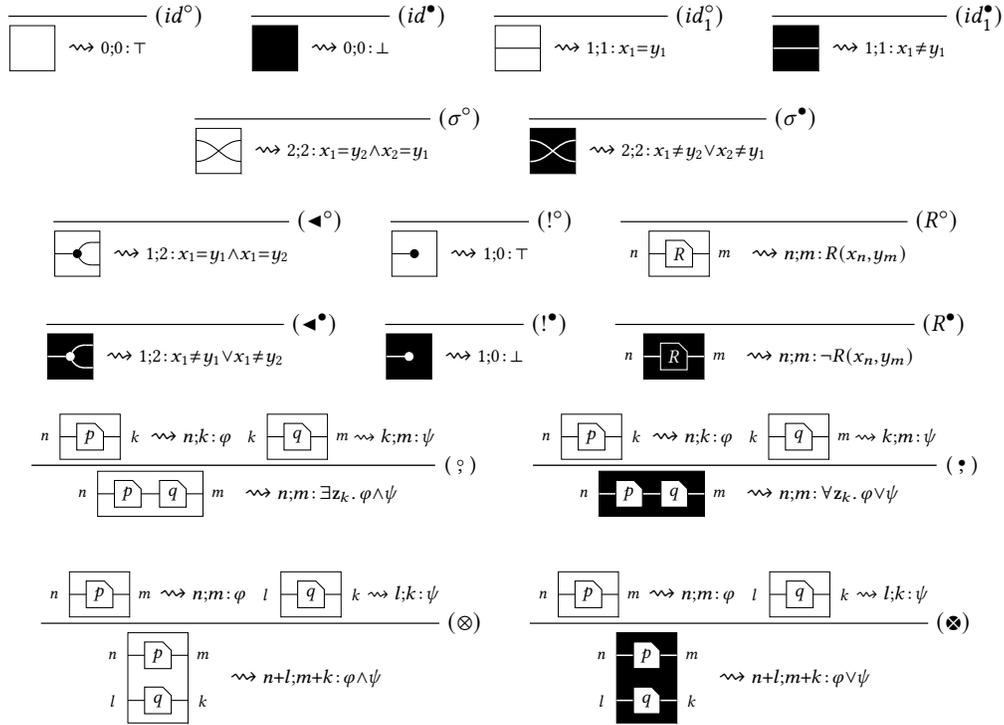

  \centering
  $
  \begin{array}{c}
    \begin{prooftree}
      \justifies
      \emptyCirc[+] \divider \scriptstyle \folFormula{0; 0}{\top}%
      \using (\id[+])
      \end{prooftree}
      \qquad
      \begin{prooftree}
      \justifies
      \emptyCirc[-] \divider \scriptstyle \folFormula{0; 0}{\bot}%
      \using (\id[-])
      \end{prooftree}
      \qquad
      \begin{prooftree}
      \justifies
      \idCirc[+] \divider \scriptstyle \folFormula{1; 1}{x_1 = y_1}
      \using (\id[+][1])
      \end{prooftree}
      \qquad
      \begin{prooftree}
      \justifies
      \idCirc[-] \divider \scriptstyle \folFormula{1; 1}{x_1 \neq y_1} %
      \using (\id[-][1])
      \end{prooftree}
      \\[20pt]
      \begin{prooftree}
        \justifies
        \symmCirc[+] \divider \scriptstyle \folFormula{2; 2}{x_1=y_2 \wedge x_2=y_1} %
        \using (\symm[+])
        \end{prooftree}
        \qquad
        \begin{prooftree}
        \justifies
        \symmCirc[-] \divider \scriptstyle \folFormula{2; 2}{x_1 \neq y_2 \vee x_2 \neq y_1} %
        \using (\symm[-])
        \end{prooftree}
        \\[20pt]
      \begin{prooftree}
        \justifies
        \copierCirc[+] \divider \scriptstyle \folFormula{1; 2}{x_1=y_1 \wedge x_1=y_2} %
        \using (\copier[+])
        \end{prooftree}
        \qquad
        \begin{prooftree}
        \justifies
        \discardCirc[+] \divider \scriptstyle \folFormula{1; 0}{\top} %
        \using (\discard[+])
        \end{prooftree}
        \qquad
        \begin{prooftree}
        \justifies
        \boxCirc[+]{R}[n][m] \divider \scriptstyle \folFormula{n; m}{R(x_n, y_m)} %
        \using (R^{\switchLabelS{+}})
        \end{prooftree}
        \\[20pt]
      \begin{prooftree}
        \justifies
        \copierCirc[-] \divider \scriptstyle \folFormula{1; 2}{x_1\neq y_1 \vee x_1\neq y_2} %
        \using (\copier[-])
        \end{prooftree}
        \qquad
        \begin{prooftree}
        \justifies
        \discardCirc[-] \divider \scriptstyle \folFormula{1; 0}{\bot} %
        \using (\discard[-])
        \end{prooftree}
        \qquad
        \begin{prooftree}
        \justifies
        \boxCirc[-]{R}[n][m] \divider \scriptstyle \folFormula{n; m}{\neg R(x_n, y_m)} %
        \using (R^{\switchLabelS{-}})
        \end{prooftree}
        \\[20pt]
        \inferrule*[right=({\!\!\! \seqOld[+] \!\!\!})]{\boxCirc[+]{p}[n][k] \!\!\!\divider \scriptstyle \folFormula{n;k}{\varphi}
    \;
    \boxCirc[+]{q}[k][m] \!\!\!\divider \scriptstyle \folFormula{k;m}{\psi}}{\seqCirc[+]{p}{q}[n][m] \divider \scriptstyle \folFormula{n;m}{\exists \mathbf{z}_k.\, \varphi \wedge \psi}}
    \qquad
    \inferrule*[right=({\!\!\! \seqOld[-] \!\!\!})]{\boxCirc[+]{p}[n][k] \!\!\!\divider \scriptstyle \folFormula{n;k}{\varphi}
    \;
    \boxCirc[+]{q}[k][m] \!\!\!\divider \scriptstyle \folFormula{k;m}{\psi}}{
        \seqCirc[-]{p}{q}[n][m] \divider \scriptstyle \folFormula{n;m}{\forall \mathbf{z}_k.\, \varphi \vee \psi}
    }
    \\[20pt]
    \begin{prooftree}
      \boxCirc[+]{p}[n][m] \!\!\!\divider \scriptstyle \folFormula{n;m}{\varphi}
      \;
      \boxCirc[+]{q}[l][k] \!\!\!\divider \scriptstyle \folFormula{l;k}{\psi} \vspace*{0.4em}
      \justifies
      \tensorCirc[+]{p}{q}[n][m][l][k] \divider \scriptstyle \folFormula{n+l;m+k}{\varphi \wedge \psi}
      \using (\tensor[+])
      \end{prooftree}
      \qquad
      \begin{prooftree}
      \boxCirc[+]{p}[n][m] \!\!\!\divider \scriptstyle \folFormula{n;m}{\varphi}
      \;
      \boxCirc[+]{q}[l][k] \!\!\!\divider \scriptstyle \folFormula{l;k}{\psi} \vspace*{0.4em}
      \justifies
      \tensorCirc[-]{p}{q}[n][m][l][k] \divider \scriptstyle \folFormula{n+l;m+k}{\varphi \vee \psi}
      \using (\tensor[-])
      \end{prooftree}
  \end{array}
$
\caption{Encoding of $\NPR$ diagrams as $\FOL$ formulas.}
\label{fig:fol reverse encoding}
\end{figure*}

Note that in general terms of $\NPR$ feature ``dangling'' wires both on the left and on the right of a term. While this is inconsequential from the point of view of expressivity, since terms can always be ``rewired'' using the compact closed structure of cartesian bicategories, this separation is convenient for composing terms in a flexible manner. Therefore, in the translation in Figure~\ref{fig:fol reverse encoding}, we keep two separate lists of free variables in the context, denoted as $n;m$, where $n$ and $m$ are the lenghts of the two lists.

\section{Proofs of Section~\ref{sec:cartesianbi}}\label{app:cb}

\begin{lemma}\label{lemma:cb maps} Let $(\Cat{C}, \copier[+], \cocopier[+])$ be a cartesian bicategory. The following holds
    \begin{enumerate}
    \item For all objects $X$, $\id[+][X]\colon X \to X$, $\copier[+][X]\colon X \to X \tensor[+] X$ and $\discard[+][X]\colon X \to \unittensor$ are maps;
    \item For maps $a$ and $b$ properly typed, $a \seq[+] b$ and $a \tensor[+] b$ are maps;
    \item If $a \colon \unittensor \to \unittensor $ is a map, then $a=\id[+][\unittensor]$;
    \item If $a \colon \unittensor \to X\tensor[+] Y $ is a map, then there exist maps $c\colon \unittensor \to X$ and $d\colon \unittensor \to Y$ such that $a=c\tensor[+]d$.
    \end{enumerate}
\end{lemma}
\begin{proof}
    See Theorem 1.6 in~\cite{carboni1987cartesian}.
\end{proof}

\begin{proof}[Proof of Proposition~\ref{prop:opcartesianfunctor}]
    See Theorem 2.4 in~\cite{carboni1987cartesian}.
\end{proof}

\begin{lemma}\label{lm:opfunctor}
    Let $\mathcal{F}\colon \Cat{C_1} \to \Cat{C_2}$ be a morphism of cartesian bicategories. Then, for all $c\colon X \to Y$, $\op{\mathcal{F}(c)} = \mathcal{F}(\op{c}).$
\end{lemma}
\begin{proof}
    See Remark 2.9 in~\cite{carboni1987cartesian}.
\end{proof}

\begin{proof}[Proof of Proposition~\ref{prop:map adj}]
    See Lemma 2.5 in~\cite{carboni1987cartesian}.
\end{proof}

The following generalises the well-known fact that $R$ is a function iff it is left adjoint to $\op{R}$.

\begin{proposition}\label{prop:map adj}
In a cartesian bicategory,  an arrow $c\colon X \to Y$ is a map iff $\op{c} \vdash c$.
\end{proposition}

\section{Proofs of Section \ref{sec:linbic}}\label{app:linearbic}
Several results stated in \S\ref{sec:linbic} (e.g., Lemmas \ref{lemma:uniquenessla}, \ref{lm:residuation} and \ref{lm:adjfunctor}) are well-known from \cite{cockett2000introduction}. However, for convenience of the reader, we group in this appendix the proofs of all the results stated in \S\ref{sec:linbic}.

\begin{proof}[Proof of Lemma \ref{lm:mix cat}]
	The proof of $(1)$ is on the left and $(2)$ on the right:
	
	\noindent\begin{minipage}{0.48\linewidth}
		\begin{align*}
			\id[-][\unittensor] &= \id[-][\unittensor] \seq[+] \id[+][\unittensor] \\
					  &= \id[-][\unittensor] \seq[+] (\id[-][\unittensor] \seq[-] \id[+][\unittensor]) \\ 
					  &\leq (\id[-][\unittensor] \seq[+] \id[-][\unittensor]) \seq[-] \id[+][\unittensor] \tag{\ref{ax:leftLinDistr}} \\ 
					  &= (\id[-][\unittensor] \tensor[+] \id[-][\unittensor]) \seq[-] \id[+][\unittensor] \tag{SMC} \\ 
					  &\leq (\id[-][\unittensor] \tensor[-] \id[-][\unittensor]) \seq[-] \id[+][\unittensor] \tag{\ref{ax:tensorPlusIdMinus}}\\ 
					  &= \id[+][\unittensor]
		\end{align*}
		\end{minipage}\quad \vline
		\begin{minipage}{0.48\linewidth}
		\begin{align*}
			&a \tensor[+] b \\
			=\; & (a \seq[-] \id[-]) \tensor[+] (b \seq[-] \id[-]) \\
			\leq\; & (a \tensor[-] b) \seq[-] (\id[-] \tensor[+] \id[-]) \tag{\ref{ax:linStrn2}} \\
			\leq\; & (a \tensor[-] b) \seq[-] (\id[-] \tensor[-] \id[-]) \tag{\ref{ax:tensorPlusIdMinus}} \\
			=\; & a \tensor[-] b
		\end{align*}
	\end{minipage}

	The proof of $(3)$ is given diagrammatically as follows:
	\input{tikz/proofs/linearlyDistrCat.tex}
\end{proof}

\begin{proof}[Proof of Lemma \ref{lemma:uniquenessla}] 
By the following two derivations.

\noindent\noindent\begin{minipage}{0.4\linewidth}
\begin{align*}
		b & =  b \seq[+] \id[+][X] \\
		 &\leq b \seq[+] (a \seq[-] c) \tag{$c \Vdash a$}\\
		 &\leq (b \seq[+] a )\seq[-] c \tag{\ref{ax:leftLinDistr}}\\
		 & \leq  \id[-][Y] \seq[-] c \tag{$b \Vdash a$}\\
		 &= c
\end{align*}
\end{minipage}\qquad \vline
\begin{minipage}{0.48\linewidth}
\begin{align*}
		c & =  c \seq[+] \id[+][X] \\
		 &\leq c \seq[+] (a \seq[-] b) \tag{$b \Vdash a$}\\
		 &\leq (c \seq[+] a )\seq[-] b \tag{\ref{ax:leftLinDistr}}\\
		 & \leq  \id[-][Y] \seq[-] b \tag{$c \Vdash a$}\\
		 &= b
\end{align*}
\end{minipage}

\end{proof}

\begin{proof}[Proof of Lemma \ref{lm:residuation}]
In the leftmost derivation we prove $a \leq b \Rightarrow \id[+][X] \leq b \seq[-] \rla{a} $ and in the rightmost $a \leq b \Leftarrow \id[+][X] \leq b \seq[-] \rla{a} $

\noindent\noindent\begin{minipage}{0.4\linewidth}
\begin{align*}
		\id[+][X] & \leq   a \seq[-] \rla{a} \tag{$\rla{a} \Vdash a$} \\
		 &\leq b \seq[-] \rla{a}  \tag{$a \leq b$}
\end{align*}
\end{minipage}\quad \vline
\begin{minipage}{0.5\linewidth}
\begin{align*}
		a & =  \id[+][X] \seq[+]  a  \\
		 &\leq (b \seq[-] \rla{a} ) \seq[+] a \tag{$\id[+][X] \leq b \seq[-] \rla{a} $}\\
		 &\leq b \seq[-] ( \rla{a}  \seq[+] a )  \tag{\ref{ax:rightLinDistr}}\\
		 & \leq  b \seq[-] \id[-][Y]   \tag{$\rla{a} \Vdash a$}\\
		 &= b
\end{align*}
\end{minipage}

\end{proof}

\begin{lemma}\label{lm:adjfunctor}
Let $\mathcal{F}\colon \Cat{C_1} \to \Cat{C_2}$ be a morphism of closed linear bicategories. Then, for all $a\colon X \to Y$,
$\rla{\mathcal{F}(a)} = \mathcal{F}(\rla{a}) %
$.
\end{lemma}	
\begin{proof}

Consider the following two derivations witnessing that $F(\rla{a})$ is right linear adjoint to $F(a)$.

\noindent\begin{minipage}{0.48\linewidth}
\begin{align*}
		\id[+][X] & =  F(\id[+][X]) \\
		 &\leq F(a \seq[-] \rla{a}) \tag{$\rla{a} \Vdash a$}\\
		 &= F(a) \seq[-] F(\rla{a}) 
\end{align*}
\end{minipage}\quad \vline
\begin{minipage}{0.48\linewidth}
\begin{align*}
		 &F(\rla{a}) \seq[+] F(a)   \\
		  = \;\; &F( \rla{a} \seq[+] a )  \\
		 \leq\;\; &F( \id[-][Y] ) \tag{$\rla{a} \Vdash a$}\\
		 = \;\;&\id[-][Y]
\end{align*}
\end{minipage}

Thus, by Lemma \ref{lemma:uniquenessla}, $\rla{(F(a))} = F(\rla{a})$. %
\end{proof}

\begin{proof}[Proof of Proposition \ref{prop:rlamorphism}]\label{lm:linadjmonoidal}\label{lm:linadjfunct}

First, we prove that for all $a,b\colon X \to Y$ it holds
\begin{enumerate}\setcounter{enumi}{-1}
\item if $a\leq b$ then $\rla{a} \geq \rla{b}$
\end{enumerate}

The proof is illustrated below.
\begin{enumerate}\setcounter{enumi}{-1}
\item \begin{align*}
		\rla{b} & =   \rla{b}\seq[+]\id[+][Y]  \\
		 & \leq \rla{b}\seq[+] (a \seq[-] \rla{a} )  \tag{$\rla{a}\Vdash a$} \\
		 & \leq ( \rla{b}\seq[+] a ) \seq[-] \rla{a}  \tag{\eqref{ax:leftLinDistr}}  \\
		 & \leq ( \rla{b}\seq[+] b ) \seq[-] \rla{a}  \tag{$a \leq b$} \\
		 &  \leq  \id[-][Y] \seq[-] \rla{a} \tag{$\rla{b}\Vdash b$} \\
		 &  =  \rla{a}  \\
\end{align*}
\end{enumerate}

We next illustrate that for all $a\colon X \to Y$ and $b\colon Y \to Z$
\begin{enumerate}
\item $\rla{(\id[+][X])}=\id[-][X]$ 
\item $\rla{(\id[-][X])}=\id[+][X]$
\item $\rla{(a \seq[+] b)} = \rla{b} \seq[-] \rla{a}$ 
\item $\rla{(a \seq[-] b)} = \rla{b} \seq[+] \rla{a}$
\end{enumerate}
The proofs are dispayed below.

\begin{enumerate}

\item Observe that $\id[+][X] = \id[+][X] \seq[-] \id[-][X]$ and $\id[-][X] \seq[+] \id[+][X] = \id[-][X]$. Thus, by Lemma \ref{lemma:uniquenessla}, $\rla{(\id[+][X])}=\id[-][X]$.

\item Similarly, $\id[+][X] = \id[-][X] \seq[-] \id[+][X]$ and $\id[+][X] \seq[+] \id[-][X] = \id[-][X]$. Again, by Lemma \ref{lemma:uniquenessla}, $\rla{(\id[-][X])}=\id[+][X]$.

\item The following two derivations

\noindent\begin{minipage}{0.54\linewidth}
\begin{align*}
		&\id[+][X] \\
		\leq \;  &a \seq[-] \rla{a} \tag{$\rla{a}\Vdash a$} \\
		 =\;  &(a \seq[+] \id[+][Y]) \seq[-] \rla{a} \\
		 \leq\;  &(a \seq[+] (b \seq[-]\rla{b}) ) \seq[-] \rla{a}  \tag{$\rla{b}\Vdash b$} \\
		 \leq\; &( (a \seq[+] b) \seq[-]\rla{b} ) \seq[-] \rla{a} \tag{\ref{ax:leftLinDistr}} \\
		 =\;  &(a \seq[+] b) \seq[-] (\rla{b}  \seq[-] \rla{a})
\end{align*}
\end{minipage}\quad \vline
\begin{minipage}{0.46\linewidth}
\begin{align*}
		 &(\rla{b}  \seq[-] \rla{a}) \seq[+] (a \seq[+] b)   \\
		  =\; &( (\rla{b}  \seq[-] \rla{a}) \seq[+] a) \seq[+] b  \\
		 \leq\;  &( \rla{b}  \seq[-] ( \rla{a} \seq[+] a)) \seq[+] b \tag{\ref{ax:rightLinDistr}}\\
		 \leq\;  &( \rla{b}  \seq[-] \id[-][Y]) \seq[+] b \tag{$\rla{a}\Vdash a$} \\
		 =\;  &\rla{b}  \seq[+] b \\
		 \leq\; &\id[-][Z] \tag{$\rla{b}\Vdash b$}
 \end{align*}
\end{minipage}
show that $(\rla{b}  \seq[-] \rla{a}) \Vdash (a \seq[+] b)$. Thus, by Lemma \ref{lemma:uniquenessla}, $\rla{(a \seq[+] b)} = \rla{b} \seq[-] \rla{a}$.

\item The following two derivations

\noindent\begin{minipage}{0.53\linewidth}
\begin{align*}
		&\id[+][X] \\
		 \leq\;&   a \seq[-] \rla{a} \tag{$\rla{a}\Vdash a$} \\
		 =\;  &a \seq[-] (\id[+][Y] \seq[+] \rla{a}) \\
		 \leq\; &  a \seq[-] ((b \seq[-]\rla{b})  \seq[+] \rla{a})  \tag{$\rla{b}\Vdash b$} \\
		 \leq\; &  a \seq[-] (b \seq[-] (\rla{b}  \seq[+] \rla{a})) \tag{\ref{ax:rightLinDistr}} \\
		 =\; &  (a \seq[-] b) \seq[-] (\rla{b}  \seq[+] \rla{a})
\end{align*}
\end{minipage}\; \vline
\begin{minipage}{0.47\linewidth}
\begin{align*}
		 &(\rla{b}  \seq[+] \rla{a}) \seq[+] (a \seq[-] b)   \\
		=\;  &\rla{b}  \seq[+] (\rla{a} \seq[+] (a \seq[-] b))  \\
		=\;  &\rla{b}  \seq[+] (( \rla{a} \seq[+] a) \seq[-] b ) \tag{\ref{ax:leftLinDistr}} \\
		\leq\;  &\rla{b}  \seq[+] ( \id[-][Y]) \seq[-] b) \tag{$\rla{a}\Vdash a$} \\
		=\; &   \rla{b}  \seq[+] b \\
		\leq\; & \id[-][Z] \tag{$\rla{b}\Vdash b$}
 \end{align*}
\end{minipage}
show that $(\rla{b}  \seq[+] \rla{a}) \Vdash (a \seq[-] b)$. Thus, by Lemma \ref{lemma:uniquenessla}, $\rla{(a \seq[-] b)} = \rla{b} \seq[+] \rla{a}$.
\end{enumerate}

Next, we illustrate that for all $a\colon X_1 \to Y_1$ and $b\colon X_2 \to Y_2$
\begin{enumerate}\setcounter{enumi}{4}
\item $\rla{(a \tensor[+] b )} = \rla{a} \tensor[-] \rla{b}$ 
\item $\rla{(a \tensor[-] b )} = \rla{a} \tensor[+] \rla{b}$
\item $\rla{(\symm[+])} = \symm[-]$
\item $\rla{(\symm[-])} = \symm[+]$
\end{enumerate}
The proofs are shown below.

\begin{enumerate}\setcounter{enumi}{4}
\item The following two derivations

\noindent\begin{minipage}{0.5\linewidth}
\begin{align*}
		    & \id[+][X_1 \tensor[+] X_2] \\
		  = & \id[+][X_1]\tensor[+] \id[+][X_2] \\
	   \leq & (a \seq[-] \rla{a}) \tensor[+] (b \seq[-] \rla{b}) \tag{$\rla{a}\Vdash a \; ,\; \rla{b}\Vdash b$} \\
	   \leq & (a \tensor[+] b) \seq[-] (\rla{a} \tensor[-] \rla{b}) \tag{\ref{ax:linStrn4}}
\end{align*}
\end{minipage}\; \vline
\begin{minipage}{0.5\linewidth}
\begin{align*}
		 &(\rla{a} \tensor[-] \rla{b}) \seq[+] (a \tensor[+] b)  \\
	\leq &(\rla{a} \seq[+] a) \tensor[-] (\rla{b} \seq[+] b) \tag{\ref{ax:linStrn3}} \\
    \leq &\id[-][Y_1]\tensor[-] \id[-][Y_2] \tag{$\rla{a}\Vdash a \; ,\; \rla{b}\Vdash b$} \\
	 =   &\id[-][Y_1 \tensor[-] Y_2]
 \end{align*}
\end{minipage}
show that $(\rla{a} \tensor[-] \rla{b}) \Vdash (a \tensor[+] b)$. Thus, by Lemma \ref{lemma:uniquenessla}, $\rla{(a \tensor[+] b)} = \rla{b} \tensor[-] \rla{a}$.

\item The following two derivations

\noindent\begin{minipage}{0.5\linewidth}
\begin{align*}
		 &\id[+][X_1 \tensor[+] X_2] \\
		=&\id[+][X_1]\tensor[+] \id[+][X_2] \\
	\leq &(a \seq[-] \rla{a}) \tensor[+] (b \seq[-] \rla{b}) \tag{$\rla{a}\Vdash a \; ,\; \rla{b}\Vdash b$} \\
	\leq &(a \tensor[-] b) \seq[-] (\rla{a} \tensor[+] \rla{b}) \tag{\ref{ax:linStrn2}}
\end{align*}
\end{minipage}\; \vline
\begin{minipage}{0.5\linewidth}
\begin{align*}
		 &(\rla{a} \tensor[+] \rla{b}) \seq[+] (a \tensor[-] b)   \\
	\leq &(\rla{a} \seq[+] a) \tensor[-] (\rla{b} \seq[+] b) \tag{\ref{ax:linStrn3}} \\
	\leq &\id[-][Y_1]\tensor[-] \id[-][Y_2] \tag{$\rla{a}\Vdash a \; ,\; \rla{b}\Vdash b$} \\
	=    &\id[-][Y_1 \tensor[-] Y_2]
 \end{align*}
\end{minipage}
show that $(\rla{a} \tensor[+] \rla{b}) \Vdash (a \tensor[-] b)$. Thus, by Lemma \ref{lemma:uniquenessla}, $\rla{(a \tensor[-] b)} = \rla{b} \tensor[+] \rla{a}$.

\item By axioms \eqref{ax:tauSymmPlus} and \eqref{ax:gammaSymmPlus}.
\item By axioms \eqref{ax:tauSymmMinus} and \eqref{ax:gammaSymmMinus}.

\end{enumerate}

\end{proof}

\section{Proof of Section \ref{sec:fobic}}

\subsection{Proofs of Proposition \ref{prop:opfunctor}}
In this appendix we illustrate several results to prove Proposition \ref{prop:opfunctor}. We first focus on $\op{(\cdot)} \colon \Cat{C} \to \opposite{\Cat{C}}$ (Lemma \ref{lemma:opfunctor}) and then $\rla{(\cdot)}\colon \Cat{C} \to \opposite{(\co{\Cat{C}})}$ (Lemma \ref{lemma:rlafunctor}).

In order to prove that $\op{(\cdot)} \colon \Cat{C} \to \opposite{\Cat{C}}$ is a morphism of fo-bicategories, it is convenient to define, for all arrows $c\colon X \to Y$, $\opp{c} \colon Y \to X$ as
\[\opp{c} \defeq \daggerCirc[-]{c}[Y][X] \] 
The assignment $c \mapsto \opp{c}$ gives rise to an identity on object functor  $\opp{(\cdot)} \colon \Cat{C} \to \opposite{\Cat{C}}$ which preserves the stucture of \emph{co}cartesian bicategories.

\begin{proposition}\label{prop:oppcartesianfunctor}
$\opp{(\cdot)}\colon \Cat{C} \to \opposite{\Cat{C}}$ is an isomorphism of cocartesian bicategories, that is the rules in the first three rows of Table \ref{table:oppproperties} hold.
\end{proposition}
\begin{proof}
	See Theorem 2.4 in~\cite{carboni1987cartesian}.
\end{proof}

\begin{table}[!htb]
		\caption{Properties of $\opp{\cdot} \colon \Cat{C} \to \opposite{\Cat{C}}$}\label{table:oppproperties}
        \begin{tabular}{c}
	{\tiny
	$ 
	\begin{array}{@{}cccc@{}}
		\toprule
		\multicolumn{2}{c}{
			\text{if }c\leq d\text{ then }\opp{c} \leq \opp{d}
		}
		&
		\multicolumn{2}{c}{
			\opp{(\opp{c})}= c
		}
		\\
	\opp{(c \seq[+] d)} = \opp{d} \seq[+] \opp{c}
	&\opp{(\id[+][X])}=\id[+][X] 
	&\opp{(\cocopier[+][X])}= \copier[+][X]
	& \opp{(\codiscard[+][X])}= \discard[+][X]
	\\
	\opp{(c \tensor[+] d)} = \opp{c} \tensor[+] \opp{d} 
	& \opp{(\symm[+][X][Y])} = \symm[+][Y][X] 
	& \opp{(\copier[+][X])}= \cocopier[+][X]
	& \opp{(\discard[+][X])}= \codiscard[+][X]
	\\
	\midrule
	\opp{(c \seq[-] d)} = \opp{d} \seq[-] \opp{c}
	& \opp{(\id[-][X])} = \id[-][X]
	&\opp{(\cocopier[-][X])}= \copier[-][X]
	& \opp{(\codiscard[-][X])}= \discard[-][X]
	\\
	\opp{(c \tensor[-] d)} = \opp{c} \tensor[-] \opp{d} 
	& \opp{(\symm[-][X][Y])} = \symm[-][Y][X] 
	& \opp{(\copier[-][X])}= \cocopier[-][X]
	& \opp{(\discard[-][X])}= \codiscard[-][X] 
	\\
	\bottomrule
	\end{array}
	$}
\end{tabular}
\end{table}

\begin{lemma}\label{lm:mixed spiders}
	The following hold: 
	\[ (1)\; \copierCirc[-] = 
    \InputIfFileExists{mixedSpiders.tikz}{}{\input{tikz/mixedSpiders.tikz}}
 \quad  (2)\; 
    \InputIfFileExists{axioms/linadj/comPmon.tikz}{}{\input{tikz/axioms/linadj/comPmon.tikz}}
 = \idCirc[-] \quad  (3)\; 
    \InputIfFileExists{axiomsNEW/linadj2/comPmon2.tikz}{}{\input{tikz/axiomsNEW/linadj2/comPmon2.tikz}}
 = \idCirc[-] \]
\end{lemma}
\begin{proof}
Point $(1)$ is proved by the following derivation: 
\[\copierCirc[-] \stackrel{\eqref{ax:comPlusUnit}}{=} 
    \InputIfFileExists{mixedSpiders2.tikz}{}{\input{tikz/mixedSpiders2.tikz}}
 \stackrel{\eqref{ax:bwFrob2}}{=} 
    \InputIfFileExists{mixedSpiders.tikz}{}{\input{tikz/mixedSpiders.tikz}}
.\]

For point $(2)$ observe that the left to right inclusion is \eqref{ax:gammaCocopierPlus} and the other inclusion is proved as follows:
\input{tikz/proofs/mixedSpecFrob.tex}

Proof of point $(3)$ is analogous to the one above, except that one exploits $\eqref{ax:rightLinDistr}$ and $\eqref{ax:gammaCocopierMinus}$.
\end{proof}

\begin{lemma}\label{lm:id dagger}
The following hold: $\id[-][X] = \op{(\id[-][X])}$ and $\id[+][X] = (\id[+][X])^\ddagger$ 
\end{lemma}
\begin{proof}
	Here we show only $\id[-][X] = \op{(\id[-][X])}$, the other follows a similar reasoning.
	\input{tikz/proofs/idDagger}
\end{proof}

\begin{lemma}\label{lm:daggeradj}
  For all $a \colon X \to Y$ it holds $\rla{(a^\dagger)} = (\rla{a})^\ddagger$
\end{lemma}
\begin{proof}
  The proof follows from the fact that $(\copier[+], \discard[+])$ is right linear adjoint to $(\cocopier[-], \codiscard[-])$, Proposition~\ref{prop:rlamorphism} and the definition of $(\cdot)^\dagger$ and $(\cdot)^\ddagger$.
\end{proof}

\begin{lemma}\label{lm:dagger}
  For all $a \colon X \to Y$ it holds $a^\dagger = a^\ddagger$
\end{lemma}

\begin{proof}
	We prove the inclusion $a^\dagger \leq a^\ddagger$ (left) by means of Lemma~\ref{lm:residuation} and the other inclusion (right) directly: \\
	\begin{minipage}{0.54\linewidth}
		\begin{align*}
				    &(a^\ddagger \seq[-] \rla{(a^\dagger)}) \\
			 =\;    & a^\ddagger \seq[-] (\rla{a})^\ddagger \tag{Lemma~\ref{lm:daggeradj}}\\
			 =\;    &(\rla{a} \seq[-] a)^\ddagger  \tag{Table~\ref{table:oppproperties}}   \\
			 \geq\; &(\id[+][Y])^\ddagger   \tag{$\rla{a}\Vdash a$}\\
			=\;     &\id[+][Y] \tag{Lemma \ref{lm:id dagger}}
		\end{align*}
	\end{minipage}
	\; \vline
	\begin{minipage}{0.46\linewidth}
		\begin{align*}
			&\opp{a} \\
			=\; &\opp{(\op{(\op{a})})} \tag{$\op{(\cdot)}$ is an iso} \\
			\leq\; &\opp{(\opp{(\op{a})})} \tag{$\op{a} \leq \opp{a}$} \\
			=\; &\op{a} \tag{$\opp{(\cdot)}$ is an iso}
		\end{align*}
	\end{minipage}
	\end{proof}

\begin{lemma}\label{lemma:opfunctor}
$\op{(\cdot)} \colon \Cat{C} \to \opposite{\Cat{C}}$ is an isomorphisms of fo-bicategories, namely all the laws in Table \ref{table:daggerproperties}.(a) hold.
\end{lemma}
\begin{proof}
Follows from Lemma~\ref{lm:dagger} and the fact that $\op{(\cdot)}$ preserves the positive structure (Proposition \ref{prop:opcartesianfunctor})  and  $(\cdot)^\ddagger$ preserve the negative structure (Proposition \ref{prop:oppcartesianfunctor}). For instance,
to prove that  $\op{(a \seq[-] b)} = \op{b} \seq[-] \op{a}$, it is enough to observe that  $\op{(a \seq[-] b)} = (a \seq[-]b)^\ddagger$ and that $ (a \seq[-]b)^\ddagger = b^\ddagger \seq[-] a^\ddagger$.
\end{proof}

\begin{lemma}\label{lemma:doublelinearadjoint}
For all $a\colon X \to Y$
\begin{enumerate}
\item $\rla{(\rla{a})}=a$
\end{enumerate}
\end{lemma}
\begin{proof}
The following two derivations

\noindent\begin{minipage}{0.5\linewidth}
\begin{align*}
		&\id[+][Y] \\
		 = &\op{(\id[+][Y])} \tag{Proposition \ref{prop:opcartesianfunctor}} \\
		 \leq &\op{( \op{a} \seq[-]\rla{(\op{a})})} \tag{$\rla{(\op{a})} \Vdash \op{a}$}\\
		 = &\op{( \op{a} \seq[-]\op{(\rla{a})})} \tag{Corollary \ref{cor:oplinearadjoint}} \\
		 =  &\op{(\op{( \rla{a} \seq[-]a)})}  \tag{Lemma \ref{lemma:opfunctor}} \\
		=  &\rla{a} \seq[-] a  \tag{Proposition \ref{prop:opcartesianfunctor}}
\end{align*}
\end{minipage}\; \vline
\begin{minipage}{0.5\linewidth}
\begin{align*}
		&\id[-][X] \\
	= &\op{(\id[-][X])} \tag{Lemma \ref{lemma:opfunctor}} \\
	\geq &\op{( \rla{(\op{a})} \seq[+]\op{a})} \tag{$\rla{(\op{a})} \Vdash \op{a}$}\\
	= &\op{( \op{(\rla{a})}  \seq[+] \op{a}  )} \tag{Corollary \ref{cor:oplinearadjoint}} \\
	=  &\op{(\op{( a \seq[+] \rla{a} )})}  \tag{Proposition \ref{prop:opcartesianfunctor}} \\
	=  &a \seq[+] \rla{a}  \tag{Proposition \ref{prop:opcartesianfunctor}}
\end{align*}
\end{minipage}

prove that the right linear adjoint of $\rla{a}$ is $a$. Thus, by Lemma \ref{lemma:uniquenessla}, $\rla{(\rla{a})}=a$.

\end{proof}

\begin{lemma}\label{lemma:rlafunctor}
$\rla{(\cdot)}\colon \Cat{C} \to \opposite{(\co{\Cat{C}})}$ is an isomorphisms of fo-bicategories, namely all the laws in Table \ref{table:rlaproperties}.(b) hold.
\end{lemma}

\begin{proof}
By  Proposition \ref{prop:rlamorphism}, $\rla{(\cdot)}\colon \Cat{C} \to \opposite{(\co{\Cat{C}})}$ is a morphism of linear bicategories.
Observe that $\opposite{(\co{\Cat{C}})}$ carries the structure of a cartesian bicategory where the positive comonoid is $(\cocopier[-], \codiscard[-])$ and the positive monoid is $(\copier[-], \discard[-])$. By Definition \ref{def:linear bicategory}.4, one has that $\rla{(\copier[+])} = \cocopier[-]$, $\rla{(\discard[+])}=\codiscard[-]$ and $\rla{(\cocopier[+])} = \copier[-]$, $\rla{(\codiscard[+])}=\discard[-]$. Thus $\rla{(\cdot)}\colon \Cat{C} \to \opposite{(\co{\Cat{C}})}$ is a morphism of cartesian bicategories.

By Lemma \ref{lemma:doublelinearadjoint}, we also immediately know that $\rla{(\copier[-])} = \cocopier[+]$, $\rla{(\discard[-])}=\codiscard[+]$ and $\rla{(\cocopier[-])} = \copier[+]$, $\rla{(\codiscard[-])}=\discard[+]$. Thus, $\rla{(\cdot)}\colon \Cat{C} \to \opposite{(\co{\Cat{C}})}$ is a morphism of cocartesian bicategories. Thus, it is a morphism of fo-bicategories.

The fact that it is an isomorphism is immediate by Lemma \ref{lemma:doublelinearadjoint}.
\end{proof}

\begin{proof}[Proof of Proposition \ref{prop:rlafunctor}]
By Lemmas \ref{lemma:opfunctor} and \ref{lemma:rlafunctor}.
\end{proof}

\subsection{Proofs of the other results}

\begin{proof}[Proof of Corollary \ref{cor:oplinearadjoint}]
$\rla{(\op{c})} = \op{(\rla{c})}$ is immediate from Proposition \ref{prop:opfunctor} and Lemma \ref{lm:adjfunctor}. The other rules follow from the definitions of $\sqcap$,  $\top$, $\sqcup$, $\bot$ in \eqref{eq:def:cap} and \eqref{eq:def:cap}, and the laws in Tables \ref{table:daggerproperties}.(a) and  \ref{table:rlaproperties}.(b). For instance $\rla{(\bot)} = \rla{( \discard[-] \seq[-] \codiscard[-])} = \rla{(\codiscard[-])} \seq[+] \rla{(\discard[-])} = \discard[+] \seq[+] \codiscard[+] = \top$.
\end{proof}

\begin{proof}[Proof of Proposition \ref{prop:maps}]
Recall that, by Proposition \ref{prop:map adj} an arrow $f\colon X \to Y$ is a map iff it is a left adjoint, namely
\begin{equation}\label{eq:mapsleftadjoint}
\id[+][X] \leq f \seq[+] \op{f} \qquad \op{f} \seq[+] f \leq \id[+][Y]
\end{equation}
 
The following two derivations prove the two inclusion.

\noindent\begin{minipage}{0.5\linewidth}
\begin{align*}
		     &f \seq[+] c \\
		=\;    &\id[+][X] \seq[+] f \seq[+]c   \\
		\leq\; &(\rla{(\op{f})}\seq[-] \op{f}) \seq[+] f \seq[+] c  \tag{$\op{f} \Vdash \rla{(\op{f})}$}\\
		\leq\; &\rla{(\op{f})} \seq[-] (\op{f} \seq[+] f \seq[+] c) \tag{\ref{ax:rightLinDistr}} \\
		\leq\; &\rla{(\op{f})} \seq[-] (\id[+][Y] \seq[+] c)   \tag{\ref{eq:mapsleftadjoint}} \\
		=\;    &\rla{(\op{f})} \seq[-]  c  
\end{align*}
\end{minipage}\; \vline
\begin{minipage}{0.5\linewidth}
\begin{align*}
		     &f \seq[+] c \\
		=\;    &f \seq[+] ( \id[-][X] \seq[-] c)   \\
	    \geq\; &f \seq[+] ( (\op{f} \seq[+] \rla{(\op{f})} ) \seq[-] c)  \tag{$(\op{f}) \Vdash \rla{(\op{f})}$}\\
		\geq\; &f \seq[+]  \op{f} \seq[+] (\rla{(\op{f})}  \seq[-] c ) \tag{\ref{ax:leftLinDistr}} \\
		\geq\; &\id[+][X] \seq[+] (\rla{(\op{f})}  \seq[-] c)    \tag{\ref{eq:mapsleftadjoint}} \\
		=\;    &\rla{(\op{f})} \seq[-]  c  
\end{align*}
\end{minipage}
Note that $\op{f} \Vdash \rla{(\op{f})}$ holds since, by Proposition \ref{prop:rlafunctor}, in any fo-bicategory left and right linear adjoint coincide (namely $\rla{(\rla{a})}=a$).
 
To check the four equivalences, first observe that
\[c \seq[+] \op{f} = \op{ (f \seq[+] c )} = \op{(\rla{(\op{f})} \seq[-] c)}= c \seq[-] \rla{f}\] 
and conclude by taking as map $f$ either $\copier[+]$ or $\discard[+]$.
\end{proof}

\begin{lemma}\label{lm:seq ditr}
	Let $\Cat{C}$ be a fo-bicategory. Then, $(\Cat{C}, \seq[+], \tensor[+])$ and $(\Cat{C}, \seq[-], \tensor[-])$ are monoidally enriched over $\sqcup$-semilattices with $\bot$ and $\sqcap$-semilattices with $\top$, respectively. Namely, the following hold:
	\begin{enumerate}
		\item $a \seq[+] (b \sqcup c) = (a \seq[+] b) \sqcup (a \seq[+] c)$ and $(b \sqcup c) \seq[+] a = (b \seq[+] a) \sqcup (c \seq[+] a)$
		\item $a \seq[-] (b \sqcap c) = (a \seq[-] b) \sqcap (a \seq[-] c)$ and $(b \sqcap c) \seq[-] a = (b \seq[-] a) \sqcap (c \seq[-] a)$
		\item $a \seq[+] \bot = \bot = \bot \seq[+] a$
		\item $a \seq[-] \top = \top = \top \seq[-] a$
		\item $a \tensor[+] (b \sqcup c) = (a \tensor[+] b) \sqcup (a \tensor[+] c)$ and $(b \sqcup c) \tensor[+] a = (b \tensor[+] a) \sqcup (c \tensor[+] a)$
		\item $a \tensor[-] (b \sqcap c) = (a \tensor[-] b) \sqcap (a \tensor[-] c)$ and $(b \sqcap c) \tensor[-] a = (b \tensor[-] a) \sqcap (c \tensor[-] a)$
		\item $a \tensor[+] \bot = \bot = \bot \tensor[+] a$
		\item $a \tensor[-] \top = \top = \top \tensor[-] a$
	\end{enumerate}
\end{lemma}
\begin{proof}
	We prove the two inclusions of the first equation in $(1)$ separately.
	\input{tikz/proofs/seqPlusDistrUnion}

	For the second equation, namely the one with the composition on the right, it suffices to apply the properites of $\op{(\cdot)}$ in Tables~\ref{table:daggerproperties}.(a) and~\ref{table:daggerrlalattice}.(c) and the drivation above to get that:
	\begin{align*}
		(b \sqcup c) \seq[+] a &= \op{(\op{((b \sqcup c) \seq[+] a)})} = \op{(\op{a} \seq[+] (\op{b} \sqcup \op{c}))}\\
		&= \op{((\op{a} \seq[+] \op{b}) \sqcup (\op{a} \seq[+] \op{c}))} = \op{(\op{((b \seq[+] a) \sqcup (a \seq[+] c))})} \\
		&= (b \seq[+] a) \sqcup (a \seq[+] c)
	\end{align*}

	The proofs for $(2)$ are analogous to those of $(1)$.

	We prove the left to right inclusion of the first equation in $(3)$. The other inclusion holds since $\bot$ is the bottom element.
	\input{tikz/proofs/seqPlusDistrBot}

	For the second equation, namely the one with the composition on the right, it suffices to apply the properites of $\op{(\cdot)}$ in Tables~\ref{table:daggerproperties}.(a) and~\ref{table:daggerrlalattice}.(c) and the drivation above to get that:
	\[ \bot \seq[+] a = \op{(\op{(\bot \seq[+] a)})} = \op{(\op{a} \seq[+] \op{\bot})} = \op{(\op{a} \seq[+] \bot)} = \op{\bot} = \bot  \]

	The proofs for $(4)$ are analogous to those of $(3)$.

	The right to left inclusion of the first equation in $(5)$ is proved by the universal property of $\sqcup$, namely:
	if $a \tensor[+] b = a \tensor[+] (b \sqcup \bot) \leq a \tensor[+] (b \sqcup c)$ and $a \tensor[+] c = a \tensor[+] (\bot \sqcup c) \leq a \tensor[+] (b \sqcup c)$, then $(a \tensor[+] b) \sqcup (a \tensor[+] c) \leq  a \tensor[+] (b \sqcup c)$.

	For the other inclusion, the following holds:
	
	\input{tikz/proofs/tensorPlusDistrUnion.tex}

	For the second equation, namely $(b \sqcup c) \tensor[+] a = (b \tensor[+] a) \sqcup (c \tensor[+] a)$, the proof follows the exact same reasoning.

	The proofs for $(6)$ are analogous to those of $(5)$.

	We prove the left to right inclusion of the first equation in $(7)$. The other inclusion holds since $\bot$ is the bottom element.

	\input{tikz/proofs/tensorPlusDistrBot}

	For the second equation, namely $\bot = \bot \tensor[+] a$, the proof follows the exact same reasoning.

	The proofs for $(8)$ are analogous to those of $(7)$.
\end{proof}

\begin{lemma}\label{lm:id cup}
	The following hold:
	\[\cupCirc[+] \leq \cupCircIdMinus  \quad\text{ and }\quad \cupCircIdPlus \leq \cupCirc[-]\]
\end{lemma}
\begin{proof}
	\input{tikz/proofs/idCup.tex}
	The proof of the other inequality is analogous. 
\end{proof}

\begin{lemma}\label{lm:id non contr}
	The following hold:
	\[
    \InputIfFileExists{axioms/noncontr.tikz}{}{\input{tikz/axioms/noncontr.tikz}}
 \leq \bottomCirc \quad \text{ and } \quad \topCirc \leq 
    \InputIfFileExists{axioms/exclmiddle.tikz}{}{\input{tikz/axioms/exclmiddle.tikz}}
\]
\end{lemma}
\begin{proof}
	We prove it by means of Lemma~\ref{lm:residuation} as follows:
	\input{tikz/proofs/LNC_id.tex}
	The proof of the other inequality is analogous.
\end{proof}

\begin{lemma}\label{lm:wrong way}
	The following hold:
	\[
    \begin{tikzpicture}
        \begin{pgfonlayer}{nodelayer}
            \node [{boxStyle/+}] (134) at (0.5, 0.5) {$c$};
            \node [style=none] (135) at (-1.25, 0) {};
            \node [style=none] (136) at (-1.25, 1.225) {};
            \node [style=none] (137) at (-1.25, -1) {};
            \node [style=none] (138) at (1.25, -1) {};
            \node [style=none] (139) at (1.25, 1.225) {};
            \node [style={dotStyle/+}] (144) at (-0.5, 0) {};
            \node [style=none] (145) at (0.25, -0.5) {};
            \node [style=none] (146) at (1.25, 0.5) {};
            \node [style=none] (147) at (1.25, -0.5) {};
            \node [style=none] (150) at (0.25, 0.5) {};
        \end{pgfonlayer}
        \begin{pgfonlayer}{edgelayer}
            \draw [{bgStyle/+}] (138.center)
                 to (137.center)
                 to (136.center)
                 to (139.center)
                 to cycle;
            \draw [style={wStyle/+}, bend left] (145.center) to (144);
            \draw [style={wStyle/+}] (144) to (135.center);
            \draw [style={wStyle/+}] (134) to (146.center);
            \draw [style={wStyle/+}] (147.center) to (145.center);
            \draw [style={wStyle/+}] (134) to (150.center);
            \draw [style={wStyle/+}, bend right] (150.center) to (144);
        \end{pgfonlayer}
    \end{tikzpicture}
    \leq
    \begin{tikzpicture}
        \begin{pgfonlayer}{nodelayer}
            \node [{boxOpStyle/+}] (134) at (-1, -0.25) {$c$};
            \node [style={boxStyle/+}] (135) at (-3, 0.25) {$c$};
            \node [style=none] (136) at (-3.75, 1.225) {};
            \node [style=none] (137) at (-3.75, -1) {};
            \node [style=none] (138) at (-0.25, -1) {};
            \node [style=none] (139) at (-0.25, 1.225) {};
            \node [style={dotStyle/+}] (144) at (-2, 0.25) {};
            \node [style=none] (145) at (-1.25, 0.75) {};
            \node [style=none] (146) at (-0.25, 0.75) {};
            \node [style=none] (147) at (-0.25, -0.25) {};
            \node [style=none] (150) at (-1.25, -0.25) {};
            \node [style=none] (152) at (-3.75, 0.25) {};
        \end{pgfonlayer}
        \begin{pgfonlayer}{edgelayer}
            \draw [{bgStyle/+}] (138.center)
                 to (137.center)
                 to (136.center)
                 to (139.center)
                 to cycle;
            \draw [style={wStyle/+}, bend right] (145.center) to (144);
            \draw [style={wStyle/+}] (144) to (135);
            \draw [style={wStyle/+}] (134) to (150.center);
            \draw [style={wStyle/+}, bend left] (150.center) to (144);
            \draw [style={wStyle/+}] (134) to (147.center);
            \draw [style={wStyle/+}] (146.center) to (145.center);
            \draw [style={wStyle/+}] (135) to (152.center);
        \end{pgfonlayer}
    \end{tikzpicture}        
    \qquad\qquad
    \begin{tikzpicture}
        \begin{pgfonlayer}{nodelayer}
            \node [{boxOpStyle/+}] (134) at (-1, -0.25) {$c$};
            \node [style={boxStyle/+}] (135) at (-3, 0.25) {$c$};
            \node [style=none] (136) at (-3.75, 1.225) {};
            \node [style=none] (137) at (-3.75, -1) {};
            \node [style=none] (138) at (-0.25, -1) {};
            \node [style=none] (139) at (-0.25, 1.225) {};
            \node [style={dotStyle/-}] (144) at (-2, 0.25) {};
            \node [style=none] (145) at (-1.25, 0.75) {};
            \node [style=none] (146) at (-0.25, 0.75) {};
            \node [style=none] (147) at (-0.25, -0.25) {};
            \node [style=none] (150) at (-1.25, -0.25) {};
            \node [style=none] (152) at (-3.75, 0.25) {};
        \end{pgfonlayer}
        \begin{pgfonlayer}{edgelayer}
            \draw [{bgStyle/-}] (138.center)
                 to (137.center)
                 to (136.center)
                 to (139.center)
                 to cycle;
            \draw [style={wStyle/-}, bend right] (145.center) to (144);
            \draw [style={wStyle/-}] (144) to (135);
            \draw [style={wStyle/-}] (134) to (150.center);
            \draw [style={wStyle/-}, bend left] (150.center) to (144);
            \draw [style={wStyle/-}] (134) to (147.center);
            \draw [style={wStyle/-}] (146.center) to (145.center);
            \draw [style={wStyle/-}] (135) to (152.center);
        \end{pgfonlayer}
    \end{tikzpicture}   
    \leq
    \begin{tikzpicture}
        \begin{pgfonlayer}{nodelayer}
            \node [{boxStyle/+}] (134) at (0.5, 0.5) {$c$};
            \node [style=none] (135) at (-1.25, 0) {};
            \node [style=none] (136) at (-1.25, 1.225) {};
            \node [style=none] (137) at (-1.25, -1) {};
            \node [style=none] (138) at (1.25, -1) {};
            \node [style=none] (139) at (1.25, 1.225) {};
            \node [style={dotStyle/-}] (144) at (-0.5, 0) {};
            \node [style=none] (145) at (0.25, -0.5) {};
            \node [style=none] (146) at (1.25, 0.5) {};
            \node [style=none] (147) at (1.25, -0.5) {};
            \node [style=none] (150) at (0.25, 0.5) {};
        \end{pgfonlayer}
        \begin{pgfonlayer}{edgelayer}
            \draw [{bgStyle/-}] (138.center)
                 to (137.center)
                 to (136.center)
                 to (139.center)
                 to cycle;
            \draw [style={wStyle/-}, bend left] (145.center) to (144);
            \draw [style={wStyle/-}] (144) to (135.center);
            \draw [style={wStyle/-}] (134) to (146.center);
            \draw [style={wStyle/-}] (147.center) to (145.center);
            \draw [style={wStyle/-}] (134) to (150.center);
            \draw [style={wStyle/-}, bend right] (150.center) to (144);
        \end{pgfonlayer}
    \end{tikzpicture} 
\]
\end{lemma}
\begin{proof}
	The inclusion on the left is usually known as "wrong way" and it holds in any cartesian bicategory. See for example~\cite{DBLP:journals/corr/abs-1711-08699} for a detailed proof.
	The inclusion on the right is the "negated" version holding in any cocartesian bicategory.
\end{proof}

\begin{lemma}\label{lm:non contr}
	The following hold: 
	\begin{enumerate}
		\item $a \sqcap \nega{a} \leq \bot$
		\item $\top \leq a \sqcup \nega{a}$
	\end{enumerate}
\end{lemma}
\begin{proof}
	We prove $(1)$. The proof for $(2)$ is analogous.
	\input{tikz/proofs/LNC.tex}
\end{proof}

\begin{proof}[Proof of Proposition \ref{prop:enrichment}]
The enrichments have been proved in  Lemma \ref{lm:seq ditr}.

The first six laws of Boolean algebras in Table~\ref{table:daggerrlalattice}.(d) are proved below:
\[
	\begin{array}{r@{}c@{}l}
		\nega{c \sqcap d} &\,\,\stackrel{\text{Def. }\nega{(\cdot)}}{=}\,\,& \op{(\rla{(c \sqcap d)})} \stackrel{\footnotesize{\text{Cor.}~\ref{cor:interactiondaggerrlalattice}}}{=} \op{(\rla{c})} \sqcup \op{(\rla{d})} \stackrel{\text{Def. }\nega{(\cdot)}}{=} \nega{c} \sqcup \nega{d},\\[8pt]
		\nega{\top} &\,\,\stackrel{\text{Def. }\nega{(\cdot)}}{=}\,\,& \op{(\rla{\top})} \stackrel{\footnotesize{\text{Cor.}~\ref{cor:interactiondaggerrlalattice}}}{=} \bot, \\[8pt]
		\nega{c \sqcup d} &\,\,\stackrel{\text{Def. }\nega{(\cdot)}}{=}\,\,& \op{(\rla{(c \sqcup d)})} \stackrel{\footnotesize{\text{Cor.}~\ref{cor:interactiondaggerrlalattice}}}{=} \op{(\rla{c})} \sqcap \op{(\rla{d})} \stackrel{\text{Def. }\nega{(\cdot)}}{=} \nega{c} \sqcap \nega{d},\\[8pt]
		\nega{\bot} &\,\,\stackrel{\text{Def. }\nega{(\cdot)}}{=}\,\,& \op{(\rla{\bot})} \stackrel{\footnotesize{\text{Cor.}~\ref{cor:interactiondaggerrlalattice}}}{=} \top, \\[8pt]
		a\sqcup (b \sqcap c) &\stackrel{\eqref{eq:def:cap}}{=}& \copier[-]  \seq[-]  ( a  \tensor[-]  (b \sqcap c)  )  \seq[-] \cocopier[-][] \\
							 &\stackrel{\text{Table }\ref{table:enrichment}.(e)}{=}& \copier[-][]  \seq[-]   ( (a \tensor[-] b) \sqcap (a  \tensor[-]  c)  )  \seq[-]  \cocopier[-][] \\
							 &\stackrel{\text{Table }\ref{table:enrichment}.(e)}{=}& ( \copier[-][]  \seq[-]  (a \tensor[-] b) \seq[-] \cocopier[-][]) \sqcap ( \copier[-][]  \seq[-]  (a \tensor[-] c)  \seq[-]  \cocopier[-][] ) \\
							 &\stackrel{\eqref{eq:def:cap}}{=}& (a \sqcup b) \sqcap (a \sqcup c), \\[8pt]
		a\sqcap (b \sqcup c) &\stackrel{\eqref{eq:def:cap}}{=}& \copier[+]  \seq[+]  ( a  \tensor[+]  (b \sqcup c)  )  \seq[+] \cocopier[+][] \\
							 &\stackrel{\text{Table }\ref{table:enrichment}.(e)}{=}& \copier[+][]  \seq[+]   ( (a \tensor[+] b) \sqcup (a  \tensor[+]  c)  )  \seq[+]  \cocopier[+][] \\
							 &\stackrel{\text{Table }\ref{table:enrichment}.(e)}{=}& ( \copier[+][]  \seq[+]  (a \tensor[+] b) \seq[+] \cocopier[+][]) \sqcup ( \copier[+][]  \seq[+]  (a \tensor[+] c)  \seq[+]  \cocopier[+][] ) \\
							 &\stackrel{\eqref{eq:def:cap}}{=}& (a \sqcap b) \sqcup (a \sqcap c)
	\end{array}
\]
The remaining two laws are proved in Lemma~\ref{lm:non contr}.
\end{proof}

It is worth emphasising that the following result stands at the core of our proofs. Once again, the diagrammatic approach proves to be an enhancement over the classical syntax. In this specific case we are looking at five (of many) different possibilities to express the ubiquitous concept of logical entailment. (1) expresses $a$ implies $b$ as a direct rewriting of the former into the latter. We have already seen that (2) corresponds to residuation. (3) corresponds to right residuation. (4) asserts the validity of the formula $\neg a \vee b$, thus it corresponds to the classical implication. Finally, (5) may look eccentric but it is actually a closed version of (3) that comes in handy if one has to consider closed diagrams.
\begin{proof}[Proof of Lemma \ref{lm:implications}]
    $(1)$ iff $(2)$ is Lemma~\ref{lm:residuation}. 
    
    $(1)$ iff $(3)$ is proved as follows: $a \leq b$ iff $\rla{b} \leq \rla{a}$ by the property of $\rla{(\cdot)}$ in Table~\ref{table:rlaproperties}.(b). By Lemma~\ref{lm:residuation}, $\rla{b} \leq \rla{a}$ iff $\id[+][Y] \leq \rla{a} \seq[-] \rla{(\rla{b})}$ where $\rla{(\rla{b})} = b$ by the property of $\rla{(\cdot)}$ in Table~\ref{table:rlaproperties}.(b).

    $(1)$ implies $(4)$ follows from the fact that every homset carries a Boolean algebra: $\nega{a} \sqcup b \stackrel{(1)}{\geq} \nega{a} \sqcup a \stackrel{\text{Table}~\ref{table:daggerrlalattice}.(d)}{=} \top$.

    $(4)$ implies $(1)$ is proved by the following derivation: \input{tikz/proofs/implication/notAorB.tex}

	$(1)$ iff $(5)$: observe that in any fo-bicategory $\boxCirc[+]{a} \leq \boxCirc[+]{b}$ iff $\cappedCirc[+]{a} \stackrel{(\ast_1)}{\leq} \cappedCirc[+]{b} \stackrel{(\ast_2)}{\leq} \cappedCirc[-]{b}$.
	
	Where $(\ast_1)$ holds in any cartesian bicategory and $(\ast_2)$ is proved below:
	\begin{align*}
    \cappedCirc[+]{b}
    &\Lleq{\footnotesize{\stackanchor{Lemma}{\ref{lm:id cup}}}}
    \begin{tikzpicture}
        \begin{pgfonlayer}{nodelayer}
            \node [style=none] (109) at (-1.25, 0.25) {};
            \node [style=none] (122) at (-1.25, -1.125) {};
            \node [style=none] (123) at (-1.25, 1.05) {};
            \node [style=none] (124) at (3.375, 1.05) {};
            \node [style=none] (125) at (3.375, -1.125) {};
            \node [{dotStyle/-}] (128) at (2.125, -0.125) {};
            \node [style=none] (129) at (1.625, 0.25) {};
            \node [style=none] (130) at (1.625, -0.5) {};
            \node [{dotStyle/-}] (131) at (2.75, -0.125) {};
            \node [style=none] (133) at (-1.25, -0.5) {};
            \node [{boxStyle/+}] (134) at (-0.5, 0.25) {$b$};
            \node [style=none] (135) at (0.25, -0.875) {};
            \node [style=none] (136) at (0.25, 0.8) {};
            \node [style=none] (137) at (3.125, 0.8) {};
            \node [style=none] (138) at (3.125, -0.875) {};
            \node [style=none] (139) at (0.25, 0.25) {};
            \node [style=none] (140) at (0.625, 0.25) {};
            \node [style=none] (141) at (0.625, 0) {};
            \node [style=none] (142) at (0.625, 0.5) {};
            \node [style=none] (143) at (1.625, 0.5) {};
            \node [style=none] (144) at (1.625, 0) {};
            \node [style=none] (145) at (0.25, -0.5) {};
        \end{pgfonlayer}
        \begin{pgfonlayer}{edgelayer}
            \draw [{bgStyle/+}] (124.center)
                 to (123.center)
                 to (122.center)
                 to (125.center)
                 to cycle;
            \draw [{bgStyle/-}] (137.center)
                 to (136.center)
                 to (135.center)
                 to (138.center)
                 to cycle;
            \draw [{bgStyle/+}] (144.center)
                 to (143.center)
                 to (142.center)
                 to (141.center)
                 to cycle;
            \draw [{wStyle/-}, bend right] (130.center) to (128);
            \draw [{wStyle/-}, bend right] (128) to (129.center);
            \draw [{wStyle/-}] (128) to (131);
            \draw [{wStyle/+}] (109.center) to (134);
            \draw [style={wStyle/+}] (134) to (139.center);
            \draw [style={wStyle/-}] (139.center) to (140.center);
            \draw [style={wStyle/+}] (140.center) to (129.center);
            \draw [style={wStyle/+}] (145.center) to (133.center);
            \draw [style={wStyle/-}] (130.center) to (145.center);
        \end{pgfonlayer}
    \end{tikzpicture}
    \Lleq{\eqref{ax:leftLinDistr}} 
    \begin{tikzpicture}
        \begin{pgfonlayer}{nodelayer}
            \node [style=none] (109) at (-1, 0.25) {};
            \node [style=none] (122) at (-1.25, -1.125) {};
            \node [style=none] (123) at (-1.25, 1.1) {};
            \node [style=none] (124) at (3.625, 1.1) {};
            \node [style=none] (125) at (3.625, -1.125) {};
            \node [{dotStyle/-}] (128) at (2.625, -0.125) {};
            \node [style=none] (129) at (2.125, 0.25) {};
            \node [style=none] (130) at (2.125, -0.5) {};
            \node [{dotStyle/-}] (131) at (3.25, -0.125) {};
            \node [style=none] (133) at (-1, -0.5) {};
            \node [{boxStyle/+}] (134) at (-0.25, 0.25) {$b$};
            \node [style=none] (135) at (-1, -0.875) {};
            \node [style=none] (136) at (-1, 0.85) {};
            \node [style=none] (137) at (2.125, 0.85) {};
            \node [style=none] (138) at (2.125, -0.875) {};
            \node [style=none] (139) at (0.5, 0.25) {};
            \node [style=none] (141) at (0.7, 0) {};
            \node [style=none] (142) at (0.7, 0.5) {};
            \node [style=none] (143) at (1.7, 0.5) {};
            \node [style=none] (144) at (1.7, 0) {};
            \node [style=none] (145) at (0.5, -0.5) {};
            \node [style=none] (146) at (0.5, -0.75) {};
            \node [style=none] (147) at (0.5, 0.725) {};
            \node [style=none] (148) at (1.875, 0.725) {};
            \node [style=none] (149) at (1.875, -0.75) {};
            \node [style=none] (150) at (1.875, 0.25) {};
            \node [style=none] (151) at (1.875, -0.5) {};
            \node [style=none] (152) at (1.7, 0.25) {};
            \node [style=none] (153) at (0.7, 0.25) {};
            \node [style=none] (154) at (-1.25, 0.25) {};
            \node [style=none] (155) at (-1.25, -0.5) {};
        \end{pgfonlayer}
        \begin{pgfonlayer}{edgelayer}
            \draw [{bgStyle/-}] (124.center)
                 to (123.center)
                 to (122.center)
                 to (125.center)
                 to cycle;
            \draw [{bgStyle/+}] (137.center)
                 to (136.center)
                 to (135.center)
                 to (138.center)
                 to cycle;
            \draw [{bgStyle/-}] (149.center)
                 to (148.center)
                 to (147.center)
                 to (146.center)
                 to cycle;
            \draw [{bgStyle/+}] (144.center)
                 to (143.center)
                 to (142.center)
                 to (141.center)
                 to cycle;
            \draw [{wStyle/-}, bend right] (130.center) to (128);
            \draw [{wStyle/-}, bend right] (128) to (129.center);
            \draw [{wStyle/+}] (109.center) to (134);
            \draw [style={wStyle/+}] (134) to (139.center);
            \draw [style={wStyle/+}] (145.center) to (133.center);
            \draw [style={wStyle/-}] (139.center) to (153.center);
            \draw [style={wStyle/-}] (152.center) to (150.center);
            \draw [style={wStyle/+}] (129.center) to (150.center);
            \draw [style={wStyle/+}] (153.center) to (152.center);
            \draw [style={wStyle/-}] (151.center) to (145.center);
            \draw [style={wStyle/-}] (154.center) to (109.center);
            \draw [style={wStyle/-}] (155.center) to (133.center);
            \draw [style={wStyle/-}] (128) to (131);
            \draw [style={wStyle/+}] (151.center) to (130.center);
        \end{pgfonlayer}
    \end{tikzpicture}
    \\
    &\Lleq{\eqref{ax:linStrn4}}
    \begin{tikzpicture}
        \begin{pgfonlayer}{nodelayer}
            \node [style=none] (109) at (-0.9, 0.5) {};
            \node [style=none] (122) at (-1.25, -1.25) {};
            \node [style=none] (123) at (-1.25, 1.35) {};
            \node [style=none] (124) at (3.15, 1.35) {};
            \node [style=none] (125) at (3.15, -1.25) {};
            \node [{dotStyle/-}] (128) at (2.15, -0.075) {};
            \node [style=none] (129) at (1.475, 0.5) {};
            \node [style=none] (130) at (1.475, -0.625) {};
            \node [{dotStyle/-}] (131) at (2.775, -0.075) {};
            \node [style=none] (133) at (-0.9, -0.625) {};
            \node [{boxStyle/+}] (134) at (-0.25, 0.5) {$b$};
            \node [style=none] (141) at (-0.9, -0.1) {};
            \node [style=none] (142) at (-0.9, 1.1) {};
            \node [style=none] (143) at (1.2, 1.1) {};
            \node [style=none] (144) at (1.2, -0.1) {};
            \node [style=none] (151) at (1.2, -0.625) {};
            \node [style=none] (152) at (1.2, 0.5) {};
            \node [style=none] (154) at (-1.25, 0.5) {};
            \node [style=none] (155) at (-1.25, -0.625) {};
            \node [style=none] (156) at (-0.9, -1) {};
            \node [style=none] (157) at (-0.9, -0.275) {};
            \node [style=none] (158) at (1.2, -0.275) {};
            \node [style=none] (159) at (1.2, -1) {};
            \node [style=none] (160) at (0.15, -0.85) {};
            \node [style=none] (161) at (0.15, -0.425) {};
            \node [style=none] (162) at (1, -0.425) {};
            \node [style=none] (163) at (1, -0.85) {};
            \node [style=none] (164) at (0.15, -0.625) {};
            \node [style=none] (165) at (1, -0.625) {};
        \end{pgfonlayer}
        \begin{pgfonlayer}{edgelayer}
            \draw [{bgStyle/-}] (124.center)
                 to (123.center)
                 to (122.center)
                 to (125.center)
                 to cycle;
            \draw [{bgStyle/+}] (144.center)
                 to (143.center)
                 to (142.center)
                 to (141.center)
                 to cycle;
            \draw [{bgStyle/+}] (159.center)
                 to (158.center)
                 to (157.center)
                 to (156.center)
                 to cycle;
            \draw [{bgStyle/-}] (163.center)
                 to (162.center)
                 to (161.center)
                 to (160.center)
                 to cycle;
            \draw [{wStyle/-}, bend right] (130.center) to (128);
            \draw [{wStyle/-}, bend right] (128) to (129.center);
            \draw [style={wStyle/-}] (154.center) to (109.center);
            \draw [style={wStyle/-}] (155.center) to (133.center);
            \draw [style={wStyle/-}] (128) to (131);
            \draw [style={wStyle/+}] (109.center) to (134);
            \draw [style={wStyle/+}] (134) to (152.center);
            \draw [style={wStyle/-}] (152.center) to (129.center);
            \draw [style={wStyle/-}] (151.center) to (130.center);
            \draw [style={wStyle/-}] (164.center) to (165.center);
            \draw [style={wStyle/+}] (165.center) to (151.center);
            \draw [style={wStyle/+}] (164.center) to (133.center);
        \end{pgfonlayer}
    \end{tikzpicture}                                           
    \structuralcong
    \cappedCirc[-]{b}     
\end{align*}
	Thus, we conclude from $(1)$ iff $(3)$ and $\rla{\scalebox{0.8}{\cappedCirc[+]{a}}} = \scalebox{0.8}{\cuppedCirc[-]{a}[b]}$.

\end{proof}

\subsection{Proofs of Section \ref{sec:freely}}

\begin{proof}[Proof of Proposition \ref{prop:soundness}]
Let $\interpretation =(X,\rho)$ be an interpretation of $\Sigma$. Recall that $\syninclusion$ is defined as $\pcong{\mathbb{FOB}}$. We prove by induction on the rules in \eqref{eq:pc}, that 
\begin{center}
if $c\syninclusion d$ then $\interpretationFunctor (c) \subseteq \interpretationFunctor (d)$.
\end{center}
By definition of $\seminclusion$, the above statement is equivalent to the proposition.

The proof for the rules $(r)$ and $(t)$ is trivial. For the rule ($\seq[][][]$), suppose that $c=c_1 \seq[] c_2$ and $d=d_1 \seq[] d_2$ with $c_1 \syninclusion d_1$ and $c_2 \syninclusion d_2$. Then
\begin{align*}
\interpretationFunctor (c) &= \interpretationFunctor (c_1 \seq[] c_2) \\
&= \interpretationFunctor (c_1) \seq[] \interpretationFunctor (c_2) \tag{\ref{fig:semantics}} \\
&\subseteq  \interpretationFunctor (d_1) \seq[] \interpretationFunctor (d_2) \tag{ind. hyp.} \\
&= \interpretationFunctor (d_1 \seq[] d_2) \tag{\ref{fig:semantics}} \\
&= \interpretationFunctor (d)  \\
\end{align*}
The proof for ($\tensor[][][]$) is analogous to the one above. The only interesting case is the rule $(id)$: we should prove that if $(c,d)\in \mathbb{FOB}$, then $\interpretationFunctor (c) \subseteq \interpretationFunctor (d)$. However, we have already done most of the work: since all the axioms in $\mathbb{FOB}$ -- with the only exception of the four stating $R^\bullet \Vdash R^\circ \Vdash R^\bullet$ (axioms \eqref{ax:tauRPlus}, \eqref{ax:gammaRPlus}, \eqref{ax:tauRMinus} and \eqref{ax:gammaRMinus}  in Figure \ref{fig:closed lin axioms})  -- are those of fo-bicategories and since $\Rel$ is a fo-bicategory, it only remains to show the soundness of those stating $R^\bullet \Vdash R^\circ \Vdash R^\bullet$. Note however that this is trivial by definition of $\interpretationFunctor (R^\bullet)$ as $\rla{\rho(R)}=\rla{(\interpretationFunctor (R^\circ))}$.
\end{proof}

In order to prove Proposition \ref{prop:LCBfo} is convenient to use the following function on diagrams and then prove that it maps every diagram in its  right (Lemma \ref{lemma:alphar}) and left (Lemma \ref{lemma:alphal}) linear adjoint.

\begin{definition}\label{defalpha}
The function $\alpha\colon \NPR \to \NPR$ is inductively defined as follows.
\[
	\begin{array}{llll}
		\alpha{(\id[+][0])}\defeq\id[-][0]
		&
		\alpha{(\id[+][1])}\defeq\id[-][1]
		&
		\alpha{(R^\circ}) \defeq R^\bullet
		&
		\alpha{(\symm[+][1][1])} \defeq \symm[-][1][1]
		\\
		\alpha{(\copier[+][1])}\defeq \cocopier[-][1]
		&
 		\alpha{(\discard[+][1])}\defeq \codiscard[-][1] 
 		&
		\alpha{(\cocopier[+][1])}\defeq \copier[-][1] 
		& 
		\alpha{(\codiscard[+][1])}\defeq \discard[-][1]
		\\
		\multicolumn{2}{l}{
			\alpha{(c \seq[+] d)} \defeq \alpha{(d)} \seq[-] \alpha{(c)}
		}
		&
		\multicolumn{2}{l}{
			\alpha{(c \tensor[+] d)} \defeq \alpha{(c)} \tensor[-] \alpha{(d)}
		}
		\\[1.2em]
		\alpha{(\id[-][0])}\defeq\id[+][0]
		&
		\alpha{(\id[-][1])}\defeq\id[+][1]
		&
		\alpha{(R^\bullet}) \defeq R^\circ
		&
		\alpha{(\symm[-][1][1])} \defeq \symm[+][1][1]
		\\
		\alpha{(\copier[-][1])}\defeq \cocopier[+][1]
		&
 		\alpha{(\discard[-][1])}\defeq \codiscard[+][1] 
 		&
		\alpha{(\cocopier[-][1])}\defeq \copier[+][1] 
		& 
		\alpha{(\codiscard[-][1])}\defeq \discard[+][1]
		\\
		\multicolumn{2}{l}{
			\alpha{(c \seq[-] d)} \defeq \alpha{(d)} \seq[+] \alpha{(c)}
		}
		&
		\multicolumn{2}{l}{
			\alpha{(c \tensor[-] d)} \defeq \alpha{(c)} \tensor[+] \alpha{(d)}
		}
	\end{array}
\]
\end{definition}

\begin{lemma}\label{lemma:alphar}
For all terms $c\colon n \to m$ in $\NPR$, $\id[+][n] \syninclusion c \seq[-]\alpha(c)$ and $\alpha(c) \seq[+] c \syninclusion \id[-][m]$.
\end{lemma}
\begin{proof}
The proof goes by induction on $c$.
For the base cases of black and white (co)monoid, it is immediate by the axioms in the first block of Figure \ref{fig:fo bicat axioms}.%
For $R^\circ$, $R^\bullet$, $\symm[+]$ and $\symm[-]$, it is immediate by the axioms in the bottom Figure \ref{fig:closed lin axioms}.
For $\id[+]$ and $\id[-]$ is trivial. For the inductive cases of $\seq[+]$, $\seq[-]$, $\tensor[+]$ and $\tensor[-]$ one can reuse exactly the proof of Proposition \ref{prop:rlamorphism}.
\end{proof}

\begin{lemma}\label{lemma:alphainv}
For all term $c\colon n \to m$ in $\NPR$, $\alpha(\alpha(c))=c$.
\end{lemma}
\begin{proof}
The proof goes by induction on $c$. For the base cases, it is immediate by Definition \ref{defalpha}. For the inductive cases, one have just to use the definition and the inductive hypothesis. For instance $\alpha(\alpha(a\seq[+] b))$ is, by Definition \ref{defalpha}, $\alpha (\alpha(a) \seq[-] \alpha(b))$ which, by Definition \ref{defalpha}, is $\alpha(\alpha(a)) \seq[+] \alpha(\alpha(b))$ that, by induction hypothesis, is $a \seq[+] b$.
\end{proof}

\begin{lemma}\label{lemma:alphaco}
For all terms $c,d\colon n \to m$ in $\NPR$, if $c \syninclusion d$, then $\alpha(d) \syninclusion \alpha (c)$.
\end{lemma}
\begin{proof}
Observe that the axioms in Figures \ref{fig:cb axioms}, \ref{fig:cocb axioms}, \ref{fig:closed lin axioms} and \ref{fig:fo bicat axioms} are closed under $\alpha$, namely if $c\leq d$ is an axiom also $\alpha(d) \leq \alpha(c)$ is an axiom.
\end{proof}

\begin{lemma}\label{lemma:alphal}
For all terms $c\colon n \to m$ in $\NPR$, $\id[+][m] \syninclusion \alpha(c) \seq[-] c$ and $c  \seq[+] \alpha(c) \syninclusion \id[-][n]$.
\end{lemma}
\begin{proof}
By Lemma \ref{lemma:alphar}, it holds that \[ \id[+][n] \syninclusion c \seq[-]\alpha(c) \text{ and } \alpha(c) \seq[+] c \syninclusion \id[-][m]\text{.}\]
By Lemma \ref{lemma:alphaco}, one can apply $\alpha$ to all the sides of the two inequalities to get
 \[   \alpha( c \seq[-]\alpha(c) ) \syninclusion \alpha( \id[+][n])  \text{ and }   \alpha( \id[-][m] ) \syninclusion \alpha( \alpha(c) \seq[+] c)  \text{.}\]
That, by Definition \ref{defalpha} gives exactly
 \[   \alpha(\alpha(c) )  \seq[+] \alpha( c)  \syninclusion  \id[-][n]  \text{ and }   \id[+][m] \syninclusion  \alpha (c) \seq[-]  \alpha( \alpha(c))  \text{.}\]
By Lemma \ref{lemma:alphainv}, one can conclude that
 \[   c  \seq[+] \alpha( c)  \syninclusion  \id[-][n]  \text{ and }   \id[+][m] \syninclusion  \alpha (c) \seq[-] c  \text{.}\]
\end{proof}

\begin{proof}[Proof of Proposition \ref{prop:LCBfo}]
By Lemmas \ref{lemma:alphar} and \ref{lemma:alphal}, the diagram $\alpha(c)$ is both the right and the left linear adjoint of any diagram $c$. Thus $\LCB$ is a closed linear bicategory.

Next, we show that $(\LCB^\circ, \copier[+], \cocopier[+])$ is a cartesian bicategory: for all objects $n\in \nat$, $\copier[+][n]$, $\discard[+][n]$, $\cocopier[+][n]$ and $\codiscard[+][n]$ are inductively defined as in Table \ref{fig:sugar}. Observe that such definitions guarantees that the coherence conditions in Definition \ref{def:cartesian bicategory}.(5) are satisfied. The conditions in Definition \ref{def:cartesian bicategory}.(1).(2).(3).(4) are the axioms in Figure \ref{fig:cb axioms} (and appear in the term version in Figure \ref{fig:textual axioms}) that we have used to generate $\syninclusion$. %

Similarly, $(\LCB^\bullet, \copier[-], \cocopier[-])$ is a cocartesian bicategory: for all objects $n\in \nat$, $\copier[-][n]$, $\discard[-][n]$, $\cocopier[-][n]$ and $\codiscard[-][n]$ are inductively defined as in Table \ref{fig:sugar}. Again,  the coherence conditions are satisfied by construction. The other conditions are  the axioms in Figure \ref{fig:cocb axioms} (and appear in the term version in Figure \ref{fig:textual axioms}) that, by construction, are in $\syninclusion$.
To conclude that $\LCB$ is a first order bicategory we have to check that the conditions in Definition \ref{def:fobicategory}.(4),(5). But these are exactly the axioms in Figure \ref{fig:fo bicat axioms} (and appear in the term version in Figure \ref{fig:textual axioms}).
\end{proof}

\begin{proof}[Proof of Proposition \ref{prop:free}]
Observe that the rules in \eqref{fig:semantics} defining  $\interpretationFunctor\colon \LCB \to \Rel$ also defines $\interpretationFunctor\colon \LCB \to \Cat{C}$ for an interpretation $\interpretation$ of $\sign$ in $\Cat{C}$ by fixing $\interpretationFunctor(R^\bullet)= \rla{(\interpretationFunctor(R^\circ))}$. To prove that $\interpretationFunctor$ preserve the ordering, one can use exactly the same proof of Proposition \ref{prop:soundness}. All the structure of (co)cartesian bicateries and linear bicategories is preserved by definition of $\interpretationFunctor$. %
Thus, $\interpretationFunctor\colon \LCB \to \Cat{C}$ is a morphism of fo-bicategories. By definition, it also holds that $\interpretationFunctor(1)= X$ and $\interpretationFunctor(R^\circ)= \rho(R)$.

To see that it is unique, observe that a morphism $\mathcal{F} \colon  \LCB \to \Cat{C}$ should map the object $0$ into $\unittensor$ (the unit object of $\tensor$) and any other natural number $n$ into $\mathcal{F}(1)^n$. Thus the only degree of freedom for the objects is the choice of where to map the natural number $1$. Similarly, for arrows, the only degree of freedom is where to map $R^\circ$ and $R^\bullet$. However, the axioms in $\mathbb{FOB}$ obliges $R^\bullet$ to be mapped into the right linear adjoint of $R^\circ$. Thus, by fixing $\mathcal{F}(1)=X$ and $\mathcal{F}(R^\circ)=\rho (R)$, $\mathcal{F}$ is forced to be $\interpretationFunctor$.
\end{proof}

\section{Proofs of Section \ref{sec:theories}}\label{app:theories}

\begin{proof}[Proof of Proposition \ref{prop:soundnessoftheories}]
By induction on \eqref{eq:pc}. For the rule $(id)$, we have two cases: either $(c,d) \in \syninclusion$ or $(c,d)\in \wtrel$. For $\syninclusion$, we conclude immediately by Proposition \ref{prop:soundness}. For  $(c,d)\in \wtrel$, the inclusion $\interpretationFunctor(c) \subseteq \interpretationFunctor(d)$ holds by definition of model. The proofs for the other rules are trivial. 
\end{proof}

\begin{lemma}\label{lemma:contraddictoryimpliestrivial}
Let $\T{T}$ be a theory. If $\T{T}$ is contradictory then it is trivial.
\end{lemma}
\begin{proof}
Assume $\T{T}$ to be contradictory and consider the following derivation.
\begin{align*}
\codiscard[+][1] & = \id[+][0] \seq[+] \codiscard[+][1]   \\
& \leq \id[-][0] \seq[+] \codiscard[+][1] \tag{$\T{T}$ contradictory} \\
& = \id[-][0] \seq[-] \codiscard[-][1] \tag{Proposition \ref{prop:maps}} \\
& = \codiscard[-][1] \\
\end{align*}
\end{proof}

\begin{proof}[Proof of Lemma \ref{lemma:trivalallequal}]
    \input{tikz/proofs/generalTrivial.tex}
The proof for $\discard[+][n+1] \seq[+] \codiscard[+][m] \syninclusionT{\T{T}} \, d \, \syninclusionT{\T{T}}  \discard[-][n+1] \seq[-] \codiscard[-][m]$ follows a similar reasoning.
\end{proof}

\subsection{Theories in $\FOL$ and $\NPR$}\label{app:closedtheories}

Once a first order alphabet is fixed, a theory in $\FOL$ is usually defined as a set $\mathcal{T}$ of \emph{closed} formulas that must be considered true. Intuitively, closed formulas corresponds in our language to diagrams $d$ of type $0\to 0$. Indeed the semantics $\interpretationFunctor$ assigns to such diagrams a relation $R\subseteq \singleton \times \singleton$: either $\{(\star, \star)\}$ (i.e., $\id[+][\singleton]$) representing true or $\varnothing$ (i.e., $\id[-][\singleton]$) representing false. The fact that $d$ must hold in any model is forced by requiring $(\id[+][0],d) \in \mathbb{I}$. This motivates the following definition.

\begin{definition}
A theory $\T{T} = (\sign, \T{I})$ is said to be \emph{closed} if all the pairs $(c,d) \in \T{I}$ are of the form $(\id[+][0], d)$. %
\end{definition}
For instance, the theory of sets and the theory of non-empty sets in Example \ref{ex:lin ord} are closed, while the third theory -- the one of order -- is not closed. By means of Lemma~\ref{lm:implications}, one can always translate an arbitrary theory $\T{T} = (\sign, \T{I})$ into a closed theory $\T{T}^c = (\sign, \T{I}^c)$ where
\[\T{I}^c \defeq \left\{ \left( \; \emptyCirc[+] \; , \scalebox{0.75}{\circleCirc{c}{d}[b]} \; \right) \mid (c,d) \in \T{I} \right\}\text{.}\]

\begin{proposition}\label{prop:closed}
Let $\T{T} = (\sign, \T{I})$ be a theory and $a,b\colon n \to m$ be diagrams in $\LCB$. Then $a \precongR{\T{T}} b$ iff $a \precongR{\T{T}^c} b$.
\end{proposition}
\begin{proof} %
    By induction on the rules in $\eqref{eq:pc}$. %
    The base case $(id)$ is given by means of Lemma~\ref{lm:implications} and in particular from the fact that:
    \[ \boxCirc[+]{a} \precongR{\T{T}} \boxCirc[+]{b} \;\text{ iff }\; \emptyCirc[+] \precongR{\T{T}} \scalebox{0.75}{\circleCirc{a}{b}[b]} \; \text{ for any $(a,b) \in \T{I}$.} \]

 The base case $(r)$ and the inductive cases are trivial.
\end{proof}

This result allows us to safely restrict our attention to closed theories, but this fact is not used in our proof of completeness. More interestingly, it tells us that while theories as introduced in \S \ref{sec:theories} appear to be rather different from the usual $\FOL$ theories, they can always be translated into closed theories which are essentially the same as the $\FOL$ ones. Indeed from a closed theory $\T{I}$, one can obtain the of set of closed formulas $\{d \mid (\id[+][0],d) \in \T{I} \}$ and, from a set of closed formulas $\mathcal{T}$ one can take $\T{I}$ as $\{(\id[+][0],d) \mid d\in \mathcal{T}\}$. %

The fact that a closed formula $d$ is derivable in $\mathcal{T}$, usually written as $\mathcal{T}\vdash d$, translates in $\NPR$ to $\id[+][0]  \;\precongR{\T{T}}\; d$.
In particular, when $d$ is an implication $c \Rightarrow b$, we have $\id[+][0] \;\precongR{\T{T}}\; b \seq[-] \rla{c}$ that, by Lemma \ref{lm:residuation}, is equivalent to $c  \;\precongR{\T{T}}\; b$. In $\FOL$ it is trivial -- by modus ponens -- that if  $\mathcal{T}\vdash c \Rightarrow b$ then $\mathcal{T}\cup \{c\} \vdash b$. In $\NPR$, this fact follows by transitivity of $\precongR{\T{T}}$: fix $\T{T}'=(\sign, \T{I} \cup \{(\id[+],c)\})$ and observe that $\id[+][0]  \;\precongR{\T{T}'}\; c \;\precongR{\T{T}'}\; b $.
The converse implication, namely  if $\mathcal{T}\cup \{c\} \vdash b$ then $\mathcal{T}\vdash c \Rightarrow b$, is known in $\FOL$ as deduction theorem. It can be generalised in $\NPR$ as in Theorem~\ref{th:deduction}.

\subsection{Deduction Theorem}

\begin{proof}[Proof of Theorem~\ref{th:deduction}]
The base cases are trivial. We show the case for $(\seq[+])$ in the main text. We show here the remaining inductive cases:
  \begin{enumerate}
      \item[$(t)$] Assume $a \;\precongR{\T{T'}}\; d$ and $d \;\precongR{\T{T'}}\; b$ for some $d \colon n \to m$. Observe that $a \;\precongR{\T{T'}}\; b$ by $(t)$ and $c \tensor[+] \id[+][n] \;\precongR{\T{T}}\; d \seq[-] \rla{a}$ and $c \tensor[+] \id[+][n] \;\precongR{\T{T}}\; b \seq[-] \rla{d}$  by inductive hypothesis. To conclude we need to show:
      \input{tikz/proofs/deductionTheorem/transitivity.tex}

      \item[$({\seq[-]})$] Assume $a_1 \;\precongR{\T{T'}}\; b_1$ and $a_2 \;\precongR{\T{T'}}\; b_2$ such that $a = a_1 \seq[-] a_2$ and $b = b_1 \seq[-] b_2$ for some $a_1, b_1 \colon n \to l, a_2, b_2 \colon l \to m$. Observe that $a_1 \seq[-] a_2 \;\precongR{\T{T'}}\; b_1 \seq[-] b_2$ by $(\seq[-])$ and $c \tensor[+] \id[+][n] \;\precongR{\T{T}}\; b_1 \seq[-] \rla{a_1}$ and $c \tensor[+] \id[+][n] \;\precongR{\T{T}}\; b_2 \seq[-] \rla{a_2}$  by inductive hypothesis. To conclude we need to show:
      \input{tikz/proofs/deductionTheorem/seqMinus.tex}

      \item[$({\tensor[+]})$] Assume $a_1 \;\precongR{\T{T'}}\; b_1$ and $a_2 \;\precongR{\T{T'}}\; b_2$ such that $a = a_1 \tensor[+] a_2$ and $b = b_1 \tensor[+] b_2$ for some $a_1,b_1 \colon n' \to m', a_2,b_2 \colon n'' \to m''$. Observe that $a_1 \tensor[+] a_2 \;\precongR{\T{T'}}\; b_1 \tensor[+] b_2$ by $(\tensor[+])$ and $c \tensor[+] \id[+][n] \;\precongR{\T{T}}\; b_1 \seq[-] \rla{a_1}$ and $c \tensor[+] \id[+][n] \;\precongR{\T{T}}\; b_2 \seq[-] \rla{a_2}$  by inductive hypothesis. To conclude we need to show:
      \input{tikz/proofs/deductionTheorem/tensorPlus.tex}

      \item[$({\tensor[-]})$] Assume $a_1 \precongR{\T{T'}} b_1$ and $a_2 \;\precongR{\T{T'}}\; b_2$ such that $a = a_1 \tensor[-] a_2$ and $b = b_1 \tensor[-] b_2$ for some $a_1,b_1 \colon n' \to m', a_2,b_2 \colon n'' \to m''$. Observe that $a_1 \tensor[-] a_2 \;\precongR{\T{T'}}\; b_1 \tensor[-] b_2$ by $(\tensor[-])$ and $c \tensor[+] \id[+][n] \;\precongR{\T{T}}\; b_1 \seq[-] \rla{a_1}$ and $c \tensor[+] \id[+][n] \;\precongR{\T{T}}\; b_2 \seq[-] \rla{a_2}$  by inductive hypothesis. To conclude we need to show:
      \input{tikz/proofs/deductionTheorem/tensorMinus.tex}
    \end{enumerate}
\end{proof}

\begin{proof}[Proof of Corollary \ref{cor:deduction}]
Suppose that $\T{T}'$ is contradictory, namely $\id[+][0] \syninclusionT{\T{T}'} \id[-][0]$. By the deduction theorem (Theorem \ref{th:deduction}), $\nega{c} \syninclusionT{\T{T}}  \id[-][0]$ and thus $\nega{\id[-][0]}  \syninclusionT{\T{T}}  \nega{\nega{c}}$, that is $\id[+][0] \syninclusionT{\T{T}} c$. 
The the other direction is trivial: since $\id[+][0] \syninclusionT{\T{T}'} c$ and $\id[+][0] \syninclusionT{\T{T}'} \nega{c}$, then $\id[+][0] \syninclusionT{\T{T}'} c \sqcap \nega{c} \syninclusionT{\T{T}'}  \bot = \id[-][0] $. 
\end{proof}

\subsection{Proofs of Section \ref{sec:funsem}}

\begin{proof}[Proof of Proposition \ref{prop:modelfactor}]
First, observe that a simple inductive argument allows to prove that, for all diagrams $c$ in $\LCB$, 
\begin{equation}\label{eq:functor1}
\mathcal{Q}_{\T{T}}^\sharp(c)=[c]_{\synequivalenceT{\T{T}}}\text{.}
\end{equation}

Now, suppose that there exists $\interpretationFunctor_{\T{T}} \colon \LCB[\T{T}] \to \Cat{C}$ making commutes the following diagram 
\[\scalebox{0.8}{\xymatrix{\LCB \ar[rd]|{\interpretationFunctor} \ar[r]^{\mathcal{Q}_{\T{T}}^\sharp}& \LCB[\T{T}] \ar@{.>}[d]^{\interpretationFunctor_{\T{T}}}\\
& \Cat{C}}}
\]
and consider $(c,d)\in \T{I}$. 
By definition, $c \syninclusionT{\T{T}} d$ and, by \eqref{eq:functor1}, 
\begin{equation}\label{eq:functor2}
\mathcal{Q}_{\T{T}}^\sharp (c) \syninclusionT{\T{T}} \mathcal{Q}_{\T{T}}^\sharp (d) \text{.}
\end{equation}
Then, the following derivation confirms that $\interpretation$ is a model of $\T{T}$ in $\Cat{C}$.
\begin{align*}
\interpretationFunctor(c) &= \interpretationFunctor_{\T{T}}( \mathcal{Q}_{\T{T}}^\sharp (c)) \tag{$\interpretationFunctor = \mathcal{Q}_{\T{T}}^\sharp ; \interpretationFunctor_{\T{T}}$} \\ 
&\leq \interpretationFunctor_{\T{T}}( \mathcal{Q}_{\T{T}}^\sharp (d)) \tag{\eqref{eq:functor2} and $ \interpretationFunctor_{\T{T}}$ is a morphism}\\
& = \interpretationFunctor(d) \tag{$\interpretationFunctor = \mathcal{Q}_{\T{T}}^\sharp ; \interpretationFunctor_{\T{T}}$} 
\end{align*}

Viceversa, suppose that $\interpretation$ is a model of $\T{T}$ in $\Cat{C}$. Then by definition of model, for all $(c,d)\in \T{I}$,  $\interpretationFunctor(c) \leq \interpretationFunctor(d)$. A simple inductive argument on the rules in \eqref{eq:pc} confirms that, for all diagrams $c,d$ in $\LCB$,
\begin{center}
if $c \syninclusionT{\T{T}} d$ then $\interpretationFunctor(c) \leq \interpretationFunctor(d)$.
\end{center}
In particular, if $c \synequivalenceT{\T{T}} d$ then $\interpretationFunctor(c) = \interpretationFunctor(d)$. Therefore, we are allowed to define $\interpretationFunctor_{\T{T}} ([c]_{\synequivalenceT{\T{T}}}) \defeq \interpretationFunctor (c)$ for all arrows $[c]_{\synequivalenceT{\T{T}}}$ of $\LCB[\T{T}]$ and $\interpretationFunctor_{\T{T}}(n)\defeq\interpretationFunctor(n)$ for all objects $n$ of $\LCB[\T{T}]$. The fact that $\interpretationFunctor_{\T{T}}$ preserves the ordering follows immediately from the above implication. The fact that   $\interpretationFunctor_{\T{T}}$ preserves the structure of fo-bicategories follows easily from the fact that $\interpretationFunctor$ is a morphism. Therefore $\interpretationFunctor_{\T{T}}$ is a morphism of fo-bicategories. The fact that the above diagram commutes is obvious by definition of $\interpretationFunctor_{\T{T}}$ and \eqref{eq:functor1}.

Uniqueness follows immediately from the fact that $\mathcal{Q}_{\T{T}}^\sharp \colon \LCB \to \LCB[\T{T}]$ is an epi, namely all objects and arrows of $\LCB[\T{T}]$ are in the image of $\mathcal{Q}_{\T{T}}^\sharp$.
\end{proof}

\begin{proof}[Proof of Corollary \ref{corollarymodelfunctor}]
To go from models to morphisms we use the assignment $\interpretation \mapsto \interpretationFunctor_{\T{T}}$ provided by Proposition \ref{prop:modelfactor}.

To transform morphisms into models, we need a slightly less straightforward assignment. Take a morphism of fo-bicategories $\mathcal{F}\colon \LCB[\T{T}] \to \Cat{C}$ and consider $\mathcal{Q}_{\T{T}}^\sharp ; \mathcal{F}\colon \LCB \to \Cat{C}$. This gives rise to the interpretation $\interpretation_\mathcal{F}$ defined as 
\begin{center} the domain $X$ is $\mathcal{Q}_{\T{T}}^\sharp ; \mathcal{F} (1)$ and $\rho(R)$ is $\mathcal{Q}_{\T{T}}^\sharp ; \mathcal{F}(R^\circ)$ for all $R \in \sign$.\end{center} 
By Proposition \ref{prop:free}, $\interpretation_\mathcal{F}^\sharp =\mathcal{Q}_{\T{T}}^\sharp ; \mathcal{F}$ and thus, by Proposition \ref{prop:modelfactor}, $\interpretation_\mathcal{F}$ is a model.

Since $\interpretation_\mathcal{F}^\sharp =\mathcal{Q}_{\T{T}}^\sharp ; \mathcal{F}$, by the uniqueness provided by Proposition \ref{prop:modelfactor}, $(\interpretation_{\mathcal{F}})_{\T{T}}^\sharp = \mathcal{F}$.

To conclude, we only need to prove that $\interpretation_{(\interpretationFunctor_{\T{T}}) } = \interpretation$. Since $\mathcal{Q}_{\T{T}}^\sharp  ; \interpretationFunctor_{\T{T}} = \interpretationFunctor$, then $\interpretation_{(\interpretationFunctor_{\T{T}}) } (R^\circ) = \mathcal{Q}_{\T{T}}^\sharp  ; \interpretationFunctor_{\T{T}} (R^\circ) = \interpretationFunctor(R^\circ) = \rho(R)$ for all $R\in \sign$. Similarly for the domain $X$.

\end{proof}

\begin{proof}[Proof of Lemma \ref{lemma:largertheories}]
By Proposition \ref{prop:modelfactor}, it is enough to give a model of $\T{T}$ in $\LCB[\T{T}']$. Define the interpretation $\interpretation$ having as domain $X$ the object $1$ of $\LCB[\T{T}']$ and $\rho(R) \defeq [R^\circ]_{\synequivalenceT{\T{T}'}}$ for each $R\in \sign$. A simple inductive arguments confirms that $\interpretationFunctor (c) = [c]_{\synequivalenceT{\T{T}'}}$ for all diagrams $c$ in $\LCB$. Since $\T{I} \subseteq \T{I'}$ is obvious that, for all $(c,d)\in \T{I}$, $\interpretationFunctor (c) \syninclusionT{\T{T}'} \interpretationFunctor (d)$. Thus $\interpretation$ is a model of  $\T{T}$ in $\LCB[\T{T}']$.
\end{proof}

\section{Proofs of Section \ref{sec:completeness}} \label{app:proofCompl}

\begin{proposition}\label{prop:nary maps}
    In any cartesian bicategory an $n$-ary map $\nterm{k} \colon 0 \to n$ can always be decomposed as:
    \[\nterm{k} = k_1 \tensor[+] k_2 \tensor[+] \hdots \tensor[+] k_n \quad \text{where each $k_i \colon 0 \to 1$ is a map.} \]
\end{proposition}
\begin{proof}
    Follows from Lemma~\ref{lemma:cb maps}.(4).
\end{proof}

\begin{lemma}\label{lm:hen dagger adj}
    For any $c \colon n \to m$ in $\LCB$ the following hold
    \[\HenFunctor(\op{c}) = \op{(\HenFunctor(c))}, \quad \HenFunctor(\rla{c}) = \rla{(\HenFunctor(c))}, \quad \HenFunctor(\nega{c}) = \nega{(\HenFunctor(c))}\]
\end{lemma}
\begin{proof}
    Since $\HenFunctor$ is a morphism of fo-bicategory the proof for $\op{(\cdot)}$ and $\rla{(\cdot)}$ follows from Lemma~\ref{lm:adjfunctor} and Lemma~\ref{lm:opfunctor}.

    Negation is preserved as well, since $\nega{(\cdot)} = \rla{(\op{\cdot})}$.

\end{proof}

\begin{proposition}\label{prop:non-contraddiction}\label{prop:non-trivialchain}
Let $I$ be a linearly ordered set and for all $i \in I$ let $\mathbb{T}_i=(\Sigma_i, \mathbb{I}_i)$ be first order theories such that if $i\leq j$, then $\Sigma_i \subseteq \Sigma_j$ and $\mathbb{I}_i \subseteq \mathbb{I}_j$. Let $\mathbb{T}$ be the theory $(\bigcup_{i\in I}\Sigma_i, \bigcup_{i\in I}\mathbb{I}_i)$. 
\begin{enumerate}
\item If all $\mathbb{T}_i$ are non-contradictory, then $\mathbb{T}$ is non-contradictory.
\item If all $\mathbb{T}_i$ are non-trivial, then $\mathbb{T}$ is non-trivial.
\end{enumerate}
\end{proposition}
\begin{proof}
By using the well-known fact that $\pcong{\cdot}$ preserves   chains, one can easily see that
\begin{equation}\label{eq:bla}
\syninclusionT{\mathbb{T}} = \bigcup_{i \in I} \syninclusionT{\mathbb{T}_i}
\end{equation}
The interested reader can find all the details in Appendix \ref{app:onprec}, Lemma \ref{lemma:noncontalltogether}.
\begin{enumerate}
\item Suppose that $\mathbb{T}$ is contradictory. By definition $\id[+][0] \syninclusionT{\mathbb{T}} \id[-][0]$ and then, by \eqref{eq:bla}, $(\id[+][0], \id[-][0]) \in \bigcup_{i \in I} \syninclusionT{\mathbb{T}_i}$. Thus there exists an $i\in I$ such that $\id[+][0] \syninclusionT{\mathbb{T}_i} \id[-][0]$. Against the hypothesis.
\item Suppose that $\mathbb{T}$ is trivial. By definition $\codiscard[+][1] \syninclusionT{\mathbb{T}} \codiscard[-][1]$ and then, by \eqref{eq:bla}, $(\codiscard[+][1], \codiscard[-][1]) \in \bigcup_{i \in I} \syninclusionT{\mathbb{T}_i}$. Thus there exists an $i\in I$ such that $\codiscard[+][1] \syninclusionT{\mathbb{T}_i} \codiscard[-][1]$. Against the hypothesis.
\end{enumerate}
\end{proof}

\begin{proposition}\label{prop:syntacticallycomplete}
Let $\mathbb{T}=(\Sigma, \mathbb{I})$ be a non-contradictory theory. There exists a theory $\mathbb{T'}=(\Sigma, \mathbb{I}')$ that is syntactically complete, non-contradictory and $\mathbb{I}\subseteq \mathbb{I}'$.
\end{proposition}
\begin{proof}[Proof of Proposition \ref{prop:syntacticallycomplete}]
The proof of this proposition relies on Zorn Lemma~\cite{zorn1935remark} which states that if, in a non empty poset poset $L$ every chain has a least upper bound, then $L$ has at least one maximal element.

We consider the set $\Gamma$ of all non-contradictory theories on $\Sigma$ that include $\mathbb{I}$, namely 
\[\Gamma \defeq \{\mathbb{T}=(\Sigma, \mathbb{J})  \mid \mathbb{T} \text{ is non-contradictory and }\mathbb{I}\subseteq \mathbb{J}\}\text{.}\]
Observe that the set $\Gamma$ is non empty since there is at least $\mathbb{T}$ which belongs to $\Gamma$.

Let $\Lambda \subseteq \Gamma$ be a chain, namely $\Lambda = \{\mathbb{T}_i=(\Sigma, \mathbb{J}_i) \in \Gamma \mid i \in I\}$ for some linearly ordered set $I$ and if $i\leq j$, then $\mathbb{J}_i \subseteq \mathbb{J}_j$. By Proposition \ref{prop:non-contraddiction}, the theory $(\Sigma, \bigcup_{i\in I}\mathbb{J}_i)$ is non-contradictory and thus it belongs to $\Gamma$.

We can thus use Zorn Lemma: the set $\Gamma$ has a maximal element $\mathbb{T'}=(\Sigma, \mathbb{I}')$. By definition of $\Gamma$, $\mathbb{I}\subseteq\mathbb{I'}$ and, moreover,  $\mathbb{T'}$ is non-contradictory.

We only need to prove that $\mathbb{T'}$ is syntactically complete, i.e., for all $c \colon 0 \to 0$, either $\id[+][0] \precongR{\T{T'}} c$ or $\id[+][0] \precongR{\T{T'}} \overline{c}$.
Assume that $\id[+][0] \not \hspace{-0.2cm}\precongR{\T{T}} c$. Thus $\mathbb{I}'$ is \emph{strictly} included into $\mathbb{I}' \cup \{(\id[+][0], c) \}$. By maximality of $\mathbb{T'}$ in $\Gamma$, we have that the theory $\mathbb{T''}=(\Sigma, \mathbb{I}' \cup \{(\id[+][0], c) \})$ is contradictory, i.e., $\id[+][0]\precongR{\T{T}''} \id[-][0]$. By the deduction theorem (Theorem \ref{th:deduction}), $c \precongR{\T{T'}} \id[-][0]$. Therefore $\id[+][0] \precongR{\T{T'}} \nega{c}$.
\end{proof}

\subsection{Proofs for Lemma \ref{lemma:addingHenkin} and Theorem \ref{thm:nontrivialALL}}

In order to prove Lemma \ref{lemma:addingHenkin} and then Theorem \ref{thm:nontrivialALL}, we
need to showing that \emph{adding} constants to a non-trivial theory results in a non-trivial theory. To do this, it is useful to have a procedure for \emph{erasing} constants. This is defined as follows.

\begin{definition}\label{def:phi}
    Let $\sign$ be a signature and $\sign' = \sign \;\cup\; \{ k \colon 0 \to 1 \}$. The function $\phi \colon \LCB[\sign'][n,m] \to \LCB[\sign][1+n,m]$ is inductively defined as follows:
    \[
    \begin{array}{r@{\;}c@{\;}l@{\qquad\quad}r@{\;}c@{\;}l}
        \phi(k^\circ) &\defeq& \idCirc[+] & \phi(k^\bullet) &\defeq& \idCirc[-] \\[10pt]
        \phi(g^\circ) &\defeq& \phiCirc[+]{g} & \phi(g^\bullet) &\defeq& \phiCirc[-]{g^\bullet}  \\[10pt]
        \phi(c \seq[+] d) &\defeq& \phiSeqCirc[+]{c}{d} & \phi(c \seq[-] d) &\defeq& \phiSeqCirc[-]{c}{d} \\[10pt]
        \phi(c \tensor[+] d) &\defeq& \phiTensorCirc[+]{c}{d} & \phi(c \tensor[-] d) &\defeq& \phiTensorCirc[-]{c}{d}
    \end{array}
    \]
where $g^\circ \in \{\copier[+][1], \discard[+][1], R^\circ, \codiscard[+][1], \cocopier[+][1], \id[+][0], \id[+][1], \symm[+][1][1] \}$ and $g^\bullet \in \{\copier[-][1], \discard[-][1], R^\bullet, \codiscard[-][1], \cocopier[-][1], \id[-][0], \id[-][1], \symm[-][1][1] \}$. 
\end{definition}

\begin{lemma}\label{lm:phi characterization}
    Let $c \colon n \to m$ be a diagram of $\LCB$, then $\phi(c) = \phiCirc[+]{c}$.
\end{lemma}
\begin{proof}
    \input{tikz/proofs/phiCharacterization.tex}
\end{proof}

\begin{lemma}[Constant Erasion]\label{lemma:phi}
    Let $\T{T} = (\sign, \T{I})$ be a theory and $\T{T'} = (\sign', \T{I'})$ be the theory where $\sign' = \sign \cup \{ k \colon 0 \to 1 \}$ and $\T{I'} = \T{I} \cup \TMAP{k}$.
    Then, for any $c,d \colon n \to m$ in $\LCB[\sign ']$ if $c \syninclusionT{\T{T'}} d$ then $\phi(c) \syninclusionT{\T{T}} \phi(d)$.
\end{lemma}
\begin{proof}
    The proof goes by induction on the rules in~\eqref{eq:pc}.

For the rule $(id)$ we have three cases: either $(c,d) \in \T{I}$ or $(c,d) \in \syninclusionT{\sign'}$ or $(c,d) \in \TMAP{k}$. 

If $(c,d) \in \T{I}$ then, by Lemma~\ref{lm:phi characterization}, $\phi(c) = \phiCirc[+]{c} \syninclusionT{\T{T}} \phiCirc[+]{d} = \phi(d)$.

If $(c,d) \in \syninclusionT{\sign'}$ then $(c,d)$ has been obtained by instantiating the axioms in Figures~\ref{fig:cb axioms},\ref{fig:cocb axioms} and \ref{fig:closed lin axioms} with diagrams containing $k$. Therefore, we need to show that $\phi$ preserves these axioms. In the following we show only $\eqref{ax:comPlusLaxNat}, \eqref{ax:discPlusLaxNat}, \eqref{ax:tauRPlus}, \eqref{ax:gammaRPlus}, \eqref{ax:leftLinDistr}$ and $\eqref{ax:linStrn2}$. The remaining ones follow similar reasonings.
\input{tikz/proofs/phiLemma/copier.tex}

\input{tikz/proofs/phiLemma/discard.tex}

\input{tikz/proofs/phiLemma/linAdjRel.tex}

\input{tikz/proofs/phiLemma/linDistr.tex}

\input{tikz/proofs/phiLemma/linStr.tex}

Similar to the previous argument, if $(c,d) \in \TMAP{k}$ then it is enough to show that $\phi$ preserves the axioms in $\TMAP{k}$.

\input{tikz/proofs/phiLemma/copierK.tex}

\input{tikz/proofs/phiLemma/discardK.tex}

The base case $(r)$ is trivial, while the proof for the remaining rules follows a straightforward inductive argument.
\end{proof}

\begin{proof}[Proof of Lemma \ref{lemma:addingHenkin}]

We prove that if $\T{T'}$ is trivial, then also $\T{T}$ is trivial. Let $\T{T''} = \{ \Sigma \cup k, \mathbb{I} \cup \TMAP{k}\}$ and assume $\T{T'}$ to be trivial, i.e.\ $\scalebox{0.8}{\codiscardCirc[+]} \!\!\syninclusionT{\T{T'}}\!\! \scalebox{0.8}{\codiscardCirc[-]}$, then:
\begin{enumerate}
    \item by the Deduction Theorem (\ref{th:deduction}) we have $\scalebox{0.8}{
    \InputIfFileExists{proofs/nonTrivialCompleteness/newAxiom.tikz}{}{\input{tikz/proofs/nonTrivialCompleteness/newAxiom.tikz}}
} \precongR{\T{T''}} \scalebox{0.8}{
    \InputIfFileExists{axioms/cb/minus/codiscDisc.tikz}{}{\input{tikz/axioms/cb/minus/codiscDisc.tikz}}
}$;
    \item thus, by Lemma \ref{lemma:phi}, $\phi( \scalebox{0.8}{
    \InputIfFileExists{proofs/nonTrivialCompleteness/newAxiom.tikz}{}{\input{tikz/proofs/nonTrivialCompleteness/newAxiom.tikz}}
} ) \,\syninclusionT{\T{T}}\, \phi( \scalebox{0.8}{
    \InputIfFileExists{axioms/cb/minus/codiscDisc.tikz}{}{\input{tikz/axioms/cb/minus/codiscDisc.tikz}}
} )$;
    \item and, by Def.~\ref{def:phi} and Lemma \ref{lm:phi characterization},
$\scalebox{0.8}{
    \InputIfFileExists{proofs/nonTrivialCompleteness/phiNewAxiom.tikz}{}{\input{tikz/proofs/nonTrivialCompleteness/phiNewAxiom.tikz}}
} \!\syninclusionT{\T{T}}\! \scalebox{0.8}{
    \InputIfFileExists{proofs/nonTrivialCompleteness/phiBone.tikz}{}{\input{tikz/proofs/nonTrivialCompleteness/phiBone.tikz}}
}$.
\end{enumerate}
To conclude, apply Lemma~\ref{lm:implications} %
and observe:
\input{tikz/proofs/nonTrivialCompleteness/mainProof.tex}

which, by Lemma~\ref{lm:implications} again, is exactly that
$\scalebox{0.8}{\codiscardCirc[+]} \precongR{\T{T}} \scalebox{0.8}{\codiscardCirc[-]}$. Namely $\T{T}$ is trivial.

Note that in the step $(\ast)$ above we used the following derivation which holds for any $c \colon 0 \to 1$:
\input{tikz/proofs/nonTrivialCompleteness/boneDerivation.tex}
\end{proof}

\begin{proof}[Proof of Theorem \ref{thm:nontrivialALL}]
    This proof reuses the well-known arguments reported e.g. in \cite{lascar2001mathematical}.
    
    We first illustrate a procedure to add Henkin witnesses without losing the property of being non-trivial.
    
    Take an enumeration of diagrams in $\LCB[\Sigma] [1,0]$ and write $c_i$ for the $i$-th diagram.
    
    For all natural numbers $n \in \nat$, we define 
    \[
        \begin{array}{c@{\qquad}c}
            \Sigma^n \defeq \Sigma \cup \{k_i\colon 0 \to 1 \mid i \leq n\}  &  \mathbb{I}^n \defeq \mathbb{I} \cup \TMAP{k_i} \cup \bigcup_{i \leq n} \mathbb{W}_{k_i}^{c_i} \\
            \multicolumn{2}{c}{\T{T}^n \defeq (\Sigma^n, \mathbb{I}^n)}
        \end{array}
    \]
    By applying Lemma \ref{lemma:addingHenkin} $n$-times, one has that $\T{T}^n$ is non-trivial. %
    Define now
    \[ \Sigma_0 \defeq \bigcup_{i\in \nat}\Sigma^i  \qquad  \mathbb{I}_0 \defeq \bigcup_{i \in \nat} \mathbb{I}^j  \qquad \T{T}_0 \defeq (\Sigma_0, \mathbb{I}_0)\]
    
    Since $\T{T}^0 \subseteq \T{T}^1 \subseteq \dots \subseteq \T{T}^n \subseteq \dots$ are all non-trivial, then by Proposition \ref{prop:non-trivialchain}.2, we have that $\T{T}_0$ is non-trivial. One must not jump to the conclusion that $\T{T}_0$ has Henkin witnesses: all the diagrams in $\LCB[\Sigma] [1,0]$ have Henkin witnesses, but in $\T{T}_0$ we have more diagrams, since we have added the constants $k_i$ to $\Sigma_0$.
    
    We thus repeat the above construction, but now for diagrams in  $\LCB[\Sigma_0] [1,0]$. We define
    \[
        \begin{array}{c@{\qquad}c}
            \Sigma_1 \defeq \Sigma_0 \cup \{k_c \mid c \in \LCB[\Sigma_0] [1,0]\} &  \mathbb{I}_1 \defeq \mathbb{I}_0 \cup \TMAP{{k_c}} \cup \mathbb{W}_{k_c}^{c} \\
            \multicolumn{2}{c}{\T{T}_1 \defeq (\Sigma_1, \mathbb{I}_1)}
        \end{array}
    \]
    The theory $\T{T}_1$ is non-trivial but has Henkin witnesses only for the diagrams in $\LCB[\Sigma_0]$.

    Thus, for all natural numbers $n \in \nat$, we define 
    \[
        \begin{array}{c@{\qquad}c}
            \Sigma_{n+1} \defeq \Sigma_{n} \cup \{k_c \mid c \in \LCB[\Sigma_n] [1,0]\} &  \mathbb{I}_{n+1} \defeq \mathbb{I}_{n} \cup \TMAP{{k_c}} \cup \mathbb{W}_{k_c}^{c} \\
            \multicolumn{2}{c}{\T{T}_{n+1} \defeq (\Sigma_{n+1}, \mathbb{I}_{n+1})}
        \end{array}
    \]
    and
    \[ \Sigma' \defeq \bigcup_{i\in \nat}\Sigma_i  \qquad  \mathbb{I}' \defeq \bigcup_{i \in \nat} \mathbb{I}_i  \qquad \T{T}' \defeq (\Sigma ', \mathbb{I}')\]
    Since $\T{T}_0 \subseteq \T{T}_1 \subseteq \dots \subseteq \T{T}_n \subseteq \dots$ are all non-trivial, then by Proposition \ref{prop:non-trivialchain}.2, we have that $\T{T}'$ is also non-trivial. Now $\T{T}'$ has Henkin witnesses: if $c\in \LCB[\Sigma'] [0,1]$, then there exists $n \in \nat$ such that $c\in \LCB[\Sigma_n] [0,1]$. By definition of $\mathbb{I}_n$, it holds that $\mathbb{W}_{k_c}^{c} \subseteq \mathbb{I}_{n+1}$ and thus $\mathbb{W}_{k_c}^{c} \subseteq \mathbb{I}'$. 
    
    Summarising, we manage to build a theory $\T{T}'=(\Sigma',\mathbb{I}')$ that has Henkin witnesses and it is non-trivial. By Lemma \ref{lemma:contraddictoryimpliestrivial}, $\T{T}'$ is non-contradictory. We can thus use Proposition \ref{prop:syntacticallycomplete}, to obtain a theory $\T{T}''=(\Sigma',\mathbb{I}'')$ that is syntactically complete and non-contradictory. Observe that $\T{T}''$ has Henkin witnesses, since the signature $\Sigma'$ is the same as in $\T{T}'$ and $\mathbb{I}'\subseteq \mathbb{I}''$.
    \end{proof}

\subsection{Proofs for Proposition~\ref{prop:henkin}}\label{app:HenkinModel}
Proposition~\ref{prop:henkin} is the second key to prove G\"odel completeness. Before illustrating its proof, we need an additional lemma.

\begin{lemma}\label{lemma:henkinn}
    Let $\T{T}$ be a theory with Henkin witnesses. For all $c \colon n \to 0$ there is a map $\nterm{k} \colon n \to 1$ s.t.
    $
    \InputIfFileExists{henkinWit1.tikz}{}{\input{tikz/henkinWit1.tikz}}
 \syninclusionT{\T{T}} 
    \InputIfFileExists{henkinWitn.tikz}{}{\input{tikz/henkinWitn.tikz}}
$.
\end{lemma}
\begin{proof}
The proof goes by induction on $n$. For $n=0$, take $\id[+][0]$ as $\nterm{k}$. For $n+1$, we have the following:

\input{tikz/proofs/henkinn.tex}

\end{proof}

\begin{proof}[Proof of Proposition \ref{prop:henkin}] The proof goes by induction on $c$.
    The white base cases are easy, we show three representative cases below. %

    \input{tikz/proofs/henkinProp/reducedBaseCases.tex}

    For the inductive case $c \seq[+] d$ we prove the two inclusions separately. 
    Suppose $c \colon n \to o$ and $d \colon o \to m$, then

    \[
    \allowdisplaybreaks
    \begin{array}{r@{}c@{\!\!\!\!\!\!\!}l}
        &\HenFunctor(\scalebox{0.8}{\seqCirc[+]{c}{d}})& \\[5pt]
                                                &\Leq{\text{Def. }\HenFunctor}&  \HenFunctor(\scalebox{0.8}{\boxCirc[+]{c}}) \seq[+] \HenFunctor(\scalebox{0.8}{\boxCirc[+]{d}}) \\[8pt]
                                                &\Leq{\text{Ind. hyp.}}&   \begin{aligned}
                                                    &\{ (\nterm{k},\nterm{t}) \in X^n\times X^o \mid \scalebox{0.8}{\emptyCirc[+]} \precongR{\T{T}} \scalebox{0.8}{\closedFormulaCirc{\nterm{k}}{c}{\nterm{t}}} \}
                                                    \\\seq[+] &\{ (\nterm{t},\nterm{l}) \in X^o\times X^m \mid \scalebox{0.8}{\emptyCirc[+]} \precongR{\T{T}} \scalebox{0.8}{\closedFormulaCirc{\nterm{t}}{d}{\nterm{l}}} \}
                                                \end{aligned} \\[12pt]
                                                &\Leq{\eqref{eq:seqRel}}& \begin{aligned}
                                                    \{ (\nterm{k},\nterm{l}) \in X^n\times X^m \mid \exists \nterm{t} \; &\scalebox{0.8}{\emptyCirc[+]} \precongR{\T{T}} \scalebox{0.8}{\closedFormulaCirc{\nterm{k}}{c}{\nterm{t}}} \\
                                                    \wedge &\scalebox{0.8}{\emptyCirc[+]} \precongR{\T{T}} \scalebox{0.8}{\closedFormulaCirc{\nterm{t}}{d}{\nterm{l}}} \}
                                                \end{aligned} \\[-5pt]
                                                &\Leq{\footnotesize{\stackanchor{\eqref{ax:comPlusLaxNat}}{\eqref{ax:discPlusLaxNat}}}}&                 \{ (\nterm{k},\nterm{l}) \in X^n\times X^m \mid \exists \nterm{t} \; \scalebox{0.8}{\emptyCirc[+]} \precongR{\T{T}} \scalebox{0.8}{
    \InputIfFileExists{henkinTensor.tikz}{}{\input{tikz/henkinTensor.tikz}}
} \} \\[10pt]
                                                &\structuralcong&                           \{ (\nterm{k},\nterm{l}) \in X^n\times X^m \mid \exists \nterm{t} \; \scalebox{0.8}{\emptyCirc[+]} \precongR{\T{T}} \scalebox{0.8}{
    \InputIfFileExists{henkinSeq.tikz}{}{\input{tikz/henkinSeq.tikz}}
} \} \\
                                                &\stackrel{\text{Prop. }~\ref{prop:map adj}}{\subseteq}&                \{ (\nterm{k},\nterm{l}) \in X^n\times X^m \mid \scalebox{0.8}{\emptyCirc[+]} \precongR{\T{T}} \scalebox{0.8}{
    \InputIfFileExists{henkinSeq2.tikz}{}{\input{tikz/henkinSeq2.tikz}}
} \}
    \end{array}
\]

For the other inclusion the following holds:

\[
    \allowdisplaybreaks
    \begin{array}{r@{\!\!\!\!\!\!\!\!\!\!\!\!\!\!\!\!\!\!\!\!\!\!\!\!\!\!\!\!\!\!\!\!\!\!\!\!}c@{\!\!\!\!\!\!\!\!\!\!\!\!\!\!\!\!\!\!\!\!\!\!\!\!\!\!\!\!\!\!\!\!\!\!\!\!\!\!\!\!\!\!\!\!\!\!\!\!\!\!\!\!\!\!\!\!\!\!}l}
        &\qquad\qquad\qquad\{ (\nterm{k},\nterm{l}) \in X^n\times X^m \mid \scalebox{0.8}{\emptyCirc[+]} \precongR{\T{T}} \scalebox{0.8}{
    \InputIfFileExists{henkinSeq2.tikz}{}{\input{tikz/henkinSeq2.tikz}}
} \}& 
        \\[5pt]
        &\stackrel{\eqref{eqdagger}}{=}& \{ (\nterm{k},\nterm{l}) \in X^n\times X^m \mid \scalebox{0.8}{\emptyCirc[+]}
        \precongR{\T{T}}
        \;\scalebox{0.8}{\begin{tikzpicture}
            \begin{pgfonlayer}{nodelayer}
                \node [style=none] (122) at (2.75, 1.35) {};
                \node [style=none] (123) at (2.75, -1.4) {};
                \node [style=none] (124) at (-2.2, -1.4) {};
                \node [style=none] (125) at (-2.2, 1.35) {};
                \node [{boxStyle/+}] (127) at (0.675, -0.65) {$d$};
                \node [{funcOpStyle/+}] (133) at (1.95, -0.65) {$\nterm{l}$};
                \node [{boxOpStyle/+}] (134) at (0.675, 0.6) {$c$};
                \node [{funcOpStyle/+}] (135) at (1.95, 0.6) {$\nterm{k}$};
                \node [style={dotStyle/+}] (136) at (-0.25, 0) {};
                \node [style={dotStyle/+}] (137) at (-1.25, 0) {};
            \end{pgfonlayer}
            \begin{pgfonlayer}{edgelayer}
                \draw [{bgStyle/+}] (124.center)
                     to (123.center)
                     to (122.center)
                     to (125.center)
                     to cycle;
                \draw [style={wStyle/+}] (127) to (133);
                \draw [style={wStyle/+}] (134) to (135);
                \draw [style={wStyle/+}] (136) to (137);
                \draw [style={wStyle/+}, bend left] (136) to (134);
                \draw [style={wStyle/+}, bend right] (136) to (127);
            \end{pgfonlayer}
        \end{tikzpicture}    
        } \} \\[12pt]
        &\stackrel{\text{Lemma }\ref{lemma:henkinn}}{\subseteq}& \{ (\nterm{k},\nterm{l}) \in X^n \times X^m \mid \exists \nterm{t} \; \scalebox{0.8}{\emptyCirc[+]}
        \precongR{\T{T}} \scalebox{0.8}{\begin{tikzpicture}
            \begin{pgfonlayer}{nodelayer}
                \node [style=none] (122) at (2.75, 1.35) {};
                \node [style=none] (123) at (2.75, -1.4) {};
                \node [style=none] (124) at (-2.2, -1.4) {};
                \node [style=none] (125) at (-2.2, 1.35) {};
                \node [{boxStyle/+}] (127) at (0.675, -0.65) {$d$};
                \node [{funcOpStyle/+}] (133) at (1.95, -0.65) {$\nterm{l}$};
                \node [{boxOpStyle/+}] (134) at (0.675, 0.6) {$c$};
                \node [{funcOpStyle/+}] (135) at (1.95, 0.6) {$\nterm{k}$};
                \node [style={dotStyle/+}] (136) at (-0.25, 0) {};
                \node [style={funcStyle/+}] (137) at (-1.25, 0) {$\nterm{t}$};
            \end{pgfonlayer}
            \begin{pgfonlayer}{edgelayer}
                \draw [{bgStyle/+}] (124.center)
                     to (123.center)
                     to (122.center)
                     to (125.center)
                     to cycle;
                \draw [style={wStyle/+}] (127) to (133);
                \draw [style={wStyle/+}] (134) to (135);
                \draw [style={wStyle/+}] (136) to (137);
                \draw [style={wStyle/+}, bend left] (136) to (134);
                \draw [style={wStyle/+}, bend right] (136) to (127);
            \end{pgfonlayer}
        \end{tikzpicture}    
        } \} \\[12pt]
        &\stackrel{\eqref{TMAPk}}{=}& \{ (\nterm{k},\nterm{l}) \in X^n \times X^m \mid  \exists \nterm{t} \;  \scalebox{0.8}{\emptyCirc[+]}
        \precongR{\T{T}} \scalebox{0.8}{
            \begin{tikzpicture}
                \begin{pgfonlayer}{nodelayer}
                    \node [{boxOpStyle/+}] (107) at (-0.025, 0.65) {$c$};
                    \node [{funcStyle/+}] (117) at (-1.25, 0.65) {$\nterm{t}$};
                    \node [style=none] (122) at (2, 1.4) {};
                    \node [style=none] (123) at (2, -1.4) {};
                    \node [style=none] (124) at (-1.95, -1.4) {};
                    \node [style=none] (125) at (-1.95, 1.4) {};
                    \node [{funcOpStyle/+}] (126) at (1.25, 0.65) {$\nterm{k}$};
                    \node [{boxStyle/+}] (127) at (0, -0.65) {$d$};
                    \node [{funcStyle/+}] (128) at (-1.225, -0.65) {$\nterm{t}$};
                    \node [{funcOpStyle/+}] (133) at (1.275, -0.65) {$\nterm{l}$};
                \end{pgfonlayer}
                \begin{pgfonlayer}{edgelayer}
                    \draw [{bgStyle/+}] (124.center)
                         to (123.center)
                         to (122.center)
                         to (125.center)
                         to cycle;
                    \draw [style={wStyle/+}] (107) to (126);
                    \draw [style={wStyle/+}] (117) to (107);
                    \draw [style={wStyle/+}] (127) to (133);
                    \draw [style={wStyle/+}] (128) to (127);
                \end{pgfonlayer}
            \end{tikzpicture}                    
        } \} \\
        &\Leq{\footnotesize{\stackanchor{\eqref{ax:comPlusLaxNat}}{\eqref{ax:discPlusLaxNat}}}}&  
        \begin{aligned}
            \{ (\nterm{k},\nterm{l}) \in X^n \times X^m \mid \exists \nterm{t} \; &\scalebox{0.8}{\emptyCirc[+]} \precongR{\T{T}} \scalebox{0.8}{\closedOpFormulaCirc{\nterm{k}}{c}{\nterm{t}}} \\
            \wedge &\scalebox{0.8}{\emptyCirc[+]} \precongR{\T{T}} \scalebox{0.8}{\closedFormulaCirc{\nterm{t}}{d}{\nterm{l}}} \}
        \end{aligned} \\[15pt]
        &\Leq{\eqref{eq:seqRel}}& \begin{aligned}
            &\{ (\nterm{k},\nterm{t}) \in X^n \times X^o \mid \scalebox{0.8}{\emptyCirc[+]} \precongR{\T{T}} \scalebox{0.8}{\closedOpFormulaCirc{\nterm{k}}{c}{\nterm{t}}} \} \\
            \seq[+] &\{ (\nterm{t},\nterm{l}) \in X^o \times X^m \mid \scalebox{0.8}{\emptyCirc[+]} \precongR{\T{T}} \scalebox{0.8}{\closedFormulaCirc{\nterm{t}}{d}{\nterm{l}}} \}
        \end{aligned} \\[15pt]
        &\stackrel{\text{Table }\ref{table:daggerproperties}.(a)}{=}&  \begin{aligned}
            &\{ (\nterm{k},\nterm{t}) \in X^n \times X^o \mid \scalebox{0.8}{\emptyCirc[+]} \precongR{\T{T}} \scalebox{0.8}{\closedFormulaCirc{\nterm{k}}{c}{\nterm{t}}} \} \\
            \seq[+] &\{ (\nterm{t},\nterm{l}) \in X^o \times X^m \mid \scalebox{0.8}{\emptyCirc[+]} \precongR{\T{T}} \scalebox{0.8}{\closedFormulaCirc{\nterm{t}}{d}{\nterm{l}}} \}
        \end{aligned} \\[15pt]
        &\Leq{\text{Ind. hyp.}}&     \HenFunctor(\scalebox{0.8}{\boxCirc[+]{c}}) \seq[+] \HenFunctor(\scalebox{0.8}{\boxCirc[+]{d}}) \\
        &\Leq{\text{Def. }\HenFunctor}& \HenFunctor(\scalebox{0.8}{\seqCirc[+]{c}{d}})
    \end{array}
\]

    The inductive case $c \tensor[+] d$ is proved as follows:

    Suppose $c \colon n \to m$ and $d \colon o \to p$, then

{\allowdisplaybreaks
\[
    \begin{array}{r@{\;}c@{\;}l}
        &\HenFunctor(\scalebox{0.8}{\tensorCirc[+]{c}{d}})& \\[20pt]
        &\Leq{\text{Def. }\HenFunctor}&  \HenFunctor(\scalebox{0.8}{\boxCirc[+]{c}}) \tensor[+] \HenFunctor(\scalebox{0.8}{\boxCirc[+]{d}}) \\[15pt]
        &\Leq{\text{Ind. hyp.}}& \begin{aligned}
            &\{ (\nterm{k_1},\nterm{l_1} \in X^n \times X^m \mid \scalebox{0.8}{\emptyCirc[+]} \precongR{\T{T}} \scalebox{0.8}{\closedFormulaCirc{\nterm{k_1}}{c}{\nterm{l_1}}} \} \\
            \tensor[+] &\{ (\nterm{k_2},\nterm{l_2}) \in X^o \times X^p \mid \scalebox{0.8}{\emptyCirc[+]} \precongR{\T{T}} \scalebox{0.8}{\closedFormulaCirc{\nterm{k_2}}{d}{\nterm{l_2}}} \}
        \end{aligned} \\[20pt]
        &\Leq{\eqref{eq:tensorREL}}&    \begin{aligned}
            \{ ( \Bigl( \begin{smallmatrix} \nterm{k_1} \\ \nterm{k_2} \end{smallmatrix} \Bigr), \Bigl( \begin{smallmatrix} \nterm{l_1} \\ \nterm{l_2} \end{smallmatrix} \Bigr) ) \in X^{n+o} \times X^{m+p} \mid &\scalebox{0.8}{\emptyCirc[+]} \precongR{\T{T}} \scalebox{0.8}{\closedFormulaCirc{\nterm{k_1}}{c}{\nterm{l_1}}} \\
            \wedge &\scalebox{0.8}{\emptyCirc[+]} \precongR{\T{T}} \scalebox{0.8}{\closedFormulaCirc{\nterm{k_2}}{d}{\nterm{l_2}}} \}
        \end{aligned} \\
        &\Leq{\footnotesize{\stackanchor{\eqref{ax:comPlusLaxNat}}{\eqref{ax:discPlusLaxNat}}}}& \{ ( \Bigl( \begin{smallmatrix} \nterm{k_1} \\ \nterm{k_2} \end{smallmatrix} \Bigr), \Bigl( \begin{smallmatrix} \nterm{l_1} \\ \nterm{l_2} \end{smallmatrix} \Bigr) ) \in X^{n+o} \times X^{m+p} \mid \scalebox{0.8}{\emptyCirc[+]} \precongR{\T{{T}}} \scalebox{0.8}{\closedDecomposedTensorCirc{\nterm{k_1}}{\nterm{k_2}}{c}{d}{\nterm{l_1}}{\nterm{l_2}}} \} 
        \\[20pt]
        &\Leq{\text{Prop. }~\ref{prop:nary maps}}& \begin{aligned}
            \{ (\nterm{k}, \nterm{l}) \in X^{n+o} \times X^{m+p} \mid &\scalebox{0.8}{\emptyCirc[+]} \precongR{\T{T}} \scalebox{0.8}{\closedTensorCirc{k}{c}{d}{l}} \, , \\
            &\qquad\nterm{k} = \nterm{k_1} \tensor[+] \nterm{k_2}, \nterm{l} = \nterm{l_1} \tensor[+] \nterm{l_2} \}
        \end{aligned}      
    \end{array}
\]
}

    The inductive case $c \seq[-] d$ is proved as follows:

    Suppose $c \colon n \to o$ and $d \colon o \to m$, then

    \input{tikz/proofs/henkinProp/seqMinus.tex}

    The proof above relies on Lemma~\ref{lm:hen dagger adj} and the previous inductive case of $c \seq[+] d$. The case of $c \tensor[-] d$ follows the exact same reasoning but, as expected, this time one has to exploit the proof of $c \tensor[+] d$.
\end{proof}

\subsection{Proofs from G\"odel completeness to Theorem \ref{thm:completeness}}
After having proved \eqref{thm:Godel}, we show how to obtain a proof for Theorem \ref{thm:completeness}. 
The main step is to prove \eqref{thm:trivialcom}.

\begin{lemma}\label{lemma:Henkinboolean}
Let $\T{T}=(\Sigma,\mathbb{I})$ be a theory that is trivial and non-contradictory and let $\booleaninterpretation$ be the Henkin interpretation of $\Sigma$.
Then, the domain $X$ of $\booleaninterpretation$ is $\varnothing$ and $\rho(R) = \{(\star,\star)\}$ if $\id[+][0] \precongR{\T{T}} R^\circ$ and $\varnothing$ otherwise.
\end{lemma}
\begin{proof}%
Recall by Definition \ref{def:henkin struct}, that the domain $X$ of $\booleaninterpretation$ is defined as the set $\texttt{Map}(\LCB[\T{T}])[0,1]$. This set should be necessarily empty since, if there exists some map $k\colon 0 \to 1$, then by \eqref{eq:remcounter}, $\T{T}$ would be contradictory, against the hypothesis. Thus $\texttt{Map}(\LCB[\T{T}])[0,1]= \varnothing$. By Proposition \ref{prop:nary maps}, one has also that $\texttt{Map}(\LCB[\T{T}])[0,n+1]= \varnothing$. We thus have only one map in $\LCB[\T{T}]$, that is 
 $\id[+][0] \colon 0 \to 0$ (depicted as $\scalebox{0.8}{\emptyCirc[+]}$). 

Recall that by Definition \ref{def:henkin struct},  $\rho(R) = \{ (\nterm{k},\nterm{l}) \in X^n \times X^m \mid \scalebox{0.8}{\emptyCirc[+]} \syninclusionT{\T{T}} \scalebox{0.8}{\closedFormulaCirc{\nterm{k}}{R}{\nterm{l}}} \}$  for all $R\in \sign$.  Since our only map is  $\id[+][0] \colon 0 \to 0$, we have that 
$\rho(R) = \{ (\star,\star) \in \singleton \times \singleton \mid \id[+][0] \syninclusionT{\T{T}} R^\circ \}$.
\end{proof}

\begin{lemma}\label{lemma:interpretation}
Let $\T{T}=(\Sigma,\mathbb{I})$ be a theory and let $c\colon n \to m+1$ and $d\colon n+1 \to m$ be arrows of $\LCB[\T{T}]$. Thus $\booleaninterpretationFunctor (c) = \varnothing$ and $\booleaninterpretationFunctor (d) = \varnothing$.
\end{lemma}
\begin{proof}
Recall that for any interpretation $\interpretation$, $\interpretationFunctor(c)   \subseteq X^{n} \times X^{m+1} = X^n \times X^m \times X$. For $\booleaninterpretation$,  $X=\varnothing$ by Lemma \ref{lemma:Henkinboolean} and thus $\booleaninterpretationFunctor (c) \subseteq  \varnothing \times \varnothing^n \times \varnothing^m$, i.e., $\booleaninterpretationFunctor (c) =  \varnothing $. The proof for $\booleaninterpretationFunctor (d)$ is identical.
\end{proof}

\begin{lemma}\label{lemma:biglemmaboolean}
Let $\T{T}$ be a trivial theory that is syntactically complete. Let $c \colon 0 \to 0$ be an arrow of $\LCB[\T{T}]$. If $\booleaninterpretationFunctor (c) = \{(\star, \star)\}$ then $c =_{\T{T}} \id[+][0]$.
\end{lemma}
\begin{proof}
We proceed by induction on $c$.

For the base cases, there are only four constants $c\colon 0 \to 0$.
\begin{itemize}
\item $c=\id[+][0]$. Then, it is trivial.
\item $c=\id[-][0]$. Then $\booleaninterpretationFunctor (c) = \varnothing$ against the hypothesis.
\item  $c=R^\circ$. If $\booleaninterpretationFunctor (R^\circ) = \{(\star, \star)\}$, then by definition of $\booleaninterpretation$, $\id[+][0] =_{\T{T}} R^\circ$.
\item  $c=R^\bullet$. If $\booleaninterpretationFunctor (R^\circ) = \{(\star, \star)\}$, then by definition of $\booleaninterpretationFunctor$,  $\{(\star,\star)\}\notin \rho(R)$. Thus, by definition of $\booleaninterpretation$, $\id[+][0] \not \hspace{-0.2cm} \precongR{\T{T}} R^\circ$. Since $\T{T}$ is syntactically complete $\id[+][0] \precongR{\T{T}} R^\bullet$.
\end{itemize}

We now consider the usual four inductive cases.
\begin{itemize}
\item $c=c_1 \tensor[+] c_2$. Since $c\colon 0 \to 0$, then also $c_1$ and $c_2$ have type $0 \to 0$. By definition, $\booleaninterpretationFunctor(c) = \booleaninterpretationFunctor(c_1) \tensor[+] \booleaninterpretationFunctor(c_2)$. By definition of $\tensor[+]$ in $\Rel$ both $\booleaninterpretationFunctor(c_1)$ and $\booleaninterpretationFunctor(c_2)$ must be $\{(\star, \star)\}$. We can thus apply the inductive hypothesis to deduce that $c_1=_{\T{T}}\id[+][0]$ and  $c_2=_{\T{T}}\id[+][0]$. Therefore $c=c_1 \tensor[+] c_2=_{\T{T}} \id[+][0] \tensor[+] \id[+][0] =_{\T{T}} \id[+][0]$.

\item $c=c_1 \seq[+]c_2$. There are two possible cases: either $c_1\colon 0 \to n+1$ and $c_2\colon n+1 \to 0$, or  $c_1\colon 0 \to 0$ and $c_2\colon 0 \to 0$. In the former case, we have by Lemma \ref{lemma:interpretation}, that $\booleaninterpretationFunctor (c)= \booleaninterpretationFunctor (c_1) \seq[+] \booleaninterpretationFunctor (c_2)=\varnothing \seq[+] \varnothing = \varnothing$. Against the hypothesis. Thus the second case should hold: $c_1\colon 0 \to 0$ and $c_2\colon 0 \to 0$. In this case we just observe that $c_1 \seq[+] c_2$ is, by the laws of symmetric monoidal categories, equal to $c_1 \tensor[+] c_2$. We can thus reuse the proof of the point above.

\item $c=c_1 \tensor[-] c_2$. Since $c\colon 0 \to 0$, then also $c_1$ and $c_2$ have type $0 \to 0$.  Consider the case where $\booleaninterpretationFunctor (c_1)= \varnothing  = \booleaninterpretationFunctor (c_2)$. Thus  $\booleaninterpretationFunctor (c)= \varnothing$, against the hypothesis. Therefore either $\booleaninterpretationFunctor (c_1)= \{ (\star,\star)\}$ or  $\booleaninterpretationFunctor (c_2)= \{ (\star,\star)\}$.  If $\booleaninterpretationFunctor (c_1)= \{ (\star,\star)\}$, then by induction hypothesis $c_1=_{\T{T}} \id[+][0]$. Therefore $c=c_1 \tensor[-]c_2 = c_1 \sqcup c_2 =_{\T{T}} \id[+][0] \sqcup c_2 =_{\T{T}} \top \sqcup c_2 =_{\T{T}} \top =_{\T{T}} \id[+][0]$. The case for $\booleaninterpretationFunctor (c_2)= \{ (\star,\star)\}$ is symmetric.

\item $c=c_1 \seq[-] c_2$. There are two possible cases: either $c_1\colon 0 \to n+1$ and $c_2\colon n+1 \to 0$, or  $c_1\colon 0 \to 0$ and $c_2\colon 0 \to 0$. In the former case, we have by Lemma \ref{lemma:trivalallequal} that $c_1 =_{\T{T}} \codiscard[-][n+1]$ and $c_2=_{\T{T}} \discard[+][n+1]$. Thus $c=_{\T{T}}\codiscard[-][n+1] \seq[-] \discard[+][n+1] =_{\T{T}}\id[+][0]$. For the last equivalence observe that $\id[+][0] \syninclusionT{\T{T}} \codiscard[-][n+1] \seq[-] \discard[+][n+1]$ since $\rla{(\codiscard[-][n+1])} = \discard[+][n+1]$. The other inclusion is $\codiscard[-][n+1] \seq[-] \discard[+][n+1] \synequivalenceT{\T{T}} (\codiscard[-][n+1] \seq[-] \discard[+][n+1]) \seq[+] \id[+][0] \stackrel{\text{Def. }\discard[+]}{\synequivalenceT{\T{T}}} (\codiscard[-][n+1] \seq[-] \discard[+][n+1]) \seq[+] \discard[+][0] \stackrel{\eqref{ax:discPlusLaxNat}}{\synequivalenceT{\T{T}}} \discard[+][0] \stackrel{\text{Def. }\discard[+]}{\synequivalenceT{\T{T}}} \id[+][0]$. Consider now the case where $c_1\colon 0 \to 0$ and $c_2\colon 0 \to 0$. In this case $c_1 \seq[-] c_2$  is, by the laws of symmetric monoidal categories, equal to $c_1 \tensor[-] c_2$. We can thus reuse the proof of the point above.
\end{itemize}
\end{proof}

\begin{lemma}\label{lemma:biglemmaboolean2}
Let $\T{T}$ be a trivial theory that is syntactically complete. Let $c \colon 0 \to 0$ be an arrow of $\LCB[\T{T}]$. If $\booleaninterpretationFunctor (c) = \varnothing$ then $c =_{\T{T}} \id[-][0]$.
\end{lemma}
\begin{proof}
If $\booleaninterpretationFunctor (c) = \varnothing$, then by Lemma \ref{lm:hen dagger adj}, $\booleaninterpretationFunctor (\nega{c}) = \nega{\varnothing} =\{(\star, \star)\}$. Thus by Lemma \ref{lemma:biglemmaboolean}, $\nega{c}=_{\T{T}} \id[+][0]$ and thus $c=_{\T{T}} \id[-][0]$.
\end{proof}

\begin{proposition}\label{prop:booleanisamodel}
\!\!if $\T{T}$ is trivial, syntactically complete and non-contradictory, then $\booleaninterpretation$ is a model. Namely, for all $c,d \colon n \to m$ in $\LCB$, if $c\precongR{\T{T}} d$, then $\booleaninterpretationFunctor (c) \subseteq \booleaninterpretationFunctor (d)$.
\end{proposition}
\begin{proof}
Recall that by definition $\booleaninterpretation$ is a model iff for all $c,d \colon n \to m$ in $\LCB$, if $c\precongR{\T{T}} d$, then $\booleaninterpretationFunctor (c) \subseteq \booleaninterpretationFunctor (d)$.
We prove that if $\booleaninterpretationFunctor (c) \not \subseteq \booleaninterpretationFunctor (d)$, then $c\not \hspace{-0.2cm} \precongR{\T{T}} d$.

If $c\colon n \to m+1$ or $c \colon n+1 \to m$, then by Lemma \ref{lemma:interpretation}, $\booleaninterpretationFunctor (c)= \varnothing$ and thus it is not the case that $\booleaninterpretationFunctor (c) \not \subseteq \booleaninterpretationFunctor (d)$. Thus we need to consider only the case where $c,d\colon 0 \to 0$.

For $c,d \colon 0 \to 0$ if $\booleaninterpretationFunctor (c) \not \subseteq \booleaninterpretationFunctor (d)$, then $\booleaninterpretationFunctor (c)=\{(\star,\star)\}$ and $\booleaninterpretationFunctor (d) = \varnothing$. By Lemmas \ref{lemma:biglemmaboolean} and \ref{lemma:biglemmaboolean2}, we thus have that $c=_{\T{T}} \id[+][0]$ and $d=_{\T{T}} \id[-][0]$. Since $\T{T}$ is non-contradictory, then $c\not \hspace{-0.2cm} \precongR{\T{T}} d$.
\end{proof}

\begin{proof}[Proof of \eqref{thm:trivialcom}]
Since $\T{T}=(\Sigma, \mathbb{I})$ is non-contradictory, by Proposition \ref{prop:syntacticallycomplete} there exists a syntactically complete non-contradictory theory $\T{T'}=(\Sigma, \mathbb{I'})$ such that $\mathbb{I}\subseteq \mathbb{I'}$. Since $\codiscard[+][1] \precongR{\T{T}} \codiscard[-][1]$, then $\codiscard[+][1] \precongR{\T{T'}} \codiscard[-][1]$, $\T{T'}$ is also trivial. We can thus use Proposition \ref{prop:booleanisamodel}, to deduce that $\T{T'}$ has a model. Since  $\mathbb{I}\subseteq \mathbb{I'}$, then by Lemma~\ref{lemma:largertheories}, also $\T{T}$ has a model.
\end{proof}

\begin{proposition}\label{prop:entailment}
\eqref{cor:gencompleteness} entails Theorem \ref{thm:completeness}.
\end{proposition}
\begin{proof}
Assuming that \eqref{cor:gencompleteness} holds, one can prove that, for all theories $\T{T}=(\sign, \T{I})$ and diagrams $c\colon 0 \to 0$ in $\LCB$,
\begin{equation}\label{lemma:beforened}
\text{if, for all models }\interpretation\text{ of }\T{T}\text{, }\{(\star,\star)\} \subseteq \interpretationFunctor(c)\text{ then }\id[+][0] \syninclusionT{\T{T}} c\text{.}
\end{equation}
Suppose indeed that $\id[+][0] \not  \syninclusionT{\T{T}} c$. Then, by Corollary \ref{cor:deduction},  $\T{T}'=(\sign, \T{I}\cup \{(\id[+][0], \nega{c} ) \} )$ is non-contradictory. Thus, by  \eqref{cor:gencompleteness}, $\T{T}'$ has a model, namely, there exists a morphism of fo-bicategories $\mathcal{G} \colon \LCB[\T{T}'] \to \Rel$. By Lemma \ref{lemma:largertheories}, we have a morphism $\mathcal{F}\colon \LCB[\T{T}] \to \LCB[\T{T}']$ and thus we have a model $\mathcal{F};\mathcal{G} \colon \LCB[\T{T}] \to \Rel$. Observe that since $\mathcal{G}$ is a model of $\T{T}'$, then $\mathcal{G}(\nega{[c]_{\synequivalenceT{\T{T}'}}}) = \{(\star,\star)\}$ and, by construction of $\mathcal{F}$,  $\mathcal{F};\mathcal{G}(\nega{[c]_{\synequivalenceT{\T{T}}}})= \{(\star,\star)\}$. By Lemmas \ref{lm:opfunctor} and \ref{lm:adjfunctor},  $\mathcal{F};\mathcal{G}([c]_{\synequivalenceT{\T{T}}}) = \varnothing$. Thus, there is a model assigning $\varnothing$ to $c$, against the hypothesis of \eqref{lemma:beforened}.

By \eqref{lemma:beforened} and Lemma \ref{lm:implications} one can easily conclude Theorem \ref{thm:completeness}.

Consider a theory $\T{T}=(\sign, \emptyset)$ for some monoidal signature $\sign$. Let $c,d\colon n \to m$ be diagrams in $\LCB$. For any interpretation $\interpretation$, if $\interpretationFunctor(c) \subseteq \interpretationFunctor(d)$ then, $\Rel$ is a fo-bicategory and Lemma \ref{lm:implications}, it holds that
\[\{(\star,\star)\} \subseteq \interpretationFunctor ( \scalebox{0.8}{\emptyCirc[+]}) \subseteq \interpretationFunctor (\circleCirc{c}{d}[b])\text{.} \]
If, for all $\interpretation$,  $\interpretationFunctor(c) \subseteq \interpretationFunctor(d)$ then, by  \eqref{lemma:beforened}, $\emptyCirc[+] \syninclusionT{\T{T}} \circleCirc{c}{d}[b]$. Again, by Lemma \ref{lm:implications}, $c \syninclusionT{\T{T}} d$. Since $\T{T}=(\sign, \emptyset)$, $c \syninclusion d$.
\end{proof}

\subsection{Proofs for Corollary~\ref{cor:cr completeness}}\label{app:encoding}

\begin{proposition}\label{prop:cr interpretation}
    For all expressions $E$ of $\CRS$ and interpretations $\interpretation$, $\dsemRel{E} = \interpretationFunctor{(\enc{E})}$.
\end{proposition}
\begin{proof}%
    The proof is by induction on $E$. The base cases are trivial. The inductive cases are shown below.
    {\allowdisplaybreaks
    \begin{align*}
        \interpretationFunctor{(\enc{E_1 \seq[+] E_2})} & \stackrel{\text{Table }\ref{table:cr encoding}}{=} \interpretationFunctor{(\enc{E_1} \seq[+] \enc{E_2})}  \\
                                                        &\stackrel{\eqref{fig:semantics}}{=} \interpretationFunctor{(\enc{E_1})} \seq[+] \interpretationFunctor{(\enc{E_2})} \\
                                                        &\stackrel{\text{Ind. hyp.}}{=} \dsemRel{E_1} \seq[+] \dsemRel{E_2}  \\
                                                        &\stackrel{\eqref{eq:sematicsExpr}}{=} \dsemRel{E_1 \seq[+] E_2} 
                                                        \\[5pt]
        \interpretationFunctor{(\enc{E_1 \seq[-] E_2})} & \stackrel{\text{Table }\ref{table:cr encoding}}{=} \interpretationFunctor{(\enc{E_1} \seq[-] \enc{E_2})}  \\
                                                        &\stackrel{\eqref{fig:semantics}}{=} \interpretationFunctor{(\enc{E_1})} \seq[-] \interpretationFunctor{(\enc{E_2})} \\
                                                        &\stackrel{\text{Ind. hyp.}}{=} \dsemRel{E_1} \seq[-] \dsemRel{E_2}  \\
                                                        &\stackrel{\eqref{eq:sematicsExpr}}{=} \dsemRel{E_1 \seq[-] E_2} \\[5pt]
        \interpretationFunctor{(\enc{E_1 \cap E_2})} & \stackrel{\text{Table }\ref{table:cr encoding}}{=} \interpretationFunctor{(\copier[+][1] \! \seq[+] (\enc{E_1} \tensor[+] \enc{E_2}) \seq[+] \! \cocopier[+][1])} \\
                                                     &\stackrel{\eqref{fig:semantics}}{=} \interpretationFunctor{(\copier[+][1])} \seq[+] (\interpretationFunctor{(\enc{E_1})} \tensor[+] \interpretationFunctor{(\enc{E_2})}) \seq[+] \interpretationFunctor{(\cocopier[+][1])} \\
                                                     & \stackrel{\eqref{fig:semantics}}{=} \copier[+][X] \seq[+] (\interpretationFunctor{(\enc{E_1})} \tensor[+] \interpretationFunctor{(\enc{E_2})}) \seq[+] \cocopier[+][X] \\
                                                     &\stackrel{\text{Ind. hyp.}}{=} \copier[+][X] \seq[+] (\dsemRel{E_1} \tensor[+] \dsemRel{E_2}) \seq[+] \cocopier[+][X] \\
                                                     & \stackrel{\eqref{eq:def:cap}}{=} \dsemRel{E_1} \cap \dsemRel{E_2} \\
                                                     &\stackrel{\eqref{eq:sematicsExpr}}{=}  \dsemRel{E_1 \cap E_2} \\[5pt]
        \interpretationFunctor{(\enc{E_1 \cup E_2})} & \stackrel{\text{Table }\ref{table:cr encoding}}{=} \interpretationFunctor{(\copier[-][1] \! \seq[-] (\enc{E_1} \tensor[-] \enc{E_2}) \seq[-] \! \cocopier[-][1])} \\
                                                     &\stackrel{\eqref{fig:semantics}}{=} \interpretationFunctor{(\copier[-][1])} \seq[-] (\interpretationFunctor{(\enc{E_1})} \tensor[-] \interpretationFunctor{(\enc{E_2})}) \seq[-] \interpretationFunctor{(\cocopier[-][1])} \\
                                                     & \stackrel{\eqref{fig:semantics}}{=} \copier[-][X] \seq[-] (\interpretationFunctor{(\enc{E_1})} \tensor[-] \interpretationFunctor{(\enc{E_2})}) \seq[-] \cocopier[-][X] \\
                                                     &\stackrel{\text{Ind. hyp.}}{=} \copier[-][X] \seq[-] (\dsemRel{E_1} \tensor[-] \dsemRel{E_2}) \seq[-] \cocopier[-][X] \\
                                                     & \stackrel{\eqref{eq:def:cup}}{=} \dsemRel{E_1} \cup \dsemRel{E_2} \\
                                                     &\stackrel{\eqref{eq:sematicsExpr}}{=}  \dsemRel{E_1 \cup E_2} \\[5pt]                                                     
        \interpretationFunctor{(\enc{\op{E}})}       & \stackrel{\text{Table }\ref{table:cr encoding}}{=} \interpretationFunctor{(\op{(\enc{E})})} \\
                                                     &\stackrel{\text{Lemma }\ref{lm:opfunctor}}{=} \op{(\interpretationFunctor{(\enc{E})})} \\
                                                     &\stackrel{\text{Ind. hyp.}}{=} \op{\dsemRel{E}} \\
                                                     &\stackrel{\eqref{eq:sematicsExpr}}{=} \dsemRel{\op{E}} \\[5pt]
        \interpretationFunctor{(\enc{\nega{E}})}     & \stackrel{\text{Table }\ref{table:cr encoding}}{=} \interpretationFunctor{(\nega{(\enc{E})})} \\
                                                     &\stackrel{\text{Def. }\nega{(\cdot)}}{=}  \interpretationFunctor{(\op{(\rla{(\enc{E})})})} \\
                                                     &\stackrel{\text{Lemmas }\ref{lm:opfunctor}, \ref{lm:adjfunctor}}{=} \op{(\rla{\interpretationFunctor{(\enc{E})}})} \\
                                                     &\stackrel{\text{Ind. hyp.}}{=} \op{(\rla{(\dsemRel{E})})} \\
                                                     & \stackrel{\text{Def. }\nega{(\cdot)}}{=} \nega{\dsemRel{E}}\\
                                                     &\stackrel{\eqref{eq:sematicsExpr}}{=} \dsemRel{\nega{E}}
    \end{align*}}
\end{proof}

\begin{proof}[Proof of Corollary~\ref{cor:cr completeness}]
    \begin{align*}
        E_1 \minorExpression E_2 &\iff \forall \interpretation . \; \dsemRel{E_1} \subseteq \dsemRel{E_2} \tag{Def. of $\minorExpression$} \\
                                 &\iff \forall \interpretation . \; \interpretationFunctor{(\enc{E_1})} \subseteq \interpretationFunctor{(\enc{E_2})} \tag{Prop. \ref{prop:cr interpretation}} \\
                                 &\iff \enc{E_1} \seminclusion \enc{E_2} \tag{Def. of $\seminclusion$} \\
                                 &\iff \enc{E_1} \syninclusion \enc{E_2} \tag{Theorem \ref{thm:completeness}}
    \end{align*}
\end{proof}

\section{Some well known facts about chains in a lattice}\label{app:continuous} 
A \emph{chain} on a complete lattice $(L, \sqsubseteq)$ is a family $\{x_i\}_{i\in I}$ of elements of $L$ indexed by a linearly oredered set $I$ such that $x_i \sqsubseteq x_j$ whenever $i \leq j$.
A monotone map $f\colon L \to L$ is said to \emph{preserve chains} if
\[f(\bigsqcup_{i\in I}x_i) = \bigsqcup_{i \in I}f(x_i)\]
We write $id \colon L \to L$ for the identity function and $f \sqcup g\colon L \to L$ for the pointwise join of $f\colon L \to L$ and $g\colon L \to L$, namely $f\sqcup g (x) \defeq f(x) \sqcup g(x)$ for all $x\in L$.  For all natural numbers $n \in \nat$, we define $f^n\colon L \to L$ inductively as $f^0 =id$ and $f^{n+1} = f^n;f$. We fix $f^\omega \defeq \bigsqcup_{n\in \nat} f^n$.

\begin{lemma}\label{lemma:dchaindecomposition}
Let $f,g \colon L \to L$ be monotone maps preserving  chains. Then
\begin{enumerate}
\item $id\colon L \to L$ preserves   chains;
\item $f \sqcup g\colon L \to L$ preserves   chains;
\item $f^\omega \colon L \to L$ preserves   chains.
 \end{enumerate}
\end{lemma}
\begin{proof}
\begin{enumerate}
\item Trivial.
\item By hypothesis we have that
$f(\bigsqcup_{i\in I}x_i) = \bigsqcup_{i \in I}f(x_i)$ and $g(\bigsqcup_{i\in I}x_i) = \bigsqcup_{i \in I}g(x_i)$.
Thus
\begin{align*}
			 f \sqcup g (\bigsqcup_{i\in I}x_i) &= f(\bigsqcup_{i\in I}x_i) \sqcup g(\bigsqcup_{i\in I}x_i)   \\
			&= \bigsqcup_{i \in I}f(x_i) \sqcup \bigsqcup_{i \in I}g(x_i)  \\
			&= \bigsqcup_{i \in I}( f(x_i) \sqcup g(x_i) )  \\
			&=  \bigsqcup_{i \in I}(f\sqcup g)(x_i)
\end{align*}
\item We prove $f^n(\bigsqcup_{i\in I}x_i) = \bigsqcup_{i\in I}f^n(x_i)$ for all $n \in \nat$. We proceed by induction on $n$.

For $n=0$, $f^0(\bigsqcup_{i\in I}x_i) = \bigsqcup_{i\in I}x_i = \bigsqcup_{i\in I}f^0(x_i)$.

For $n+1$, we use the hypothesis that $f$ preserves   chain and thus
\begin{align*}
			f^{n+1}((\bigsqcup_{i\in I}x_i) &= f (\; f^{n+1}((\bigsqcup_{i\in I}x_i)\; )   \\
			&= f (\,\bigsqcup_{i \in I}f^n(x_i) \,) \tag{induction hypothesis}  \\
			&=  \bigsqcup_{i \in I}f (\, f^n(x_i) \,)\\
			&=  \bigsqcup_{i \in I}f^{n+1}(x_i)
\end{align*}
\end{enumerate}

\end{proof}

\begin{lemma}\label{lemma:fomega2}
Let $f,g\colon L \to L$ be monotone maps preserving   chains such that $g\sqsubseteq f$. Then $f^\omega ; g \sqsubseteq f^\omega$
\end{lemma}
\begin{proof}
For all $x\in L$, $f^\omega ; g (x) = g (\bigsqcup_{n\in \nat} f^n(x)) = \bigsqcup_{n \in \nat}g( f^n(x)) \sqsubseteq \bigsqcup_{n \in \nat} f^{n+1}(x) \sqsubseteq \bigsqcup_{n \in \nat} f^n(x) =f^\omega (x)$.
\end{proof}

\begin{lemma}\label{lemma:omegaomega}
Let $f\colon L \to L$ be a monotone map preserving   chains. Thus $f^\omega = f^\omega ; f^\omega$
\end{lemma}
\begin{proof}
$f^\omega = f^\omega ; id \sqsubseteq f^\omega ; f^\omega$.
For the other direction we prove that $f^\omega ; f^n \sqsubseteq f^\omega$ for all $n \in \nat$.  We proceed by induction on $n$. For $n=0$ is trivial. For $n+1$, $f^\omega ; f^{n+1} = f^\omega ; f^ n ; f \sqsubseteq f^\omega ; f \sqsubseteq f^\omega$. For the last inequality we use Lemma \ref{lemma:fomega2}.
\end{proof}

\begin{lemma}\label{lemma:fg}
Let $f,g\colon L \to L$ be monotone maps preserving   chains. Then $(f\sqcup g)^\omega = (f^\omega \sqcup g)^\omega$
\end{lemma}
\begin{proof}
Since $f=f^1 \sqsubseteq f^\omega$ and since $(\cdot)^\omega$ is monotone, it holds that $(f\sqcup g)^\omega \sqsubseteq (f^\omega \sqcup g)^\omega$.

For the other inclusion, we prove that $(f^\omega \sqcup g)^n \sqsubseteq (f\sqcup g)^\omega$ for all $n \in \nat$. We proceed by induction on $n\in \nat$.
For $n=0$, $(f^\omega \sqcup g)^0 = id \sqsubseteq (f\sqcup g)^\omega$.

For $n+1$, observe that $f^\omega \sqsubseteq (f\sqcup g)^\omega $  and than $g\sqsubseteq (f\sqcup g)^\omega$. Thus
\begin{equation}\label{eq:prooflattice}
(f^\omega \sqcup g) \sqsubseteq (f\sqcup g)^\omega
\end{equation}
We conclude with the following derivation.
\begin{align*}
(f^\omega \sqcup g)^{n+1} &= (f^\omega \sqcup g)^{n} ; (f^\omega \sqcup g) \\
& \sqsubseteq  (f\sqcup g)^\omega ; (f^\omega \sqcup g) \tag{Induction Hypothesis}\\
& \sqsubseteq  (f\sqcup g)^\omega ; (f\sqcup g)^\omega   \tag{\eqref{eq:prooflattice}} \\
& =  (f\sqcup g)^\omega \tag{Lemma \ref{lemma:omegaomega}}
\end{align*}
\end{proof}

\subsection{Some well known facts about precongruence closure}\label{app:onprec}

Let $X = \{X[n,m]\}_{n,m\in \nat}$ be a family of sets indexes by pairs of natural numbers $(n,m)\in \nat \times \nat$. A well-typed relation $\mathbb{R}$ is a family of relation $\{R_{n,m}\}_{n,m\in \nat}$ such that each $R_{n,m} \subseteq X[n,m] \times X[n,m]$. We consider the set $\WTRel_X$ of well typed relations over $X$. It is easy to see that $\WTRel_X$ forms a complete lattice with join given by union $\cup$. Hereafter we fix an arbitrary well-typed relation $\mathbb{I}$ and the well-typed identity relation $\Delta$.

We define the following monotone maps for all $\mathbb{R} \in \WTRel_X$:
\begin{itemize}
\item $(id)\colon \WTRel_X \to \WTRel_X$ defined as the identity function;
\item $(\mathbb{I}) \colon \WTRel_X \to \WTRel_X$ defined as the constant function $\mathbb{R}\mapsto\mathbb{I}$;
\item $(r) \colon \WTRel_X \to \WTRel_X$ defined as the constant function $\mathbb{R}\mapsto\Delta$;
\item $(t) \colon \WTRel_X \to \WTRel_X$ defined as $\mathbb{R}\mapsto\{(x,z) \mid \exists y. (x,y)\in \mathbb{R} \wedge (y,z)\in \mathbb{R}\}$;
\item $(s) \colon \WTRel_X \to \WTRel_X$ defined as $\mathbb{R}\mapsto\{(x,y) \mid (y,x)\in \mathbb{R}\}$;
\item $(\seq[]) \colon \WTRel_X \to \WTRel_X$ defined as $\mathbb{R}\mapsto\{(x_1\seq[]y_1, x_2 \seq[] y_2) \mid (x_1,x_2)\in \mathbb{R} \wedge (y_1,y_2)\in \mathbb{R}\}$;
\item $(\tensor[]) \colon \WTRel_X \to \WTRel_X$ defined as $\mathbb{R}\mapsto\{(x_1\tensor[]y_1, x_2 \tensor[] y_2) \mid (x_1,x_2)\in \mathbb{R} \wedge (y_1,y_2)\in \mathbb{R}\}$;
\end{itemize}

Observe that the function $(id)$, $(r)$, $(t)$, $(\seq[])$ and $(\tensor[])$ are exactly the inference rules used in the definition of $\pcong{\cdot}$ given in \eqref{eq:pc}. Indeed the function
$\pcong{\cdot}\colon \WTRel_X \to \WTRel_X$ can be decomposed as
\[\pcong{\cdot} = (\, (id) \cup (r) \cup (t) \cup (\seq[]) \cup (\tensor) \,)^\omega\]
where $f^\omega$ stands the $\omega$-iteration of a map $f$ defined in the standard way (see Appendix \ref{app:continuous} for a definition).

Similarly the congruence closure $\congr{\cdot}\colon \WTRel_X \to \WTRel_X$ can be decomposed as \[\congr{\cdot} = (\, (id) \cup (r) \cup (t) \cup (s) \cup (\seq[]) \cup (\tensor) \,)^\omega\]
These decompositions allow us to prove several facts in a modular way. For instance, to prove that $\pcong{\cdot}$ preserves   chains is enough to prove the following.

\begin{lemma}\label{lemma:componentspreserves}
The monotone maps $(id)$, $(\mathbb{I})$, $(r)$, $(s)$, $(t)$, $(\seq[])$ and $(\tensor[])$ defined above preserve   chains.
\end{lemma}
\begin{proof}
All the proofs are straightforward, we illustrate as an example the one for $(\tensor)$.

Let $I$ be a linearly ordered set and $\{\mathbb{R}_i\}_{i\in I}$ be a family of well-typed relations such that if $i\leq j$, then $R_i\subseteq R_j$.
We need to prove that $(\tensor) (\bigcup_{i\in I}\mathbb{R}_i) = \bigcup_{i \in I}(\tensor)(\mathbb{R}_i)$.

The inclusion $(\tensor) (\bigcup_{i\in I}\mathbb{R}_i) \supseteq \bigcup_{i \in I}(\tensor)(\mathbb{R}_i)$ trivially follows from monotonicity of $(\tensor)$ and the universal property of union. For the inclusion $(\tensor) (\bigcup_{i\in I}\mathbb{R}_i) \subseteq \bigcup_{i \in I}(\tensor)(\mathbb{R}_i)$, we take an arbitrary $(a,b)\in (\tensor) (\bigcup_{i\in I}\mathbb{R}_i)$. By definition of $(\tensor)$, there exist $x_1,x_2,y_1,y_2$ such that
\[a=x_1 \tensor[] y_1 \quad b = x_2 \tensor[] y_2 \quad (x_1,x_2)\in \bigcup_{i\in I}\mathbb{R}_i \quad (y_1,y_2)\in \bigcup_{i\in I}\mathbb{R}_i\]
By definition of union, there exist $i,j\in I$ such that $(x_1,y_1)\in R_i$ and $(x_2,y_2)\in R_j$.
Since $I$ is linearly ordered, there are two cases: either $i\leq j$ or $i \geq j$.

If $i\leq j$, then $R_i \subseteq R_j$ and thus $(x_1,y_1)\in R_j$. By definition of $(\tensor[])$, we have $(x_1 \tensor x_2, y_1 \tensor y_2)\in R_j$ and thus $(a,b)\in R_j$. Since $R_j \subseteq \bigcup_{i\in I}\mathbb{R}_i$, then $(a,b) \in \bigcup_{i\in I}\mathbb{R}_i$. The case for $j\leq i$ is symmetric.
\end{proof}

\begin{proposition}
The monotone maps $\pcong{\cdot},\congr{\cdot}\colon \WTRel_X \to \WTRel_X$ preserve   chains.
\end{proposition}
\begin{proof}
Follows immediately from Lemma \ref{lemma:componentspreserves} and Lemma \ref{lemma:dchaindecomposition} in Appendix \ref{app:continuous}.
\end{proof}

\begin{lemma}\label{lemma:pcJcont}
For all well-typed relations $\mathbb{J}$, the  map $\pcong{\mathbb{J} \cup \cdot}\colon \WTRel_X \to \WTRel_X$ preserves   chains.
\end{lemma}
\begin{proof}
Follows immediately from Lemma \ref{lemma:componentspreserves} and Lemma \ref{lemma:dchaindecomposition} in Appendix \ref{app:continuous}.
\end{proof}

\begin{lemma}\label{lemma:pcongdis}
For all well-typed relations $\mathbb{I}$ and $\mathbb{J}$,
$\pcong{\mathbb{I} \cup \mathbb{J}} =  \pcong{\pcong{\mathbb{I}} \cup \mathbb{J}}$
\end{lemma}
\begin{proof}
Let $(\mathbb{J}) \colon \WTRel_X \to \WTRel_X $ be the constant function to $\mathbb{J}$ and define
$f,g\colon \WTRel_X \to \WTRel_X$ as
\[f \defeq   (id) \cup (r) \cup (t) \cup (\seq[]) \cup (\tensor)  \qquad g \defeq (\mathbb{J})\]
From Lemma \ref{lemma:componentspreserves} and Lemma \ref{lemma:dchaindecomposition}, both $f$ and $g$ preserve   chains.
Observe that $f^\omega(\mathbb{I}) = \pcong{\mathbb{I}}$, that $(f\cup g)^\omega = \pcong{\mathbb{I} \cup\mathbb{J}}$ and that $(f^\omega \cup g)^\omega(\mathbb{I})=\pcong{\pcong{\mathbb{I}} \cup\mathbb{J}}$.  Conclude with Lemma \ref{lemma:fg} in Appendix \ref{app:continuous}.
\end{proof}

\begin{lemma} Let $\mathbb{T}=(\Sigma, \mathbb{I})$ be a first order theory.
Then $\syninclusionT{\mathbb{T}} = \pcong{\mathbb{FOB} \cup \mathbb{I}}$
\end{lemma}
\begin{proof}
By definition  $\syninclusionT{\mathbb{T}} = \pcong{ \syninclusion \cup \mathbb{I}}$. Recall that $\syninclusion = \pcong{\mathbb{FOB}}$. Thus $\syninclusionT{\mathbb{T}} = \pcong{ \pcong{\mathbb{FOB}} \cup \mathbb{I}}$. By Lemma \ref{lemma:pcongdis}, $\syninclusionT{\mathbb{T}} = \pcong{ \mathbb{FOB} \cup \mathbb{I}}$.

\end{proof}

\begin{lemma}\label{lemma:noncont}
Let $I$ be a linearly ordered set and, for all $i \in I$, let $\mathbb{T}_i=(\Sigma, \mathbb{I}_i)$ be first order theories such that if $i\leq j$, then $\mathbb{I}_i \subseteq \mathbb{I}_j$. Let $\mathbb{T}$ be the theory $(\Sigma, \bigcup_{i\in I}\mathbb{I}_i)$.
Then $\syninclusionT{\mathbb{T}} = \bigcup_{i \in I} \syninclusionT{\mathbb{T}_i}$.
\end{lemma}
\begin{proof}
By definition $\syninclusionT{\mathbb{T}} = \pcong{\syninclusion \cup \bigcup_{i\in I}\mathbb{I}_i}$. Since $\mathbb{I}_i $ form a   chain, by Lemma \ref{lemma:pcJcont}, $\pcong{\syninclusion \cup \bigcup_{i\in I}\mathbb{I}_i} = \bigcup_{i \in I} \pcong{\syninclusion \cup \mathbb{I}_i} $. The latter is, by definition, $\bigcup_{i \in I} \syninclusionT{\mathbb{I}_i}$.
\end{proof}

\begin{lemma}\label{lemma:noncontsyntax}
Let $I$ be a linearly ordered set and, for all $i \in I$, let $\mathbb{T}_i=(\Sigma_i, \mathbb{I})$ be first order theories such that if $i\leq j$, then $\Sigma_i \subseteq \Sigma_j$. Let $\mathbb{T}$ be the theory $(\bigcup_{i\in I} \Sigma_i, \mathbb{I})$.
Then $\syninclusionT{\mathbb{T}} = \bigcup_{i \in I} \syninclusionT{\mathbb{T}_i}$.
\end{lemma}
\begin{proof}
By Lemma \ref{lemma:componentspreserves}, the monotone map $\pcongr{\cdot} \defeq (\, (id) \cup (\mathbb{I}) \cup (t) \cup (\seq[]) \cup (\tensor) \,)^\omega$ preserves chains.
Let $\Delta_i$ be the well-typed identity relation on $\LCB[\Sigma_i]$.
Observe that $\syninclusionT{\mathbb{T}_i} = \pcongr{\Delta_i}$ and that $\syninclusionT{\mathbb{T}}= \pcongr{\bigcup_{i\in I} \Delta_i}$.  To summarise:
\begin{align*}
\syninclusionT{\mathbb{T}} & = \pcongr{\bigcup_{i\in I} \Delta_i}\\
&=\bigcup_{i\in I}  \pcongr{\Delta_i} \tag{preserve chains}\\
& = \bigcup_{i\in I}  \syninclusionT{\mathbb{T}_i}
\end{align*}
\end{proof}

\begin{lemma}\label{lemma:noncontalltogether}
Let $I$ be a linearly ordered set and, for all $i \in I$, let $\mathbb{T}_i=(\Sigma_i, \mathbb{I}_i)$ be first order theories such that if $i\leq j$, then $\Sigma_i \subseteq \Sigma_j$ and $\mathbb{I}_i\subseteq\mathbb{I}_j$. Let $\mathbb{T}$ be the theory $(\bigcup_{i\in I} \Sigma_i, \bigcup_{i\in I} \mathbb{I}_i)$.
Then $\syninclusionT{\mathbb{T}} = \bigcup_{i \in I} \syninclusionT{\mathbb{T}_i}$.
\end{lemma}
\begin{proof}
Immediate by Lemma \ref{lemma:noncontsyntax} and Lemma \ref{lemma:noncont}.
\end{proof}

\end{document}